\documentclass[useAMS,usenatbib]{mn2e}
\usepackage{amssymb,epsfig,natbib,rotating,capt-of,longtable}

\title[The death of massive stars]
      {The death of massive stars - I. Observational constraints on the progenitors of type II-P supernovae}
\author[S. J. Smartt {\rm et al.}]
{S.~J. Smartt$^1$, J.J. Eldridge$^{2}$, R.M. Crockett$^1$, J.R. Maund$^{3,4}$\\
\\
$^{1}$Astrophysics Research Centre, School of Mathematics and Physics, Queen's University Belfast, Belfast BT7 1NN, U.K.\\
$^{2}$Institute of Astronomy, The Observatories, University of Cambridge, Madingley Road, Cambridge, CB3 0HA, U.K.\\
$^{3}$Department of Astronomy and McDonald Observatory, University of Texas, 1 University Station, C1400, Austin, TX 78712, U.S.A.\\
$^{4}$Dark Cosmology Centre, Niels Bohr Institute, University of Copenhagen, Juliane Maries Vej 30, 2100 Copenhagen \O, Denmark}
\date{Released 2006 Xxxxx XX}

\pagerange{\pageref{firstpage}--\pageref{lastpage}} \pubyear{2006}

\def\LaTeX{L\kern-.36em\raise.3ex\hbox{a}\kern-.15em
  T\kern-.1667em\lower.7ex\hbox{E}\kern-.125emX}
\def \aj {AJ}
\def \mnras {MNRAS}
\def \apj {ApJ}
\def \apjl {ApJL}
\def \aap {A\&A}
\def \nat {Nature}
\def \araa {ARAA}
\def \iauc {IAUC}
\def \pasp {PASP}

\def \apjs {ApJS}
\def \aaps {A\&AS}
\def\lesssim{\mathrel{\hbox{\rlap{\hbox{\lower4pt\hbox{$\sim$}}}\hbox{$<$}}}}
\def\gtrsim{\mathrel{\hbox{\rlap{\hbox{\lower4pt\hbox{$\sim$}}}\hbox{$>$}}}}

\newcommand{\msun}{\mbox{M$_{\odot}$}}
\newcommand{\msol}{\mbox{M$_{\odot}$}}
\newcommand{\zsol}{\mbox{Z$_{\odot}$}}

\newcommand{\kms}{\mbox{$\rm{\,km\,s^{-1}}$}}
\newcommand{\logl}{\mbox{$\log L/{\rm L_{\odot}}$}}

\newcommand{\nick}{\mbox{$^{56}$Ni}}

\DeclareMathAlphabet{\mathsc}{OT1}{cmr}{m}{sc}
\def\testbx{bx}%
\DeclareRobustCommand{\ion}[2]{%
\relax\ifmmode
\ifx\testbx\f@series
{\mathbf{#1\,\mathsc{#2}}}\else
{\mathrm{#1\,\mathsc{#2}}}\fi
\else\textup{#1\,{\mdseries\textsc{#2}}}%
\fi}

\newcommand{\Hii} {\ion{H}{ii}}

\begin{document}

\label{firstpage}

\maketitle

\begin{abstract}

We present the results of a 10.5\,yr, volume limited (28\,Mpc) search
for supernova (SN) progenitor stars. In doing so we compile all SNe
discovered within this volume (132, of which 27\% are type Ia) 
and determine the relative rates of each
sub-type from literature studies. The core-collapse SNe break down
into 59\% II-P and 29\% Ib/c, with the remainder being IIb (5\%),
IIn(4\%) and II-L (3\%). There have been 20 II-P SNe with high quality
optical or near-IR pre-explosion images that allow a meaningful search
for the progenitor stars. In five cases they are clearly red
supergiants, one case is unconstrained, two fall on compact coeval
star clusters and the other twelve have no progenitor detected. We review
and update all the available data for the host galaxies and SN
environments (distance, metallicity and extinction) and determine
masses and upper mass estimates for these 20 progenitor stars using the STARS
stellar evolutionary code and a single consistent homogeneous method.
A maximum likelihood calculation suggests that the minimum stellar
mass for a type II-P to form is $m_{min}=8.5^{+1}_{-1.5}$\msol\ and
the maximum mass for II-P progenitors is $m_{max}=16.5\pm1.5$\msol, 
assuming a Salpeter initial mass function holds 
for the progenitor population (in the range $\Gamma = -1.35^{+0.3}_{-0.7}$).
The minimum mass is consistent with current estimates for
the upper limit to 
white dwarf progenitor masses, but the maximum mass does not appear
consistent with massive star populations in Local Group galaxies. 
Red supergiants in the Local Group have masses up to 25\msol\
and the minimum mass to produce 
a Wolf-Rayet star in single star evolution (between solar and LMC 
metallicity) is similarly
25-30\msol. The reason we have not detected any high mass
red supergiant progenitors above 17\msol\ is unclear, but
we estimate that it is
statistically significant at 2.4$\sigma$ confidence.  
Two simple reasons for this could be that we have systematically 
underestimated the progenitor masses due to dust extinction or that 
stars between 17-25\msol\ produce other kinds of SNe which are not II-P. 
We discuss these
possibilities and find that neither provides a satisfactory solution. 
We term this
discrepancy the ``red supergiant problem'' and speculate that these stars
could have core masses high enough to form black holes and SNe which
are too faint to have been detected. We compare the $^{56}$Ni
masses ejected in the SNe to the progenitor mass estimates and find
that low luminosity SNe with low $^{56}$Ni production are most likely to arise
from explosions of low mass progenitors near the mass threshold
that can produce a core-collapse.

\end{abstract}


\section{Introduction}

Stars which are born with masses above a critical threshold mass of
around 8\msol\ have long been thought to produce supernovae when their
cores collapse at a point when nuclear burning no longer provides
support against gravity. Supernovae were first suggested to be a new
class of astrophysical phenomena by \cite{1934PNAS...20..254B} and
since then detailed study has allowed them to be split into physical
types : the thermonuclear explosions and core-collapse supernovae
(CCSNe). The CCSNe form when their cores evolve to iron white dwarfs
and detailed stellar evolutionary models predict a minimum mass for
this to occur of between 7-12\msol
\citep{2003ApJ...591..288H,2004MNRAS.353...87E,2007A&A...476..893S,2008ApJ...675..614P}
. The CCSNe form a diverse group in terms of their spectral and
photometric properties and the classification scheme has arisen
primarily based on the appearance of their optical spectra, but also
supplemented with their photometric behaviour
\citep{1997ARA&A..35..309F}. There are the H-rich type II SNe, which
are sub-classified into II-P (plateau lightcurves), II-L (linear
decline lightcurves), IIn (narrow emission lines), and some peculiar
events, generically labelled II-pec. The H-deficient SNe are split
into Ib and Ic depending on whether He is visible and a hybrid class
of IIb (type II events which metamorphose into Ib SNe) has also been
uncovered. The evolutionary stage of the progenitor star (i.e. its
position in the HR diagram and chemical composition of its atmosphere)
very likely dictates what type of CCSN is produced.

Determining what types of stars produce which types of SNe and what 
mass range can support a CCSN is a major goal in modern studies
of these explosions. The first attempts relied primarily on 
linking the results of stellar evolutionary models to the
spectral (and photometric) evolution of the SNe. 
\citep[e.g.][]{1980ApJ...237..541A,1976ApJ...207..872C}
More recently \cite{2003ApJ...582..905H} and \cite{2003MNRAS.346...97N} 
have studied the lightcurves
and ejecta velocities of II-P SNe to estimate the mass of the 
material ejected. These studies tend to favour quite large masses
for progenitor stars, with Hamuy suggesting ranges of 10-50\msol\
and Nadyozhin 10-30\msol. 
However when SN~1987A
exploded a new opportunity arose. The SN was in a galaxy close enough
that its progenitor star could be easily identified. A
blue supergiant star of around 15-20\msol\ was identified and it
is clear that this object no longer exists
\citep{1987Natur.328..318G,1989A&A...219..229W,1992PASP..104..717P}
The 
fact that it was a compact blue supergiant was a key factor in 
enabling the community to understand the event as a whole. 
A progenitor was also detected for the next closest explosion (SN~1993J in
M81) and the binary nature of the progenitor helped understand the
physical reason behind the II-b type given to the event
\citep{1994AJ....107..662A,1993Natur.364..509P}.

Studies of the environments of CCSNe after 
discovery have been ongoing for many years with authors looking
for correlations between the ages of starformation regions and 
the type of SNe that occur.
For example \cite{1992AJ....103.1788V} and \cite{1996AJ....111.2017V}
suggested that there was no clear difference in the spatial 
distributions of type Ib/c and type II SNe compared to 
giant \Hii\ regions in their host galaxies. They concluded that they
hence arose from parent populations of similar mass. 
However recently both \cite{2006A&A...453...57J}
and \cite{2007arXiv0712.0430K} suggest that the stripped type 
Ic SNe are more likely to follow regions of either high surface
brightness or high H$\alpha$ emission than type II SNe. A larger
sample of nearby core-collapse events and H$\alpha$ correlations
has been compiled by \citep{2008MNRAS.390.1527A}
 indicating that there
is a progressive trend for Ib/c SNe to be associated with 
H$\alpha$ emission regions, in the sense that Ic show the closest association
with galactic H$\alpha$ emission, then comes the Ib and then 
type II.  While these efforts are very valuable to discern differences
in progenitor channel, it
is difficult to assign definitive mass ranges to the progenitor systems 
from spatial correlations alone.

A much more direct way to determine the type of star that exploded
is to search directly for progenitors in images of the host
galaxies taken before explosion. The ease of access to large 
telescope data archives makes this search feasible for nearby events. 
There are now a number of groups around the world that are 
competitively searching for progenitor stars in such archive
images. Particularly with the  {\em Hubble Space Telescope} (HST)
it has become possible to resolve massive stellar populations 
out to at least  20Mpc. In this case the information on each
progenitor is much more detailed and quantitative than can be 
achieved with the unresolved environment studies, but there are 
fewer events which allow such a study. 

Early work with HST concentrated on looking at the environments 
of SNe \citep{1996AJ....111.2047B,1999AJ....118.2331V}. But it
has now become possible to search directly for progenitor stars
and carry out concerted campaigns on the nearest events. Three of the first type II SNe to have excellent HST and ground-based 
pre-explosion images were the II-P SNe 1999gi, 1999em and 2001du  
\citep{2001ApJ...556L..29S,2002ApJ...565.1089S,2003MNRAS.343..735S,2003PASP..115..448V}. 
In all there was no detection of a progenitor, but meaningful limits
were derived. \cite{2003ApJ...594..247L} and \cite{2002AJ....124.2490L}
subsequently showed the importance of having reliable distances
to the host galaxies of the SN progenitors in order to constrain the
upper mass limits. 
Efforts to find progenitors continued
\citep{2003MNRAS.343..735S,2003PASP..115....1V} until the 
first confirmation of a red supergiant progenitor of a type 
II-P explosion in prediscovery HST and Gemini-North images
\citep{2004Sci...303..499S}. Unambiguous detections in HST 
images require the SN to be located on the pre-explosion
images with accuracies of around 10 milliarcseconds, which 
requires follow-up images of the SN to be taken at HST
or corrected adaptive optics ground-based resolution
\citep{2005MNRAS.360..288M,2005ApJ...630L..29G,2008MNRAS.391L...5C}
 The next
clear and unambiguous detection of a progenitor star 
was also a red supergiant in NGC5194 
\citep[SN2005cs ; ][]{2005MNRAS.364L..33M,2006ApJ...641.1060L}.
And the recent discovery of SN2008bk  in NGC7793 (3.9\,Mpc) 
has produced detections in $IJHK$ of a red progenitor star
\citep{mattila08bk}. The near-IR spectral energy (SED) distribution of the
SN2008bk progenitor is the best sampled SED yet of any red supergiant
progenitor and matches a late M4I spectral type with moderate extinction
of $A_{V} \simeq 1$ and an initial  mass around 8-9\msol. 
There have been claims of the detections of others such 
as 2004A \citep{2006MNRAS.369.1303H}, 
2004et \citep{2005PASP..117..121L}, 
2006my and 2006ov \citep{2007ApJ...661.1013L}
and we review these in this paper. 
An additional method that has much to offer this field is 
locating SNe directly coincident with compact, coeval star
clusters. If the age of the cluster can be determined then a
main-sequence turn-off age, and turn-off mass can lead directly
to an estimate of the progenitor star mass \cite[e.g. 2004dj in NCG2403
as studied by][]{2004ApJ...615L.113M}. 
The detection and characterisation of progenitor stars has the potential to 
directly link the type of star to the SN explosion characteristics, 
(e.g. the amount of $^{56}$Ni synthesised in explosive O and Si burning
and the total energy of the explosion) and also to set quantitative 
limits on progenitor mass ranges. The latter is of great interest to 
compare to the highest mass white dwarf progenitors and to the 
stars that form neutron stars and black holes after core-collapse. 

Work in this field has progressed substantially in the last 8 years
and compilations of progenitor properties have been made by 
\cite{2003MNRAS.343..735S,2007ApJ...656..372G,2007ApJ...661.1013L,2008arXiv0802.0456K}. 
However these compilations are somewhat ad-hoc, incomplete and potentially
biased as they do not define the selection criterion for inclusion
rigorously. In addition the results are based on different methods for
estimating the progenitor masses and upper limits (in terms of 
measurement and the theoretical models employed). The
goal of this series of two papers is to define the selection criteria
for inclusion (a volume and time limited survey) and to determine the
physical parameters of the progenitor stars (luminosities and masses, or
limits thereon) in a homogeneous and consistent way. Only then is 
it possible to reliably estimate the population parameters.
This paper specifically deals with the type II SNe and all but
two of the final sample of twenty have been confirmed as type II-P. 
A companion paper will discuss the stripped SNe of types IIb, Ib and Ic
which are drawn from the same volume and time limited survey
(Crockett et al. 2008, in prep). 

In addition to the supernovae with pre-explosion images we discuss in the
paper there are three SNe with progenitor detections that fall outside
either our time or volume definition. Those are SN~1987A, SN~1993J 
and SN~2005gl. Discussions of the first two are well documented in 
the literature 
\cite[e.g.][]{1989A&A...219..229W,1994AJ....107..662A,2002PASP..114.1322V,2004Natur.427..129M}; 
1987A will be discussed later in this paper and
the implications of 1993J will be discussed in the 2nd paper in 
this series. SN 2005gl is a type IIn in NGC266 at approximately 66\,Mpc, hence
although a progenitor is detected by \cite{2007ApJ...656..372G}
it does not fall within our distance limit. \cite{2007ApJ...656..372G}
suggest that this was a very massive star with $M_{V} \simeq -10.3$
and very likely a luminous blue variable. Although with only a detection
in a single filter the blue colour is not confirmed, and there is 
not enough data to determine if the source was indeed variable. 
 This is evidence that very 
massive stars do explode as bright SNe 
and is a point we will return to in the discussion. 

This paper starts with defining the sample from which the targets with 
high quality pre-explosion images are drawn. A consistent and homogeneous
analysis method is then defined and justified. We then review previous 
detections (and add some new data) to build the data required for
analysis and then statistically analyse the results. We follow this 
with an extensive discussion.

\section{The sample of Local Universe supernovae}
\label{section:sample}
The observational data for this paper are compiled from many sources in the 
recent literature but the sample selection requires some 
justification and explanation if the later comparisons and 
discussions of physical parameters are to be meaningful. 
We have selected SNe for inclusion based on the
following selection criteria, and we justify the choice of criteria where 
appropriate. 

\subsection{Definition and selection}
\label{section:defintion}

We consider all core collapse SNe discovered in the ten year
period between
1998 January 1 and 2008 June 30. The earlier date was chosen as the
sensible start point for the concerted efforts to find SNe in archive
pre-discovery images due to the fact that the local SN discovery rate
had reached a significant level  \citep{2005PASP..117..773V}, and 
we estimate the amount of imaging of nearby
galaxies in the {\em Hubble Space Telescope} (HST) archive had become rich
enough that coincidences were likely to occur. This was
also the effective date of the start of concerted efforts to search
for SN progenitors. Since the late 1990's our group
\citep{2001ApJ...556L..29S,2002ApJ...565.1089S}, and others 
\citep[e.g.][]{2003PASP..115....1V,2005ApJ...630L..29G} have been
systematically searching for pre-explosion images of core-collapse SNe
in nearby galaxies. We have further restricted our sample to galaxies
with recessional velocities less than 2000 \kms, effectively
restricting us to a volume limited sample. The recessional velocities,
corrected for the infall of the Local Group towards Virgo cluster,
of all nearby galaxies hosting SNe were taken from the 
HyperLEDA\footnote{http://leda.univ-lyon1.fr} \citep{2003A&A...412...45P}
database. We emphasise that we have used the corrected velocities to 
apply the selection criterion of 2000\kms\ and assuming $\mathrm{H_{0} = 72 km\,s^{-1}\,Mpc^{-1}}$, this local volume has a limit of 28\,Mpc.

During this period, and in this volume, there have been 138 SNe candidates
discovered and all are listed in the
Asiago\footnote{http://web.pd.astro.it/supern/snean.txt}, 
CfA\footnote{http://cfa-www.harvard.edu/cfa/ps/lists/Supernovae.html} and 
Sternberg Astronomical Institute 
\footnote{http://www.sai.msu.su/sn/sncat/} catalogues. 
Of course these are simply compilations of 
discoveries reported in the {\em International Astronomical Union}
(IAU) Circulars and these catalogues normally list 
the discovery magnitudes and types reported in the first IAU
announcement. 
It often happens that the SN classification is revised or refined 
with 
subsequent higher quality spectra, or longer monitoring. Both 
can reveal peculiarities and transformations, or simply
give a more secure classification.
In particular the sub-classification of 
II-P can be added when significant light curve information is gathered.
Hence we have carefully checked the classification of each event, and
have gone further and classified those supernovae listed as ``II''
into the subtypes II-P and II-L or IIn where possible. This was done
using the following criteria in order. Firstly the refereed literature
was searched and a classification taken from published photometric and
spectroscopic results. Secondly unpublished, professional spectra and
lightcurves where taken from reliable sources such as those of the
Carnegie Supernova Project \citep{2006PASP..118....2H} and the 
Asiago data archive \citep{2000MmSAI..71..573T}. Thirdly amateur
lightcurves available on the web were checked and, if possible, a II-P
classification was made if a clear and unambiguous plateau lasting
longer than 30 days was recorded. The vast majority of II-P SNe have 
plateau phases lasting significantly longer than 30 days, in fact most
are around 90-110 days \citep{2003PhDTPastorello,2003ApJ...582..905H,2004MNRAS.347...74P} and there is no clear evidence
that plateaus of shorter duration are particularly common. 
However to observe such a long plateau necessarily means the SN must
have been discovered close to explosion. This is often not the case, 
and we have chosen to take 30 days simply as an indicator that an
extended plateau phase is evident. 
In all cases of our
subclassifications of type II SNe, we believe the designations not to
be controversial or ambiguous and for only 9 events classed
as type II were we unable to assign a subtype. In these cases the
supernovae were generally discovered late in the nebular phase. It is
often difficult to distinguish Ib and Ic SNe from single spectra taken
at an unknown epoch, and indeed there are 6 events for which authors
have listed Ib/c classifications and we cannot improve on these
classifications.  All of the individual SNe are listed in
Table\,\ref{table:fullsample}. There are only two SNe 
which have not had a classification spectrum
reported in the literature, 1998cf and 1999gs 
and these are ignored in the following frequency comparison. 

We realise that the classifications into the standard bins is somewhat
simplistic. In particular there are some SNe that show evidence of
interaction with the circumstellar medium, which result in narrow
lines (usually of H or He) superimposed on the spectrum. When narrow
lines of H dominate the spectrum then the IIn designation is often
used by the community, but some II-P and Ibc SNe do show evidence of
this behaviour at a weaker level. 
The most striking example recently is that of the
SN2006jc-like objects and the two examples in our
sample are 2006jc and 2002ao \citep[see][]{2007Natur.447..829P,2007ApJ...657L.105F,2008arXiv0801.2277P}. 
These show broad lined spectra resembling type Ic SNe (in that
they do not exhibit H or He {\em in the high velocity ejecta}), but 
have strong and narrow He emission lines and weak 
H$\alpha$ emission. This has led \cite{2008arXiv0801.2277P}
to term this class of objects Ibn. However rather than introducing
this small and very specific class of events, we will class them 
as Ic SNe as this is a fair description of the underlying 
spectrum of the SN ejecta. Giving them a simple Ib label 
could be argued as being misleading in that they do not exhibit the 
broad, He absorption typical of this class of H-deficient events. 
The label Ibn is certainly a valid type for them but in this 
paper it is too specific to be a useful addition to the 
compiled subtypes. 
We will discuss these objects further in the second paper in this
series which concerns the stripped events (Crockett et al. in prep.).

\subsection{The relative frequencies of core-collapse SNe}
\label{section:rates}

In Table\,\ref{table:rates} we list the relative frequency of each
subtype occurring in our sample. This is a volume limited relative
frequency rate of SN types. There may well have been undiscovered
local SNe in this period e.g. dust extinguished events, or events
which exploded in solar conjunction which were missed at late
times. The distance limit imposed ($\mu$=32.3) and the
range in the absolute magnitudes of each subtype
\citep{2002AJ....123..745R} would initially suggest that 
it is unlikely that there is a serious
bias in the {\em relative number} of the different subtypes. 
 However
there are two arguments which can be put forward against this. 
The first is if there is a substantial number of intrinsically
faint SNe that may have gone undetected if they have typical
magnitudes below about $-13$. The nature of the faint optical
transient in M85 \citep{2007Natur.447..458K,2007arXiv0710.3192O} is
currently debated, and could conceivably be a core-collapse SN
\citep[][see Sect. 2.4]{2007Natur.449....1P}. 
  If a large number of intrinsically very
faint SNe are evading discovery by current pointed surveys it could
lead to a dramatic change in our understanding of the link between
progenitor star and SN explosion. 
Secondly there is an issue with the lack of type Ia
SNe discovered recently within 10\,Mpc. In the 10.5\,yr considered
there are about 13 core-collapse that have been discovered within 
10\,Mpc \cite[the exact number depends on some individual galaxy distance
estimates and how strictly one enforces the distance limit;
for a more in depth discussion see][]{2008arXiv0810.1959K}. 
But there have been no type Ia SNe discovered and with a 
relative frequency of 27\%, one might have expected to have seen
3-4 SNe. The Poisson probability of this being a statistical 
fluctuation is not zero, but is small (2-5\%) and one could invoke
the argument that there are more core-collapse SNe (perhaps of the
fainter type II)  beyond 10\,Mpc which are being missed
and hence the relative rate of CCSNe/Ia is intrinsically much
higher than we currently believe \cite[also see][]{2008arXiv0809.0510T}.

\cite{2005PASP..117..773V} have presented a homogeneous
sample of 604 recent SNe discovered (or recovered) by the Lick
Observatory Supernova Search (LOSS) with the KAIT telescope
\citep{2001ASPC..246..121F}. 
The
galaxy search sample spans a much larger volume ($cz \lesssim 10\,000$
\kms) than we are considering and a significant majority of our sample in
the overlapping time frames ($\sim$85 per cent of the core collapse
events between 1999-2004) are listed in the van den Bergh et
al. summary of the LOSS survey.  The ones which are not are
predominantly more southern than $\delta = -30^{\circ}$. Hence our
sample is very similar to that which would be obtained if one selected
a distance and time limited sample from the discovered and recovered
events of \cite{2005PASP..117..773V}.  The relative number of type Ia
SNe in the full LOSS catalogue is significantly higher than within our
smaller volume (44 per cent compared to 27 per cent that we find
here). This is very likely due to type II-P SNe going undetected at
the largest distances. At $cz \sim 10\,000$ \kms, with a moderate amount
of foreground reddening at least half of the II-P distribution of
\cite{2002AJ....123..745R} would be missed at the limiting magnitude
of $R\sim$19 of the KAIT survey. The type Ib/c SNe appear slightly
more abundant in our local sample (29 per cent of all core-collapse)
compared to the \cite{2005PASP..117..773V} frequency (25 per cent),
although the difference is not significantly greater than the expected
Poisson scatter.

\begin{table}
\caption[]{The relative frequency of SNe types discovered between 1998-2008 (10.5 yrs) in galaxies 
with recessional velocities less than 2000 kms$^{-1}$, and type taken from 
Table\,\ref{table:fullsample}.
The relative frequency of all types and the relative frequency of only core-collapse SNe are 
listed separately. }
\label{table:rates}
\begin{tabular}{rlll}
\hline\hline
                &    & Relative   &  Core-Collapse only \\\hline 
Type                & No.   & / per cent  &  / per cent\\\hline 
II-P                 &  54  &  	39.1     &  58.7	   \\
II-L                 &  2.5 &	1.8	 &   2.7	\\
IIn                  &  3.5 &   2.5      &   3.8       \\
IIb                  &  5   &	3.6	 &   5.4	\\
Ib                   &  9   &   6.5      &   9.8    \\
Ic                   & 18   &   13.0     &  19.6     \\
Ia             	     & 37   &	26.8	 &  ...       \\
LBVs 	     	     &  7   &	5.1	 &  ...         \\
Unclassified         &  2   &   1.4      &  ...  \\\hline 
Total		     & 138  &	100      &  100      \\		
Total CCSNe          & 92   &	66  	 &  100    \\\hline\hline
\end{tabular}
\end{table}

The 9 SNe which were classed as type II (and could not be further sub-classified) can be
split proportionally over the type II subtypes, which assumes that
that there was no particular bias underlying their poor observational
coverage.  As they were all discovered late in the nebular phase, this
is likely to hold. We put 8 in the II-P bin, and split the other
1 equally between the IIn and II-L, as they have equal numbers of 
confirmed types; hence the fraction which appears in table \ref{table:rates}. 
In a similar manner the 6 Ib/c SNe are split
proportionately into the Ib and Ic bins based on the measured ratio of
Ib:Ic = 7:14 (which comes from those events where a Ib or Ic classification
seems secure). 
We did not include the 2 unclassified SNe in any of the 
rate estimates.
Seven of the events originally 
announced as SNe in Table \ref{table:fullsample} have been shown to actually
be outbursts of luminous blue variables (LBVs) similar to those seen 
historically in Local Group LBVs such as $\eta$-Carinae and P-Cygni. 
These are 
1999bw \citep{1999IAUC.7152....2F},
2000ch \citep{2004PASP..116..326W},
2001ac \citep{2001IAUC.7597....3M},
2002kg or NGC2403-V37 \citep{2005A&A...429L..13W,2006MNRAS.369..390M,2006astro.ph..3025V},
2003gm \citep{2006MNRAS.369..390M}, 2006fp \citep{2006CBET..636....1B},
2007sv \citep{2007CBET.1184....1H}. 
Hence we remove these from the 
rates of CCSNe since they are not true SN explosions. 
Table \ref{table:rates} lists the relative frequencies of core-collapse SNe. 
It is clear that the 
types II-L and IIn are intrinsically quite rare and the
majority of core-collapse events are SNe II-P. Such a breakdown
of subtypes has been suggested before 
\citep{2007ApJ...661.1013L,1999A&A...351..459C} although this is
the first time quantitative volume limited statistics have been 
compiled and presented. The preliminary analysis by 
\cite{2007ApJ...661.1013L} of 68 LOSS only discovered events
(within 30Mpc) in 9 years suggests 68:26:2:4 per cent breakdown
between II:Ib/c:IIb:IIn. Perhaps the ratio of most interest is the 
Ibc/II ratio which has been used by previous studies to try to 
place constraints on progenitor populations
\citep{2008ApJ...673..999P,2003A&A...406..259P,2007RMxAC..30...35E}. 
These three studies have estimated the ratio as a function of 
metallicity finding that, at approximately solar metallicity (\zsol),
the ratio is $N_{\rm Ibc}/N_{\rm II} \simeq 0.4 \pm 0.1$ (Poissonian uncertainty)
and this goes 
down to around 0.1 at 0.3\zsol\ (although with fairly small numbers in each metallicity bin). 
The  ratios at approximately \zsol\  are  fairly similar to what we find 
($N_{\rm Ib/c}/N_{\rm II} = 0.45 \pm 0.13$) and as discussed below in 
Sect.\ref{abundances} and Sect.\ref{section:masses}
our SN population is likely drawn from metallicities in the range
0.5-1.0\zsol\ due to the fact that nearby, high starformation rate 
galaxies are those that are most frequently monitored for SNe. 
A full analysis of the chemical composition of the sites of the SNe in 
this volume limited sample would be desirable. We will discuss the 
SN rates more in Sect.\,\ref{II-P:mass-max}. 

The absolute rates of the different types in the standard supernova units 
(1SNu = 1 SN(100yr)$^{-1}$(10$^{10}L^B_{\odot}$)$^{-1}$) 
is much more difficult to assess given that detailed knowledge of 
the sampling frequency for each galaxy and search strategy is required
\citep[see][]{1999A&A...351..459C}. The LOSS team will
address this \citep{2004AAS...205.7102L} and 
will provide the best estimate of the local rates so far. While we cannot derive the 
SN rate in standard units, 
the numbers in Table\,\ref{table:rates} serve as a 
good guide to the relative numbers of SNe expected in future surveys,
when used in conjunction with absolute magnitude distributions
(assuming the local galaxy population is cosmically representative). 
If bias factors affecting the relative rate
of discovery of the different types are minor, then these rates
are a direct consequence of the 
initial mass function combined with stellar evolution which is dependent
on mass, metallicity, duplicity and initial rotation rate. 
The numbers of SNe we have tabulated give a lower limit on the number
of types per Gpc$^{-3}$yr$^{-1}$, which is at least a useful comparison to 
higher redshift estimates and also when considering the rates of
Gamma-Ray Bursts (GRBs) and X-ray Flashes (XRFs). The volume enclosed by the 28\,Mpc limit is 
$9.2\times10^{-5}$\,Gpc$^{3}$; hence the local rate of CCSN explosions
is likely  to be $\geq 9.6\times10^{4}$\,Gpc$^{-3}$yr$^{-1}$ and the 
local Ibc SN rate is 
$\geq 2.7\times10^{4}$\,Gpc$^{-3}$yr$^{-1}$. The latter is in reasonable
agreement with the $\sim 2\times10^{4}$\,Gpc$^{-3}$yr$^{-1}$
put forward by \cite{2007ApJ...657L..73G}, based on 
the \cite{1999A&A...351..459C} rates and local galaxy luminosity functions.

One further point to bear in mind when considering the relative rates
is the likely number of local SNe which are not discovered  because
they are in faint hosts which are not monitored. The Sloan 
Digital Sky Survey Data Release 5 (5713 sq degrees) contains
about 2200 faint galaxies with $M_{g} \gtrsim -17$ and 
with recessional velocities less than 2000\kms. Hence over the full
40,000 square degrees of sky, there are about 15,000 of these faint hosts
within about 28\,Mpc. These are generally not monitored by the LOSS
and the amateur efforts, who typically target the most luminous 10,000
galaxies within about 100-140\,Mpc. \cite{2008A&A...489..359Y}
estimate that such galaxies (with metallicities corresponding
roughly to oxygen abundance $<8.4$ dex) would contribute about 5-20\% of
the total star formation locally. Hence at least this fraction of 
core-collapse SNe are missing from the local samples and all of them
are within faint hosts and low metallicity progenitors. 
This could mean that very bright events like SNe 2005ap,
 2006tf and 2008es 
\citep{2007ApJ...668L..99Q,2008ApJ...686..467S,2009ApJ...690.1313G,2009ApJ...690.1303M}
found in blank-field  searches (and faint hosts) could
be missed. Although the true rate of such events appears
to be quite small and likely less than $\sim$1\% 
\citep{2009ApJ...690.1303M}. 
Discovery of these events locally should be  possible with 
future all-sky surveys such as Pan-STARRS and LSST 
\citep[for an estimate of rates see][]{2008A&A...489..359Y}.

Of the 100 core-collapse SNe and LBV classified outbursts, the host galaxies 
of 46 of them were
imaged by HST with either the Wide-Field-Planetary-Camera 2 
(WFPC2) or Advanced Camera for Surveys (ACS) {\em before} explosion. 
However given the small field of view of both of these cameras
(2.6$'$ and 3$'$ respectively) the site of the SNe did not always 
fall on the field-of-view (FOV) of the cameras. Of these 46, only
26 had images of the SN site in the camera FOV, an overall
hit rate of 26 per cent. 
A column is included in Table\,\ref{table:fullsample} which 
specifies whether or not the galaxy was observed by HST before 
explosion and if the SN falls on one of the camera FOVs. 
A further 6 have had high quality ground-based images of the
SN site taken before explosion and they are also included in this compilation. 
The observational sample for this paper, and its companion 
studying the stripped events (Crockett et al. 2008, in prep.) 
is thus all of the core-collapse SNe which fulfill the above criteria and have good quality pre-explosion imagery. 
 We have confirmed that there are no other SNe in 
Table\,\ref{table:fullsample} with HST pre-discovery images. We cannot make the 
same definitive statement about ground-based images given the 
amount of inhomogeneous imaging data around. But our manual searching 
of all well maintained large telescope archives suggests it is 
highly unlikely that further {\em high quality} images of any of these events
will surface i.e. images with sub-arcsec resolution with the depth to 
detect a large fraction of the galaxy's massive stellar population. 
As such we have a well defined sample in terms of distance and 
time. The rest of this paper focuses on the progenitor properties of the 
20 type II SNe listed in Table\,\ref{table:main}, of which 18 are
confirmed II-P and two are of uncertain subtype (1999an and 2003ie). 
The other twelve
supernovae which are likely to have had stripped progenitors:
 2000ds (Ib), 2000ew  (Ic), 2001B  (Ib), 2001ci  (Ic), 2002ap  
(Ic), 2003jg  (Ib/c), 2004gt  (Ib/c), 2005ae  (IIb), 2005V  (Ibc), 
2005cz  (Ib), 2007gr (Ic), 2008ax (IIb), 
will be discussed in a companion analysis paper (Crockett et al.
in prep.). 

\subsection{The relative frequencies of SN1987A-like events}
\label{section:87Arates}
In this volume limited sample, there is only one SN which has been 
conclusively shown to be similar to SN1987A, that is 
SN1998A \citep{2005MNRAS.360..950P}. 
SN1987A had a peculiar lightcurve and distinctly strong
Ba\,{\sc ii} lines (probably a temperature effect) 
and an asymmetric H$\alpha$ profile 
during its first $\sim$40 days of evolution and 
the community would have been very unlikely to miss such
events as they would have created great interest. There is 
one other relatively nearby event that has a SN1987A-like
appearance, which is 2000cb 
\citep{2001PhDTHamuy} in IC1158. This one 
however has a $V_{\rm vir} = 2017$\,\kms, which puts it just 
beyond our selection criteria. 
We shall see in Sect.\ref{subsect:2003ie} that 
\cite{2008arXiv0804.1939H} suggest that a single spectrum of
SN2003ie shows similarities to 1987A but it is not well studied
enough to be definitive. Hence even if we would include 2000cb
and 2003ie in our sample we can certainly say that 1987A events are
intrinsically rare
and probably less than around 3\% of all core-collapse events.

\subsection{SN2008S and the M85 and NGC300 optical transients}
\label{section:08S}
Recently three optical transients have been reported whose nature is
still ambiguous and intensely debated. The optical transient in M85
reported by \cite{2007Natur.447..458K} was suggested by the authors to
be a ``luminous red nova'' which most likely arose from a stellar
merger. However this view was challenged by \cite{2007Natur.449....1P}
who suggested a CCSN origin could not be ruled out.  Since then two
other optical transients of similar absolute magnitude have been
discovered. One has been termed a supernova
\citep[SN2008S;][]{2008CBET.1235....1S} although
\cite{2008arXiv0811.3929S}
suggest it could be a supernova imposter and the outburst of a 
moderately massive star rather than a core-collapse. 
The other, in NGC300 \citep{2009arXiv0901.0198B,2009arXiv0901.0710B},
has not yet received an official supernova designation, hence we 
refer to it as NGC300\,OT2008-1 \citep[as in][]{2009arXiv0901.0710B}
SN2008S has already been subject to a study of its
pre-explosion environment and a detection of a source in Spitzer
mid-IR images has been suggested to be a dust enshrouded red
supergiant which is visually obscured \citep{2008arXiv0803.0324P}.  A
similar dust dominated object has been found to be coincident with the
optical transient NGC300\,OT2008-1 \citep{2008arXiv0809.0510T,2009arXiv0901.0198B}.
\cite{2008arXiv0809.0510T} suggested that all three could be the
similar explosion of massive stars embedded in optically thick dust
shells. The early studies of the evolution of 
SN2008S, NGC300\,OT2008-1 and their comparisons
with M85OT2006-1 and other erupting systems have so far 
not favoured a core-collapse supernova explanation for the physical source of 
the outburst 
\citep{2008arXiv0811.3929S,2009arXiv0901.0198B,2009arXiv0901.0710B} 
The transients lack broad lines 
from high velocity ejecta; their spectra are very slowly evolving and
dominated by narrow H-emission.  Strangely they also don't appear to
be similar to the V838 Mon variable system or M31 luminous red
variable as initially suggested by both \citet{2007Natur.447..458K}
and \citet{2008IAUC.8946....2B}. 
Based on the mid-IR progenitor detections, 
\cite{2008arXiv0809.0510T} argue that the precursors may have been
going through a short evolutionary phase which ends in a weak, 
electron-capture supernova. A full multi-wavelength study of the evolution of 
SN2008S from early to late times, and comparisons with the other two
 suggest there is some evidence for the supernova
explanation (Botticella et al., in prep.; Kotak
et al. in prep).  All these studies reveal that the three objects are
incredibly similar in their properties. As  their nature is ambiguous and
currently debatable, we will not consider them further in this paper. It is certain,
however, that they are not normally type II-P SNe.

\section{The stellar evolutionary models}
\label{section:models}

\begin{figure}
    \centering
    \epsfig{file = 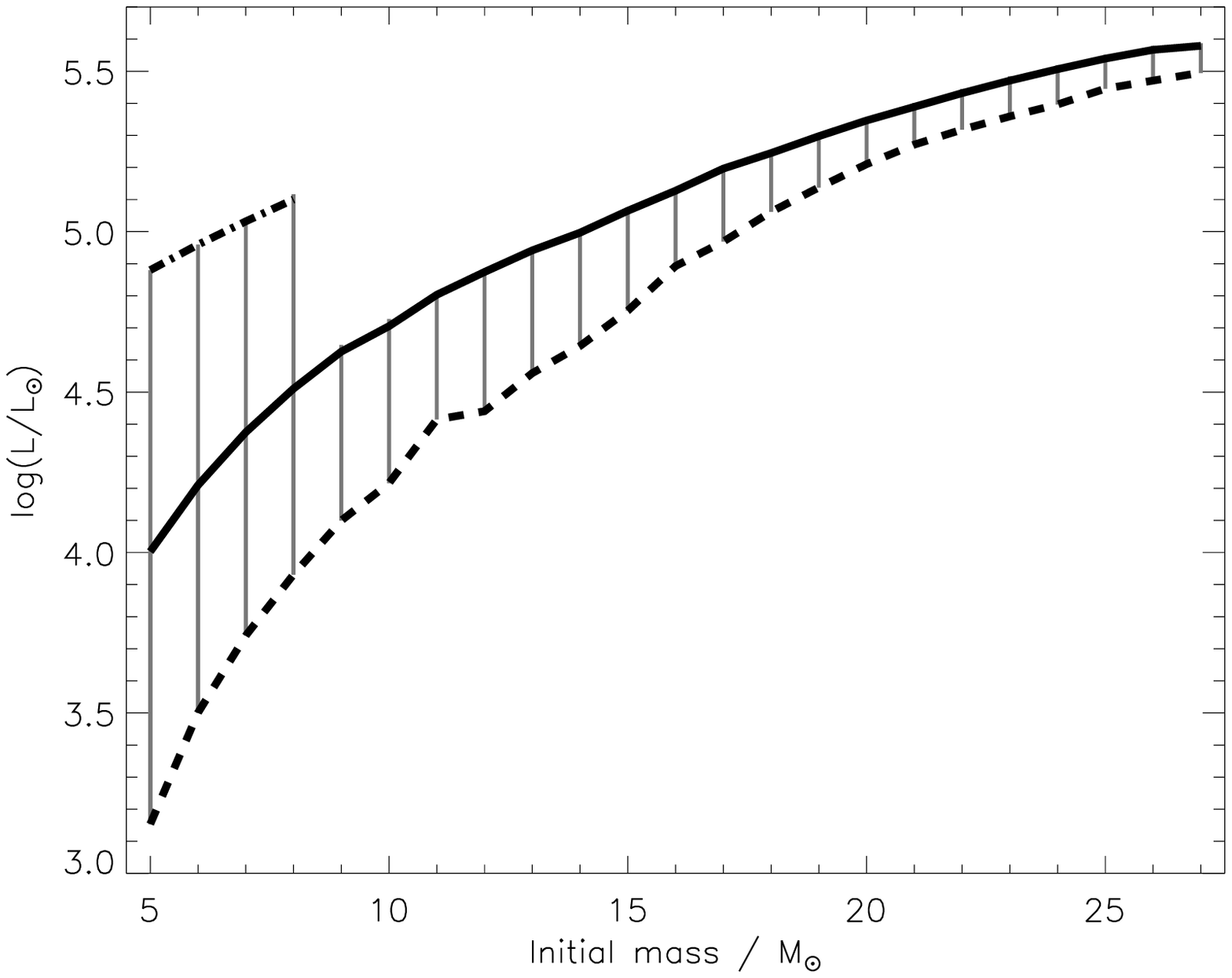, width=80mm}
     \epsfig{file = 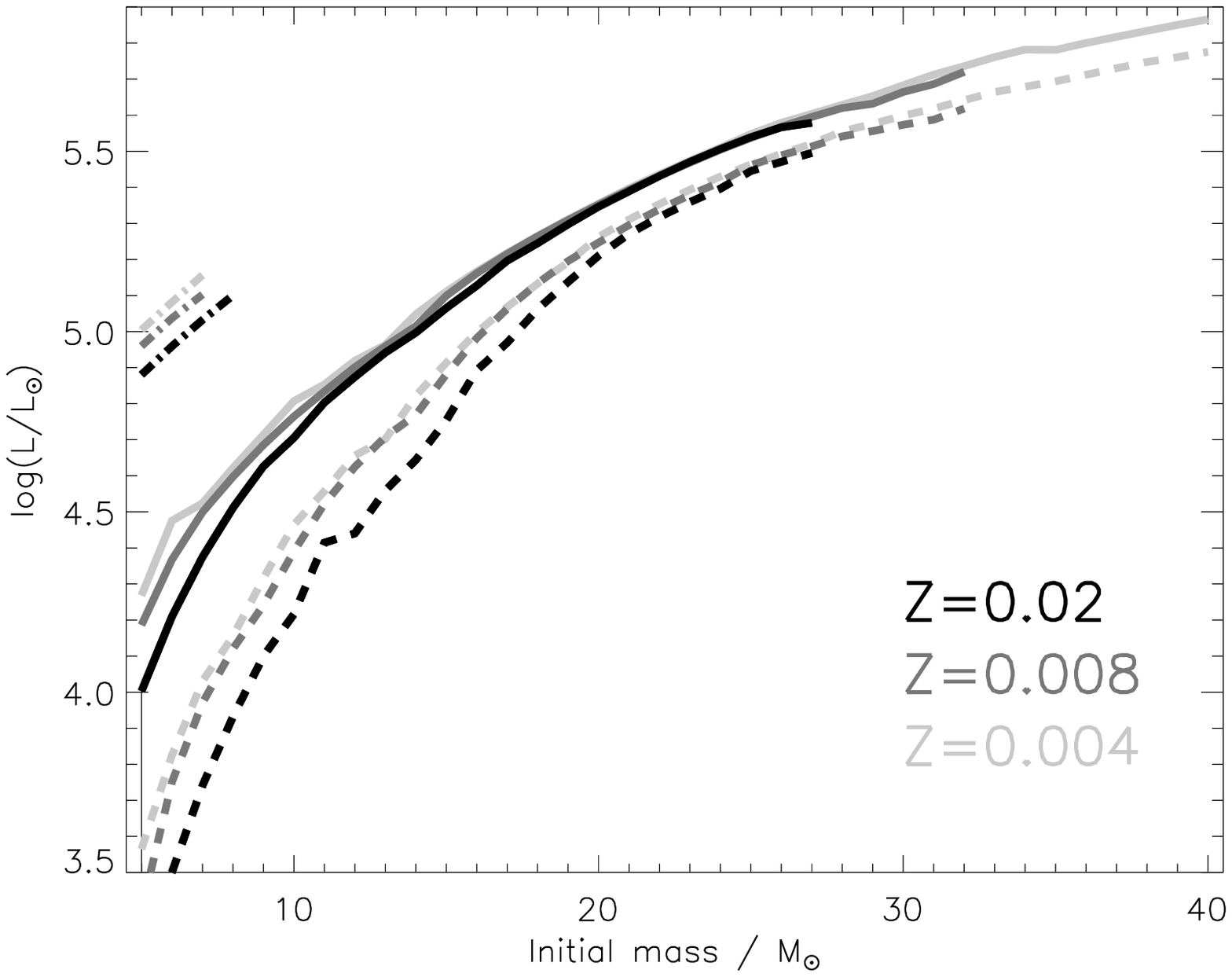, width=80mm}
    \caption[]{{\bf (a)}: 
The initial mass compared with the final luminosity of the stellar models for $Z=0.02$. 
Each mass has the luminosity range corresponding to the end of He-burning and the end of the 
model, just before core-collapse (these are the thin grey vertical lines). From a limit of luminosity an upper limit to the initial mass can be determined. The solid line is the luminosity of the model end-point, the dashed line the luminosity at the end of core helium burning and the dash-dotted line is the luminosity after second dredge-up when the lower mass stars become AGB stars. {\bf (b)}: The same as (a) but with three metallicities shown for comparison, and with the 
vertical joining bars omitted for clarity. This illustrates that the choice of metallicity for
the tracks is not critical, but we do use the most appropriate track to remove any systematic
error.}
\label{fig:mass-logL}
\end{figure}

As discussed above, observational and theoretical studies both now 
strongly suggest that the progenitors of type II-P are typically 
red supergiants. To estimate an
initial mass for observed red supergiant progenitors we require
stellar models to obtain a theoretical initial mass to final
luminosity relation, as shown in Figure \ref{fig:mass-logL}. The
stellar models we use were produced with the Cambridge stellar
evolution code, STARS, originally developed by \citet{E71} and updated
most recently by \citet{P95} and \citet{E03}. Further details can be
found at the code's web
pages\footnote{http://www.ast.cam.ac.uk/$\sim$stars}. The
models are available from the same location for download without
restriction. The models are the same as those described in
\citet{E03} but here we use every integer initial mass from $5$
to $40{\rm M}_{\odot}$ and integer steps of 5-10\msol\ above.

As will be discussed in Sect.\,\ref{abundances}, we can estimate the
metallicity of the exploding star from the nebular abundances in the
disks of the host galaxies, hence we have calculated stellar
evolutionary models for three metallicities; solar, LMC and SMC where
we assume mass fractions of Z=0.02, 0.008 and 0.004 respectively. All
the models employ our standard mass-loss prescription for
hydrogen-rich stars \citep{2004MNRAS.353...87E}: we use the rates of
\citet{dJ} except for OB stars, for which we use the theoretical rates
of \citet{VKL2001}.


In Figure \ref{fig:mass-logL} we plot the range of luminosity for a
star from the end of core helium burning to the model end point at the
beginning of core neon burning (for a solar 
metallicity model). The beginning of core neon burning is only a few years before
core-collapse and this point is likely to be
an accurate estimate of the pre-SN luminosity. 
The estimate of final mass from the observational limits will depend
on uncertainties in these stellar models, which is a systematic that is 
difficult to constrain. To allow for this we assume that the range of
reasonable luminosities for progenitor stars is somewhere between the 
end of core helium burning (dashed line in Fig.\,1a) 
to the model end point at the beginning of core neon burning 
(solid line in Fig.\,1a). For the lower mass stars that
undergo the process known as second dredge-up to become AGB stars, we
also consider the luminosity before second dredge-up occurs. After second
dredge-up the models have much higher bolometric luminosities but their
observable characteristics are quite different to red supergiants
\citep{2007MNRAS.376L..52E}. We have previously shown that in the case of SN2005cs
the progenitor could not have been a super-AGB star in the 5-8\msol\
range. Hence we assume throughout this paper that such stars are not the 
progenitors of the SNe discussed, and their positions in 
Fig.\,\ref{fig:mass-logL} are shown for completeness. A full discussion
of this is given in \citep{2007MNRAS.376L..52E}. In Fig.\,\ref{fig:mass-logL}b we
show the final luminosity ranges for the three metallicities, and the
most appropriate metallicity for each SNe is used when the initial 
masses are calculated in Sect.\,\ref{section:masses}. 

In Sect.\,5, when we estimate a progenitor initial mass from Figure
\ref{fig:mass-logL},  we will assume that
the models are reliable enough to predict correctly that SNe will
undergo core-collapse after helium burning, so we use the full range
of luminosities between the start of helium burning and the model endpoints in a
conservative way. The mass estimate should be reasonable for all cases
where the progenitor is a red supergiant.

\section{Metallicities of the progenitor stars}
\label{abundances}
As the stellar evolutionary tracks do differ slightly, to make sure that
there are no underlying systematics in our analysis we require an estimate
of the initial metallicity of the exploding star. There is good observational evidence
now to show that mass-loss from massive stars is metallicity dependent, 
and that the the lifetimes of stars in various phases as they evolve
depend on metallicity \citep[e.g.][]{2003ARA&A..41...15M,2007A&A...473..603M}.
Models predict that mass-loss and
metallicity are driving forces behind stellar evolution
\citep{2003ApJ...591..288H,MM03,2004MNRAS.353...87E}.

The most reliable determination of the metallicity of the progenitor
star would be a measurement of the interstellar medium abundance in
the galactic disk at the position of the event. In some cases the SNe
have exploded in, or very close to, a previous catalogued \Hii\
region which has published spectroscopy and emission line fluxes. For
these events we use these fluxes and calculate the nebular abundance
of oxygen, using the strong-line method of the ratio of the
[O\,II]\,$\lambda\lambda$\,3727 plus
[O\,III]\,$\lambda\lambda$\,4959,5007 to H\,$\beta$ \citep[see][for a
discussion]{2006astro.ph..8410B}.  Recently there has been much debate
in the literature over which calibration to use to determine the
nebular oxygen abundances from this method.
\cite{2006astro.ph..8410B} has shown that the calibration of
\cite{2005ApJ...631..231P} best matches the abundances determined in
nearby galaxies where it is possible to measure the strength of the
electron temperature sensitive lines and hence determine a simple
empirical calibrations for the strong-line method. The result is that
the empirical determinations of \cite{2004ApJ...615..228B} and the
calibrations of \cite{2005ApJ...631..231P} and
\cite{2004MNRAS.348L..59P} give significantly lower abundances (by a
factor 0.3-0.4\,dex) than photoionisation models \citep[of][ for
example]{2002ApJS..142...35K}.  \cite{2002A&A...395..519T} have shown
that photospheric abundances of massive stars in M31 (B-type
supergiants) are in much better agreement with the ``lower''
metallicity scales of the $P$ method employed by
\cite{2005ApJ...631..231P}, and the empirical determinations form
auroral lines. 
The recent downwards revision of the solar oxygen abundance to $8.66\pm0.05$
\citep{2004A&A...417..751A,2005ASPC..336...25A} and the agreement with all the other 
estimators of the Milky Way's 
ISM abundance at the solar radius (B-stars, young F\& G stars, 
H\,{\sc ii} regions, diffuse ISM) would also suggest that our adopted, lower,  scale is appropriate
\citep{2006A&A...448..351S, 2004ApJ...604..362D, 2001ApJ...554L.221S}. 
Hence in this paper we will favour 
the calibrations of \cite{2004ApJ...615..228B} and
\cite{2005ApJ...631..231P} to determine nebular abundances.

\cite{2008AJ....135.1136M} have investigated the metallicities at the sites
of type Ic SNe and GRB related SNe and show the importance of 
comparing abundances derived by self-consistent methods. By 
employing consistent abundance indicators they find that GRB related
SNe tend to have significantly lower metallicities within their host
galaxy environments than broad lined type Ic SNe without GRBs. Their
study illustrates the need to adopt consistent methods and compare
abundances differentially. 

We chose which stellar tracks to estimate the initial masses as
follows. We use the observed present day oxygen abundances as compiled
by \cite{2007A&A...466..277H} for the Sun, LMC and SMC (8.65, 8.35,
8.05) to guide our choice of model. For those SNe which have estimated
ISM oxygen abundances of ${\rm [O/H]} \ge 8.4$ we choose to use the
$Z=0.02$ metallicity tracks (solar).  For those in the range $8.2 \le
{\rm [O/H]} < 8.4$ we use the $Z=0.008$ tracks (LMC). We would have
used the $Z=0.004$ tracks for anything below ${\rm [O/H]} < 8.2$
(SMC), but none of our targets have such low metallicity.

\begin{table*}
\caption[]{The results of the homogeneous reanalysis of the all the SN progenitors. 
The galaxies, their class, distance and extinction along the line of sight to the 
SNe are listed. The methods employed in the literature to determine distances are 
listed (TF = Tully Fisher; Kin. = Kinematic; Cep. = Cepheid; PNLF = Planetary Nebulae
Luminosity Function; TRGB = Tip of the Red Giant Branch; Mean = mean of several methods which 
are detailed in Section 5). Kinematic distances are based on $\mathrm{H_{0} = 72 km\,s^{-1}\,Mpc^{-1}}$. 
The de-projected galactocentric radii are calculated as
well as the radius with respect to the $r_{25}$ value (the radius at which the surface brightness drops to 25
mag arcsec$^{-1}$). Oxygen abundances of the galactic ISM at the positions of the SNe are quoted  
($[O/H] = 12 + \log N_{\rm O}/N_{\rm H}$). The final estimated luminosities, or luminosity limits
are in solar luminosity units. 
The zero-age main-sequence (ZAMS) masses and upper mass limits as
discussed in Sect.\ref{section:masses} are listed in the final column. 
\label{table:main}}
\begin{tabular}{rlllllllllll}\hline 
Supernova & SN      & Galaxy     & Galaxy & \multicolumn{2}{c}{Distance}        & $A_{V}$        & $r_{\rm G}$ & $r_{\rm G}/r_{\rm 25}$  & [O/H] 
& $\log L/{\rm L_{\odot}}$ & ZAMS \\
          & Type    &            & Class  & Mpc         & Method &                & (kpc)      &                         & (dex)  & (dex) & (\msol) \\\hline
1999an  &  II       &   IC 755   & SBb  & 18.5 $\pm$ 1.5  & TF & 0.40 $\pm$ 0.19 &   4.7  & 0.82  &  8.3    & $<5.16$ &    $<18$          \\
1999br  &  II-P     &  NGC 4900   & SBc  & 14.1$\pm$ 2.6  & Kin. & 0.06 $\pm$ 0.06 &   3.1  & 0.69  &  8.4  & $<4.76$ &    $<15$          \\
1999em  &  II-P     &  NGC 1637   & SBc  & 11.7 $\pm$ 1.0 & Cep. & 0.31 $\pm$ 0.16 &   1.6  & 0.28  &  8.6  & $<4.69$ &    $<15$              \\
1999ev  &  II-P     &  NGC 4274   & SBab & 15.1$\pm$ 2.6  & Kin. & 0.47 $\pm$ 0.16 &   5.3  & 0.46  &  8.5  & $5.1\pm0.2$ &    16$^{+6}_{-4}$     \\
1999gi  &  II-P     &  NGC 3184   & SABc & 10.0$\pm$0.8   & Mean& 0.65 $\pm$ 0.16 &   3.1  & 0.30  &  8.6   & $<4.64$ &    $<14$             \\ 
2001du  &  II-P     & NGC 1365    & SBb  & 18.3 $\pm$ 1.2 & Cep. & 0.53 $\pm$ 0.28 &  14.7  & 0.53  &  8.5  & $<4.71$ &    $<15$             \\
2002hh  &  II-P     &  NGC 6946   & SABc & $5.9 \pm 0.4$  & Mean & 5.2 $\pm$  0.2  &   4.1  & 0.45  &  8.5  & $<5.10$ &    $<18$              \\ 
2003gd  &  II-P     &  NGC 628   & Sc   & 9.3 $\pm$ 1.8   & Mean & 0.43 $\pm$ 0.19  &   7.5  & 0.58  &  8.4 & $4.3\pm0.3$ & $7^{+6}_{-2}$     \\
2003ie  &  II?      &  NGC 4051  & SABb & 15.5 $\pm$ 1.2  & TF & 0.04             &  7.3   & 0.66  &  8.4   & $<5.40$ &     $<25$                  \\
2004A   &  II-P     &  NGC 6207   & Sc   & 20.3 $\pm$ 3.4 & Mean & 0.19 $\pm$ 0.09 &   6.7  & 0.79  &  8.3  & $4.5\pm0.25$ &    $7^{+6}_{-2}$      \\
2004am  &  II-P     & NGC 3034    & Sd   & $3.3 \pm 0.3$  & Cep. & 3.7 $\pm$ 2.0   &   0.64 & 0.14  &  8.7  & Cluster &    $12^{+7}_{-3}$      \\       
2004dg  &  II-P     &  NGC 5806   & SBb  & $20.0 \pm 2.6$ & Kin. & 0.74 $\pm$ 0.09 &   4.3  & 0.50  &  8.5  & $<4.45$ &    $<12$               \\
2004dj  &  II-P     &  NGC 2403   & SABc & $3.3 \pm 0.3$  & Cep. & 0.53 $\pm$ 0.06 &   3.5  & 0.37  &  8.4  & Cluster &    $15\pm3$             \\  
2004et  &  II-P     &  NGC 6946   & SABc & $5.9 \pm 0.4$  & Mean & 1.3 $\pm$ 0.2   &   8.4  & 0.92  &  8.3  & $4.6\pm0.1$ &  $9^{+5}_{-1}$  \\
2005cs  &  II-P     &  NGC 5194   & Sbc  & $8.4 \pm 1.0$  & PNLF & 0.43 $\pm$ 0.06 &   2.7  & 0.22  &  8.7  & $4.25\pm0.25$  &    $7^{+3}_{-1}$ \\
2006bc  &  II-P     &   NGC 2397 & SBb  & $14.7 \pm 2.6$  & Kin. & 0.64            &   1.4  & 0.30  &  8.5  & $<4.43$ &    $<12$                \\
2006my  &  II-P     &  NGC 4651  & Sc   & $22.3 \pm2.6$   & TF & 0.08            &   4.4  & 0.37  &  8.7    & $<4.51$ &      $<13$               \\
2006ov  &  II-P     &  NGC 4303  & SBbc & $12.6 \pm 2.4$  & TF & 0.07            &   2.3  & 0.26  &  8.9    & $<4.29$ &      $<10$              \\
2007aa  &  II-P     &  NGC 4030  & Sbc  & $20.5 \pm 2.6$  & Kin. &  0.09        & 10.3 &   0.91   & 8.4     & $<4.53$ &      $<12$           \\
2008bk  &  II-P     &  NGC 7793  & Scd  & $3.9 \pm 0.5$   & TRGB &  1.0 $\pm$ 0.5 &  3.9   & 0.66   & 8.4   & $4.6\pm0.1$    & $9^{+4}_{-1}$      \\
\hline
\end{tabular}\\
\end{table*}

\section{The masses of the progenitors of Type II-P supernovae}
\label{section:masses}

We expect that the progenitors of various subtypes of type II SNe are
hydrogen rich stars which have evolved from main-sequence stars of an 
approximate initial mass of 8\msol\ and above. If objects are detected in
pre-explosion images then their colours and luminosities can be
determined. 
If there is no star detected at the SN positions then the
sensitivity of the images can be used to determine an upper luminosity
limit and hence upper mass limit 
\citep[see for example][]{2003MNRAS.343..735S,2003PASP..115.1289V,2005MNRAS.360..288M}.  In the papers presenting the original
results, slightly different methods of determining the 
luminosity and mass limits from the prediscovery images have been adopted. 
A mixture of
3$\sigma$ and 5$\sigma$ limits have been quoted, uncertainties 
treated in varying manners, different stellar evolutionary models
adopted and distances used which were not always the most recent 
and most accurate. In order to compare
the sample as a whole, this calls for some homogenisation, and we
adopt the following method.

In the cases where there is no detection of a progenitor (13 in total)
we determine the upper luminosity limit corresponding to the 84 per cent
confidence limit.  First we take the 3$\sigma$ detection limit for
each pre-explosion image, where this is the detection magnitude in the filter
system employed.  To convert this to a bolometric
luminosity one requires a measurement of extinction, distance and
bolometric correction (with respect to the filter employed) for the
progenitor. The 1$\sigma$ uncertainties of these 
quantities are combined in quadrature to give a total
1$\sigma$ uncertainty on the upper limit. If one assumes that the progenitor 
star was a red supergiant just prior to explosion then the bolometric 
and colour corrections  for an M0 supergiant \citep{2000asqu.book.....C}
are appropriate. This assumption seems well justified as nearly all the 
SNe in our sample have been shown to type II-P, which require
extended atmospheres physically similar to red supergiants 
\citep[][]{1980ApJ...237..541A,1976ApJ...207..872C}. 
Recent detections of the UV shock breakout from two type II-P
SNe determined the radii of the progenitor stars which adds further
weight to the idea of the progenitors being red supergiants
\citep{2008Sci...321..223S,2008ApJ...683L.131G}. 
The uncertainty in the
bolometric correction is taken to be $\pm0.3^{m}$ corresponding to the
1$\sigma$ range of values for red supergiants between late-K
and late M-type supergiants \citep{2005ApJ...628..973L}.
Assuming that the uncertainties are 
representative of a normal distribution of measurements, the 
84 per cent confidence limit for the upper luminosity 
limit is 1$\sigma$ above the best estimate i.e. there is an 84 per cent chance that the progenitor 
stars have luminosities below this value given the individual uncertainties in the calculation.  
The 1$\sigma$ distance and extinction errors are taken from the quoted
sources as listed in the notes on the individual events below. 
These 84 per cent  upper luminosity limits are
then plotted on the final mass-luminosity plot discussed in 
Section\,\ref{section:models}. We determine the upper
mass limit to be the maximum mass of a star which does not
have part of its post-He burning track within the 84 per cent luminosity 
limit.

This method is equivalent to that previously employed by
\cite{2003MNRAS.343..735S,2002ApJ...565.1089S} 
and \cite{2005MNRAS.360..288M} (for example) which an exclusion region
of the HR diagram was determined as a function of effective
temperature and an upper mass limit from the red supergiant region was
determined. If an $I$-band (or $I$-band like) filter is employed both
methods have the advantage of being fairly insensitive to the
effective temperature of the assumed red supergiant progenitor as the
peak of the stellar SED at this temperature range is $\sim$8300\AA
\citep[e.g see Figs. 5 and 6 of][]{2003MNRAS.343..735S}.  In all cases
we have revised the distances to the galaxies to the most reliable, in
our opinion, and most recent in the literature. Where no other
distance is available we have calculated a kinematic distance estimate
using the host galaxy radial velocity corrected for the local group
infall into Virgo ($V_{\rm vir}$ from the
LEDA\footnote{http://leda.univ-lyon1.fr/} galaxy catalogue) and a
value of the Hubble constant of $\mathrm{H_{0} = 72
km\,s^{-1}\,Mpc^{-1}}$. In such cases we employ an uncertainty of the
local cosmic thermal velocity 187\kms \citep{2000ApJ...530..625T},
equivalent to $\pm2.6$ Mpc. If the value of $H_{\rm 0}$ adopted
were either 65 or 85\kms\,Mpc$^{-1}$, the systematic effect on the
distance scale would provide systematic luminosity differences
of  $+0.22$\,dex and $-0.36$\,dex for the five SNe which have
kinematic host galaxy distances (1999br, 1999ev, 2004dg, 2006bc, 
2007aa). We will discuss the effects of this systematic difference
in Sect.\,\ref{section:imfs}. When determining the extinction in each
wavelength band, we use the law of
\cite{1989ApJ...345..245C}.

In the cases where we have a direct detection of the progenitor (4 in total)
the uncertainties in the luminosity are trivially determined 
and discussions of the individual events are listed below. 

Two others (2004dj and 2004am) fall on bright, compact star clusters
which are not resolved into individual stars 
\citep[][Mattila et al. 2008, in prep]{2004ApJ...615L.113M,2005ApJ...626L..89W,2008arXiv0812.1589V}.
These
papers have determined the total cluster mass and age and hence the
turn off mass at the top of the main-sequence. From this the mass of
the progenitor has been determined.  Hence the stellar mass determination is
somewhat indirect and relies on the assumption that the timescale of
star formation in the cluster is significantly less than the current
estimated age. The results are based on 
population synthesis codes which use different individual stellar evolutionary
codes as input to those we have employed. The \cite{2004ApJ...615L.113M,2008arXiv0812.1589V}
results are based on the stellar synthesis codes $starburst99$
which  uses the Geneva models \citep[e.g.][]{schaller}
as input, and 
the discussion in Section\,\ref{subsect:modelerror} indicates that the choice of stellar
model does not introduce significant uncertainties. Hence although 
the analysis method, and hence mass determination, is different for 
these two events, we believe the mass estimates  
are worth including in this compilation. If they are left out, the 
main conclusions of this paper are not altered in any significant way.

\subsection{1999an}
\label{sec:1999an}

IC755 is an SBb spiral with $M_{B} = -18.85$ (from LEDA). As no direct abundance 
study of this galaxy has been done, we attempt to infer a probable abundance at the 
position of the progenitor from the ${\rm O/H} - M_{B}$ relation of
\cite{2004A&A...425..849P}. For galaxies between 
$-20 < M_{B} < -19$, the characteristic oxygen abundance
(the oxygen abundance at a galactocentric distance of $r=0.4r_{25}$)
is typically in the range $8.5\pm0.2$\,dex. The mean abundance gradient of this sample
is $-0.5\pm0.3$\,dex/$r_{25}$. Hence at a de-projected distance of 0.8$r_{25}$, the
metallicity of the progenitor star of SN1999an can be approximated
at 8.3\,dex. This is somewhat uncertain given the large 
uncertainties on the gradient and the range of characteristic abundances and the
error is likely to be $\pm$0.3\,dex. However it 
is the best estimate that can be derived with the current data in the literature. 

There is no detection of a progenitor star in the WFPC2 pre-explosion
images presented by \cite{2005MNRAS.360..288M} and
\cite{2003PASP..115....1V}. Both studies calculated similar sensitivity 
limits for the images, and we adopt the 3$\sigma$ limit of 
\cite{2005MNRAS.360..288M} of $m_{\rm F606W}=24.7$. 
Solanes et al. (2002) report a mean distance modulus for the host galaxy IC755 of 
$\mu = 31.33 \pm 0.18$ or $d = 18.45 \pm1.5 Mpc$. Applying a line of sight extinction 
of $E(B-V)=0.13 \pm 0.06$ 
\citep{2005MNRAS.360..288M} and assuming an M-type supergiant as the 
progenitor (for the colour correction between $F606W$ and Johnson $V$ ; 
see Maund \& Smartt 2005)
results in absolute upper limit of $M_{V}  = -6.46 \pm 0.26$. 
For an M0 supergiant this corresponds to an
upper luminosity limit of \logl$=5.00 \pm 0.16$, and an 84 per cent confidence limit of 
\logl$=5.16$. From Fig. \ref{fig:mass-logL} this implies an upper mass
limit of 18\msol.

\subsection{1999br}

NGC4900 is an SBc spiral, and similar to the case of  IC755 discussed above
it does not have a published abundance study. The same arguments as in
Section\,\ref{sec:1999an} can be used (NGC4900 has $M_{B}=-19.05$) to 
infer an oxygen abundance at the galactocentric radius of SN1999br ($0.69r_{\rm 25}$)
of approximately 8.4\,dex. 

With no detection of a progenitor object at the SN position, 
\cite{2005MNRAS.360..288M} place an upper limit on the magnitude of 
any progenitor of $m_{\rm F606W}=24.9$. This is significantly brighter
than that of \cite{2003PASP..115....1V} $m_{\rm F606W}=25.4$, and we adopt the 
former as the more conservative result. For the host galaxy
(NGC4900) only a kinematic distance modulus is available, with the 
Virgo infall corrected velocity giving $d = 14.1 \pm2.6$ Mpc. 
The extinction to this event appears very low 
\citep{2005MNRAS.360..288M,2003PASP..115....1V,2004MNRAS.347...74P}
and we adopt the foreground value quoted in these papers of 
$E(B-V)=0.02 \pm 0.02$. As for SN1999an, we assume that the progenitor 
was a red supergiant and apply a colour correction and 
bolometric correction to determine an upper luminosity limit of \logl$=4.55 \pm 0.21$
and an 84 per cent confidence limit of 
\logl$=4.76$. From  Fig.\ref{fig:mass-logL} this implies an upper mass
limit of 15\msol.

\subsection{1999em}

\cite{1998AJ....116.2805V} have published line strength
measurements and abundances for 15 H\,{\sc ii} regions in
NGC1637. Using the calibration of \cite{2004ApJ...615..228B} we have
redetermined the abundance gradient and at the galactocentric distance
of SN1999em, the metallicity is $8.6\pm0.1$\,dex. The nearest \Hii\
regions to 1999em are 510\,pc and 794\,pc, when de-projected, from the
site of SN1999em and have oxygen abundances of 8.5 and 8.7\,dex
respectively. Hence we adopt $8.6$\,dex.

The updated distance to NGC1637 of $d=11.7\pm1$\,Mpc is taken from the
Cepheid variable star estimate of \cite{2003ApJ...594..247L} and the
reddening value of $E(B-V)=0.1 \pm 0.05$ is adopted from
\cite{2000ApJ...545..444B}. \cite{2002ApJ...565.1089S} 
present deep ground based images of NGC1637 before explosion in $VRI$ 
filters and from these results we have determined a 
$3\sigma$ upper limit of $I=23$. The $I$-band is the most 
sensitive to red supergiant progenitors and 
between the supergiant spectral types of K2-M4 this corresponds to an
upper luminosity limit of \logl$=4.54 \pm0.15$, and an 84 per cent confidence limit of 
\logl$=4.69$. From Fig.\ref{fig:mass-logL} this implies an upper mass
limit of 15\msol.

\subsection{1999ev}

NGC4274 is an SBab spiral, and also has no abundance study of its \Hii\ regions.
The same arguments as in
Section\,\ref{sec:1999an} can be used (NGC4274 has $M_{B}=-20.18$) to 
infer an oxygen abundance at the galactocentric radius of SN1999ev ($0.46r_{\rm 25}$)
of approximately  8.5\,dex. 

SN1999ev was recovered in late, deep HST ACS images by \cite{2005MNRAS.360..288M}
and is coincident with a progenitor object found on a pre-explosion WFPC2
F555W image. Although \cite{2003PASP..115....1V} originally suggested two other
stars as possible progenitors, the HST follow-up clearly ruled this 
out and points to the object of magnitude $m_{\rm F555W}=24.64 \pm 0.17$, 
at 4.8$\sigma$ significance \citep{2005MNRAS.360..288M}. There is no 
distance measurement to the galaxy NGC4274 apart from a kinematic estimate, which 
is $d=15.14 \pm 2.6$\,Mpc, from LEDA (Virgo infall corrected). \cite{2005MNRAS.360..288M} 
determined the extinction to 
the nearby stellar population of $E(B-V)=0.15 \pm 0.05$. We again 
assume that the progenitor was a red supergiant and apply a BC of $-1.3 \pm 0.3$
to determine a final luminosity of \logl$=5.1 \pm 0.2$. The 
tracks in Fig.\ref{fig:mass-logL} imply the star would have been of mass 
16$^{+6}_{-4}$\msol.

\subsection{1999gi}

\cite{2001ApJ...556L..29S} suggested that the \Hii\ region number 3 of 
\cite{1994ApJ...420...87Z} at a position of $68''$N $0''$E is coincident with the star-forming
region, or OB association that hosted SN1999gi. The calibration of 
\cite{2004ApJ...615..228B} using the $R_{23}$ value of \cite{1994ApJ...420...87Z}
gives an oxygen abundance of 8.6\,dex.

A study of the progenitor site of SN1999gi was carried out by 
\cite{2001ApJ...556L..29S}, but the distance to this galaxy was then 
improved in a compilation study of \cite{2002AJ....124.2490L} and 
\cite{2006PhDTHendry}. Here we adopt the result in \cite{2006PhDTHendry} which 
is a mean of four estimates $d=10.0 \pm 0.8$\,Mpc, and the extinction of  
\citeauthor{2002AJ....124.2490L} $E(B-V)=0.21 \pm 0.05$.  The four 
distance methods detailed in \cite{2006PhDTHendry} are Tully Fisher, 
expanding photosphere method (EPM), kinematic and the tertiary
distance indicators of \cite{1979ApJ...227..380D}. 
The 3$\sigma$ detection 
limit determined by \cite{2001ApJ...556L..29S} is $m_{\rm F606W}=24.9$, 
which results in an upper luminosity limit for an M-type supergiant of 
\logl$=4.49 \pm 0.15$, after the colour and bolometric
corrections are applied. This gives an  84 per cent confidence limit of 
\logl$=4.64$ and, from Fig. \ref{fig:mass-logL}, this implies an upper mass
limit of 14\msol.

\subsection{2001du}

As discussed in 
\cite{2003MNRAS.343..735S} the \Hii\ region RW21 
\citep{1997MNRAS.288..715R} is $1.5''$ from SN2001du, and it is likely the metallicity of
this region is representative of the progenitor star composition. 
The calibration of \cite{2004ApJ...615..228B} to the $R_{23}$ value of 
\cite{1997MNRAS.288..715R} gives an oxygen abundance of 8.5\,dex.

The host galaxy NGC1367 was observed as part of the HST Cepheid Key Project, hence
the most accurate and recent distance estimate is
taken from \cite{2002A&A...389...19P}, $\mu = 31.31 \pm 0.15$. The
extinction toward the SN was measured by three different methods by 
\cite{2003MNRAS.343..735S} to be $E(B-V)=0.17\pm0.09$, giving 
$A_{I}=0.25 \pm 0.13$,  which is similar to that adopted 
($A_{I}\simeq0.2$) by \cite{2003PASP..115.1289V}. 
\cite{2003MNRAS.343..735S} and \cite{2003PASP..115.1289V} presented 
pre-explosion images in the WFPC2 filters F336W, 
F555W, F814W and the most sensitive of these to red supergiants is the F814W. We
determine the  $3\sigma$ upper limit from the Smartt et al. results to
be $I=24$, similar
to the sensitivity $m_{F814W}=24.25$ of \cite{2003PASP..115.1289V}. 
Between the supergiant spectral types of K2-M4 this corresponds to an
upper luminosity limit of \logl$=4.57 \pm0.14$, and an 84 per cent confidence limit of 
\logl$=4.71$. From Fig.\ref{fig:mass-logL} this implies an upper mass
limit of 15\msol.

\subsection{2002hh}
\label{sec:2002hh}
None of the 9 \Hii\ regions in NGC6946 compiled by \citet{2004A&A...425..849P}
are near the location of 2002hh. Hence we use the abundance gradient 
determined by \citet{2004A&A...425..849P} and the de-projected galactocentric 
radius of the SN position to determine the likely metallicity. As discussed above, 
the calibration of \citet{2004A&A...425..849P} is similar to the simple
linear calibration of \cite{2004ApJ...615..228B}, hence these should be on a similar scale. 
We determine an oxygen abundance of 8.5\,dex.

A deep pre-explosion $i'$-band archive image of NGC6946 from the 
{\em Isaac Newton Telescope Wide Field Camera} (INT-WFC) will be presented in a
forthcoming paper (See Sect.\ref{2004et}). Although this SN suffered
significant extinction, the proximity of the galaxy and 
the depth of the 3600s $i'$-band image still places useful restrictions on the 
progenitor star. 
The 3600s image is composed of 6$\times$600s exposures, with a final image
quality of 1$''$. There is no object visible at the position of SN2002hh, 
and the 5$\sigma$ detection limit for a point 
source was estimated to be $i_{\rm CCD}$=22.8. This instrumental 
magnitude can be converted to a standard $I$ using the well calibrated
colour transformations for the INT-WFC \citep{2001NewAR..45..105I}
\footnote{http://www.ast.cam.ac.uk/$\sim$wfcsur/index.php}. 
We employ the reddening law determined in \cite{2006MNRAS.368.1169P} to estimate the
extinction in the $I-$band of $A_{I}=2.1\pm0.3$. A distance of 
$d=5.9 \pm 0.4$\,Mpc is used which is a mean of the distance values
from the 
compilation of Botticella et al. (2009 in prep.) using the 
methods of Tully Fisher, brightest supergiants, sosies, PLNF, EMP (applied
to 1980K) and standard candle method (SCM) applied to SN2004et. 
SN2002hh appears
to be a normal II-P, but behind a large dust pocket \citep{2006MNRAS.368.1169P}, 
hence we assume the progenitor was a red supergiant of type between K0-M5. 
The falling bolometric correction combined
with the rising intrinsic $(V-I)$ between K0-M5 means that the 
bolometric luminosity limit stays approximately constant in this 
spectral range at $\log L/L_{\odot}=4.9 \pm0.2$. Hence the 
84 per cent confidence limit is  \logl$=5.1$ and
from  Fig.\ref{fig:mass-logL} this implies an upper mass limit of 18\msol.
We note that this is consistent with the progenitor mass of 16-18\msol\
estimated by \cite{2006MNRAS.368.1169P} from the 
[O\,{\sc i}] $\lambda\lambda$6300,6364\AA\ doublet.

\subsection{2003gd}

None of the previously catalogued  \Hii\ regions in NGC628 which have spectra and 
the $R_{23}$ ratio measured are particularly near the spatial position of SN2003gd. 
Hence we use the abundance gradient 
determined by \citet{2004A&A...425..849P} and the de-projected galactocentric 
radius of the SN position to determine the metallicity at this position. The
parameters derived are Table\,\ref{table:main}, with an abundance of 8.4\,dex 
derived. 

The progenitor star was detected by
\cite{2004Sci...303..499S} and \cite{2003PASP..115.1289V}, and an extensive
compilation of distance measurements to NGC628 and reddening towards the SN was
carried out in \cite{2005MNRAS.359..906H}. Those distance and extinction 
values determined were
close to those employed in \cite{2004Sci...303..499S} to estimate the 
progenitor luminosity and mass. The \cite{2005MNRAS.359..906H} distance listed in Table\,\ref{table:main} 
is a mean of the three methods : kinematic, brightest supergiants and standard
candle method (applied to SN2003gd). 
 The intrinsic
$(V-I)_0 = 2.3 \pm 0.3$ colour is consistent with a supergiant in the spectral type
range K5-M3, which \cite{2004Sci...303..499S}  used to determine a
luminosity of \logl$= 4.3 \pm 0.3$. 
In the diagram of Fig.\ref{fig:mass-logL} the best value of 4.3 is closest 
to the termination point of the 7\msol\ track, if we assume the 
progenitor did not go through 2nd dredge-up. The uncertainties would bracket
the 5\msol\ and 13\msol\ post-He burning tracks, hence we adopt the value 
$7^{+6}_{-2}$\msol. Although the most likely value is below the lowest
mass that is normally assumed possible to provide
an iron core-collapse \cite[8-10\msol; see][]{2003ApJ...591..288H,2004MNRAS.353...87E},
the range of masses comfortably brackets the theoretically predicted limits.

\subsection{2003ie}
\label{subsect:2003ie}

NGC4051 is a Seyfert 1 galaxy of morphological type SABb and has no
published abundance study of its \Hii\ regions. Hence we can
estimate a probable abundance at the position of the progenitor as in
Sect.\,\ref{sec:1999an}. The galaxy has a $M_{B} = -20.3$, using our
adopted distance and the corrected $B-$band magnitude given in
$LEDA$. At this magnitude the the characteristic oxygen abundance
(at a galactocentric distance of $r=0.4r_{25}$) is approximately
$8.5\pm0.2$\,dex \citep{2004A&A...425..849P} Again using the typical
gradient of the galaxy sample (as in Sect.\,\ref{sec:1999an} of
$-0.5\pm0.3$\,dex/$r_{25}$, the oxygen abundance at 0.66$r_{25}$ is
approximately 8.4\,dex. As stated above this is quite uncertain
($\pm$0.3\,dex) given the lack of detailed study of the galaxy 
but does show that it is unlikely to be a particularly low metallicity 
environment. The SN was not studied in great detail by any group (as far
as we know), but a single photospheric spectrum shows 
P-Cygni features of H\,{\sc i} \citep{2008arXiv0804.1939H}. 
The best match for the spectrum found by \cite{2008arXiv0804.1939H}
is that of 1998A, which itself appears like a 1987A-type event. 
Hence this event may not be a normal II-P SN and we have no 
lightcurve information to consider. As we shall see below, the 
mass limits for the progenitor are not particularly restrictive and if
the object were to be left out the conclusions of the paper would be 
unchanged. 

The pre-explosion site of SN~2003ie was recovered in archive $r'$-band 
observations of NGC4051 taken with the INT-WFC. This image
was taken on 1999 November 11, with an exposure time of 900s and image quality of 
$1.1$ arcsec.  We determined the position of SN~2003ie within an error
circle of $0.17$ arcsec (using an image of the SN from 2003
provided to us by Martin Mobberley). There is 
no progenitor object detected within this error circle and we   
derive a 3$\sigma$ detection limit of $r' = 23$. This instrumental magnitude was converted to a 
standard $R$ using the well calibrated colour transformations for the INT-WFC 
\citep{2001NewAR..45..105I}$^7$, giving $R = 22.65$. Pierce \& Tully (1988) calculate the 
distance to the Ursa Major Cluster to be $15.5 \pm 1.2$ Mpc using the Tully-Fisher method 
and we adopt this distance for NGC4051. We have no measure of the internal extinction 
towards this SN and simply adopt the Galactic extinction value of $E(B-V)=0.013$ 
\citep{1998ApJ...500..525S}. We assume once again that the progenitor is a red supergiant 
and apply appropriate bolometric and colour corrections to determine a luminosity limit of
 $\log L/L_{\odot} = 5.26 \pm0.14$ and an 84 per cent confidence limit of
 $\log L/L_{\odot} = 5.40$. From Fig.~\ref{fig:mass-logL} this implies an upper 
mass limit of 24\msol.

\subsection{2004A}
\label{subsect:04A}
There are no measurements of \Hii\ regions in NGC6207 and we use the 
arguments presented in \cite{2006MNRAS.369.1303H} to estimate the metallicity at
the galactocentric distance of SN2004A. This paper based the results
on typical abundance gradients measured by \citet{2004A&A...425..849P}
hence again the estimate is on the same scale as the rest of the 
values. We point out that there is a typographical error in that paper, where $R_{25}$ is quoted as 4\,kpc, whereas
it should be 8.6\,kpc. However, repeating the same method, this does not change the 
calculated metallicity at the position of SN2004A which we estimate as 8.3\,dex.
The \cite{2006MNRAS.369.1303H} distance listed in Table\,\ref{table:main} 
is a mean of the three methods : kinematic, brightest supergiants and standard
candle method (applied to SN2004A).

A faint object is detected at the position of SN2004A in \cite{2006MNRAS.369.1303H}, 
claimed as a 4.7$\sigma$ detection in the F814W filter, and it is 
not detected in the F435W or F555W. If we assume this detection to 
be valid, it provides a blue limit for the colour of the progenitor 
and hence a star in the spectral range G5-M5. This 
gives a bolometric luminosity in the range \logl$=4.5 \pm 0.25$. On 
Fig.\ref{fig:mass-logL} this implies a best estimate of 7\msol, and the 
errors bracket the 5\msol\ and 13\msol\ post-He burning tracks. Hence we
adopt $7^{+6}_{-2}$\msol. If the detection is not valid then the I-band 
detection sensitivity implies an 84 per cent confidence limit of \logl=4.75
and an upper limit of 13\msol.

\subsection{2004am}
\label{subsect:2004am}

There is no extensive published study of SN2004am to date, which is
surprising given its proximity and the fact it is the only optically
discovered SN in the starburst M82 (NGC3034).  However it is clearly a II-P
from the unfiltered magnitudes of \cite{2004IAUC.8297....2S} which
stay constant for 76 days, and the spectrum of
\cite{2004IAUC.8299....2M}.  Alignment of post-explosion
 near-IR images and HST pre-explosion
images shows SN2004am is spatially coincident with the well studied 
super star cluster M82-L (Mattila et al., in prep). The distance to 
M82 is assumed to be that of the M81 group, estimated
from Cepheids in M81 \cite[NGC3031;][]{2001ApJ...553...47F}. 

A new study of M82-L has recently been carried out by 
\cite{2008A&A...486..165L} who modelled the integrated near infra-red 
$0.8-2.4\mu$m spectra. They used a population synthesis code (P\'{E}GASE.2)
and a new library of red supergiant observational and theoretical spectra
to determine the age of M82-L. The fit to the overall SED and the individual 
molecular absorption features is impressive and gives an age estimate of 
$18^{+17}_{-8}$\,Myrs. With a lower than normal value of $R_{V} = 2.4-2.7$, 
\cite{2008A&A...486..165L} can also reproduce the optical SED of the cluster down 
to 6000\AA\ (the $A_{V}$ value in Table\,\ref{table:main} is taken from Lancon et al.). 
This age is somewhat younger than $65^{+70}_{-35}$\,Myrs
that was first inferred by 
\cite{2006MNRAS.370..513S} using only a limited range optical spectrum 
and the spectral synthesis code $starburst99$ \citep{1999ApJS..123....3L}.
\cite{2008A&A...486..165L} point out that by using a low value of $R_{V}$ they
can reconcile the optical SED {\em and} the NIR molecular bands with the younger
age and their updated spectral modelling technique and firmly exclude an 
age of 60\,Myr. 
The cluster age provides quite a strong constraint on the mass of the
progenitor star, assuming that the cluster formed coevally and the
progenitor's age is similar to that of the cluster. The STARS models
predict that the cluster ages correspond to lifetimes of stars of 
masses $12^{+7}_{-3}$\,\msol.  The models used in the population synthesis code
of \cite{2008A&A...486..165L} were those of \cite{1993A&AS..100..647B}, 
which give very similar age-mass relationships to the STARS code (the uncertainty
on the derived mass due to choice of code is within the error range). 

In all of the above we have assumed solar metallicity for the stellar
evolutionary tracks is appropriate. The fitting of
 \cite{2008A&A...486..165L} implies that this is appropriate. Also 
two recent papers have speculated on the abundances in the nuclear regions of
M82, in environments close to super star cluster M82-L 
\citep{2006MNRAS.370..513S,2004ApJ...606..862O}. There is some uncertainty
and difference in the abundances derived but  
the stellar abundances of red supergiants in these inner regions
from \cite{2004ApJ...606..862O} suggest a solar like oxygen abundance. The
photospheric abundance in such objects are likely to be
applicable to M82-L and the progenitor of SN2004am. 
Hence in the age estimations we chose the tracks close to 
8.7\,dex as the most appropriate.

\subsection{2004dg}
\label{subsec:04dg}
NGC5806 is an SBb spiral, and also has no abundance study of its \Hii\ regions. 
The same arguments as in Section\,\ref{sec:1999an} can be used (NGC5806 has 
$M_{B}=-19.86$) to infer an oxygen abundance at the galactocentric radius of 
SN2004dg ($0.56r_{\rm 25}$) of approximately  8.5\,dex. 

The pre-explosion site of SN~2004dg was imaged using both 
WFPC2 (2001 July 5) and ACS (2004 April 3) cameras on-board HST (the SN 
was discovered on 2004 July 31). The WFPC2 exposure times were
460s in F450W and F814W filters. For the ACS images had total exposure times 
of 700s in F658N and 120s in F814W. 
We re-observed SN2004dg on (2005 March 10) with the ACS camera (in F435W, 
F555W and F814W, as part of GO10187) and recovered the SN at 
transformed magnitudes of $B = 22.2, V=20.8, I=19.1$. 
Alignment of the two sets of images allowed us to locate the position of 
the SN on the pre-explosion images to within 0.015 arcsec
\citep[see][for details of alignment procedures]{2005MNRAS.364L..33M,2008MNRAS.391L...5C}
Within this error circle there was no detection of a progenitor star in
any of the image and filter combinations. 
A progenitor star was not detected at 
the SN position, therefore a 3$\sigma$ detection limit of $m_{F814W}= 25.0$
 was determined. There is no distance measurement to the galaxy
apart from a kinematic estimate, which is $d=20.0 \pm 2.6$\,Mpc, from
$LEDA$ (Virgo infall corrected). The total reddening towards SN~2004dg
was estimated to be $E(B-V)=0.24 \pm0.03$, giving $A_{I}=0.36 \pm
0.04$.  Again we assume a red supergiant progenitor and find a
luminosity limit of $\log L/L_{\odot} = 4.28 \pm0.17$ and an 84 per
cent confidence limit of $\log L/L_{\odot} = 4.45$. From
Fig.~\ref{fig:mass-logL} this implies an upper mass limit of 12\msol.

\subsection{2004dj}
As discussed in Sect.\,\ref{section:masses}, SN2004dj fell on the compact star 
cluster identified by \cite{2004ApJ...615L.113M} and \cite{2005ApJ...626L..89W}
and the analysis used to determine a progenitor mass is different
to the direct identification and direct upper luminosity limits for
the other SNe presented here. 
\cite{2004ApJ...615L.113M} determine an age of the compact star cluster 
of 14\,Myrs and hence a main-sequence mass of 15\msol\ for the progenitor. 
\cite{2005ApJ...626L..89W} determine an age of around 
20\,Myrs and hence a main-sequence mass of 12\msol. 
A new and
improved study by  \cite{2008arXiv0812.1589V}
using new UV observations of the cluster and 
extensive comparison of SEDs
 based on different model atmospheres and 
evolutionary tracks suggests a most likely turn off mass
(and hence progenitor ZAMS mass of between 12-20\msol. 
This age (10-16\,Myrs) is  consistent with the lack of 
H$\alpha$ emission seen in the cluster spectrum of  \cite{2008arXiv0812.1589V}
and \cite{1987AJ.....94.1156H}, as the  ionizing O-stars have died out. 
Hence we favour the older age and  will adopt $15\pm3$\msol\ as the progenitor mass
\cite[if we adopt $15^{+5}_{-3}$\msol\ as suggested by][it does not affect any
of the results below]{2008arXiv0812.1589V}. The distance to 
NGC2403 of $3.3\pm0.3$\,Mpc is from the HST Cepheid Key Project \citep{2001ApJ...553...47F}

\cite{2004ApJ...615L.113M} adopted solar abundances in using the Geneva
tracks of $starburst99$. On closer inspection this may be too high. 
We used the abundance gradient 
determined by \citet{2004A&A...425..849P} and the de-projected galactocentric 
radius of the SN position to determine the metallicity at this position
of 8.4\,dex. Although the \Hii\ regions in this galaxy have been studied extensively
the only region which is physically close to the position of 2004dj is that of 
VS44 studied by \cite{1997ApJ...489...63G}, and even that is around 600pc from the cluster that hosted
SN2004dj. The $R_{23}$ ratio provided by \cite{1997ApJ...489...63G} also gives an abundance of 
8.4\,dex with the calibration of 
\cite{2004ApJ...615..228B}, in good agreement with the abundance gradient measurement. Although 
our oxygen abundance is below that employed by \cite{2004ApJ...615L.113M}, this 
does not significantly affect the age (and hence turn-off mass) estimate when we compare 
STARS models of such different metallicities. We note that \cite{2008arXiv0812.1589V}
favour a solar metallicity in their SED fits.

\subsection{2004et}
\label{2004et}

As discussed for SN2002hh (Sect.\,\ref{sec:2002hh}), there is no \Hii\ region near the
galactic position of 2004et, which is some way from the centre of NGC6946. We 
use the same method as for 2002hh to determine a metallicity typical for 
the galactocentric radius of 2004et of 8.3\,dex. The adopted distance to NGC6946 
of $d=5.9 \pm 0.4$\,Mpc is 
discussed in Sect.\ref{sec:2002hh}

\cite{2005PASP..117..121L} presented the detection of a candidate
progenitor star of SN2004et in ground-based images from the Canada
France Hawaii Telescope (CFHT) in both $UBVR$ and $u'g'r'$ filters.
They suggested it was a yellow supergiant as the $BVR$ colours
were matched with a G-type supergiant SED. 
From these colours and the models of \cite{2001A&A...366..538L},
\cite{2005PASP..117..121L} determine a mass in the range
$15^{+5}_{-2}$\msol.

However it is now clear that the putative source 
detected by \cite{2005PASP..117..121L} was not a single star. 
Adaptive optics images of the site by  \cite{Crock04et}
using Gemini North show that the source breaks up into 
several stars. In addition $BVR$ images taken 3 years after
explosion show identical colours to the pre-explosion object, 
indicating that the progenitor star was not detected in 
the pre-explosion frame. A deep $i'$-band image (the same
image as discussed above for SN2002hh) does show a clear
detection of a progenitor compared to the late time $I$-band
image.  \cite{Crock04et} use the $i'$-band detection
(after converting to Johnson; $I = 22.06 \pm 0.12$) and
limits on the $BVR$ magnitudes to infer that the progenitor
was a red supergiant  with  {\mbox{$(R\!-\!I)_o$}} $>  
1.80\pm0.22$. This implies an M4 spectral type or later, 
giving a bolometric magnitude of  of $M_ {\rm bol} = -6.73\pm0.22$ and  a  
progenitor luminosity of \logl = 4.59$\pm$0.09.
Comparing this to  
stellar models of LMC metallicity,  we
estimate its initial mass to  
be $9^{+5}_{-1}$ M$_{\odot}$ \citep[as in][]{Crock04et}.

\subsection{2005cs}
\cite{2005MNRAS.364L..33M} have estimated the abundance at the galactocentric
radius of SN2005cs in an identical manner as we have employed consistently
in this paper, using the NGC5194 abundance gradient of \cite{2004ApJ...615..228B}, 
hence it is already on our common calibration scale. 
They determine $8.66\pm0.11$\,dex, which we adopt in this paper. 

The detection of the progenitor of SN2005cs is well documented by 
\cite{2007MNRAS.376L..52E}, \cite{2006ApJ...641.1060L} and \cite{2005MNRAS.364L..33M}
 which all give similar mass estimates in the range 7-10\msol. 
\cite{2007MNRAS.376L..52E} have recently re-analysed all of the available
photometry from the initial two discovery papers and suggested that the progenitor
could not have been a super-AGB star that has gone through 2nd dredge up. 
This analysis was done with the STARS code in an identical manner as this
study and they found likely progenitor range of $6-8$\msol
(assuming a distance of $8.4\pm1.0$\,Mpc). 
The \cite{2005MNRAS.364L..33M} luminosity estimate for the progenitor
is $\log L/L_{\odot} = 4.25\pm0.25$ and using the STARS tracks employed
here this would suggest a mass of $7^{+3}_{-1}$\msol. 
If the closer distance of $7.1\pm1.2$\,Mpc is chosen
\citep[from the mean of the compilation of][]{2006MNRAS.372.1735T} 
then the best estimate of mass would reduce slightly to around 
6\msol. This would be rather low, but the uncertainty on the upper
bound (3\msol) would still place it comfortably within the normal
theoretical ranges for core-collapse.

\subsection{2006bc}

NGC2397 is an SBb spiral, and is the final galaxy in this sample which
has no published abundance study of its \Hii\ regions. The same
arguments as in Section\,\ref{sec:1999an} are employed (NGC2397 has
$M_{B}=-19.67$) to infer an oxygen abundance at the galactocentric
radius of SN2006bc ($0.3r_{\rm 25}$) of approximately 8.5\,dex.

The pre-explosion site of SN~2006bc was imaged using 
WFPC2 (2001 November 17) on-board HST with exposure times of 
460s in each of the F450W and F814W filters (the SN position fortunately
fell on the PC1 chip). 
We re observed SN2006bc on (2006 October 14, as part of GO10498) 
with the ACS {\em Wide Field Camera} (WFC) in three filters 
F435W 1400s, F555W 1500s and F814W 1600s). 
\footnote{This constituted an ESA Photo Release : http://www.spacetelescope.org/news/html/heic0808.html}
Aligning  the before and after
explosion images allowed us to locate the position of 
the SN on the pre-explosion images to within 0.024 arcsec 
\citep[again see][for details of alignment procedures]{2005MNRAS.364L..33M,2008MNRAS.391L...5C}. 
Within this error circle there was no detection of a progenitor star in
any of the image and filter combinations. 
At the progenitor position we determined a 3$\sigma$ detection limit
of $m_{F814W}$ = 24.45. There is no distance
measurement to NGC2397 apart from a kinematic estimate, which is
$d=14.7 \pm 2.6$\,Mpc, from LEDA (Virgo infall corrected). The Galactic
extinction is estimated to be $E(B-V)=0.205$
\citep{1998ApJ...500..525S}. Assuming that the progenitor was a red
supergiant we find a luminosity limit of $\log L/L_{\odot} = 4.23
\pm0.20$ and an 84 per cent confidence limit of $\log L/L_{\odot} =
4.43$. From Fig.~\ref{fig:mass-logL} this implies an upper mass limit
of 12\msol.

\subsection{2006my}
SN2006my occurred at $r_{\rm G}/r_{\rm 25}=0.37$ almost exactly at the galactocentric radius of the characteristic oxygen abundance measured in NGC4651 by \cite{2004A&A...425..849P}. They measure a value of 8.7\,dex at this position. 

\cite{2007ApJ...661.1013L} claim the detection of a red supergiant
progenitor of SN~2006my in pre-explosion HST/WFPC2 observations of
NGC4651.  In order to determine the position of the SN on the
pre-explosion HST images \cite{2007ApJ...661.1013L} aligned these
images with ground based observations of the SN from the
Canada-France-Hawaii Telescope (CFHT).
They derived an initial mass of $M =10^{+5}_{-3}$\msol\
for the object which they find coincident with the
SN position.

However in an improved analysis, \cite{2008PASP..120.1259L} have shown
that this is unlikely to be correct and the progenitor star is most
likely not detected in the pre-explosion images. They used HST images
of much higher resolution than CFHT which allows for object positions
to be more accurately measured and ultimately leads to a more reliable
transformation between the coordinate systems of the pre- and
post-explosion images. They find that the offset between the SN and
possible progenitor position is too large to support the claim that
the two objects are associated (at about the 96\% confidence level).
In a completely independent manner, we used similar data to 
carry out the same image alignment 
and the details of this analysis are presented in our companion paper
\citep{Crock04et}. 
Using the HST post-explosion to HST pre-explosion  transformation, we also
find that the progenitor
object proposed by \cite{2007ApJ...661.1013L}  is
$\sim$74 mas from the transformed SN position. Given our total
astrometric error this is approximately a 1.8$\sigma$ separation. Hence we
also 
find that this object is unlikely to be the
progenitor of SN~2006my. Most likely it is not and the progenitor
is undetected in the images, so we derive a 3$\sigma$ detection limit
of $m_{F814W}$ = 24.8.

Solanes et al. (2002) have collected Tully-Fisher
distance estimates for NGC4651 from seven different sources
and derive a mean distance modulus $\mu
= 31.74 \pm0.25$ (or $d = 22.3 \pm2.6$ Mpc). As in
\cite{2007ApJ...661.1013L} we apply only a correction for the Galactic
extinction of $E(B-V)=0.027$. Assuming that the progenitor star was a
red supergiant we derive a luminosity limit of $\log L/L_{\odot} =
4.35 \pm0.16$ and an 84 per cent confidence limit of $\log L/L_{\odot}
= 4.51$. From Fig.~\ref{fig:mass-logL} this implies an upper mass
limit of 13\msol.

\subsection{2006ov}
\cite{2004A&A...425..849P} redetermined the 
oxygen abundance gradient, and our calibrations are on this equivalent
scale, hence we use the galactocentric radius $0.26r_{\rm 25}$ and
abundance gradient of Pilyugin et al. to determine an oxygen abundance
of 8.9\,dex. which is the highest in this sample.

\cite{2007ApJ...661.1013L} report the detection of a red supergiant progenitor of $M_{ZAMS} = 15^{+5}_{-3}$\msol\ for SN~2006ov in archival HST/WFPC2 observations of NGC4303. In this case the pre-explosion frames were aligned with HST observations of the SN in order to pinpoint the position of the progenitor on the archival images. Having performed PSF-fitting photometry using HSTphot \citep{2000PASP..112.1383D} without detecting a progenitor star, it was noticed by Li et al. that a significant point source was still visible in the residual image close to the SN site. \cite{2007ApJ...661.1013L} suggest that this object in the residual image is in fact coincident with the SN position, and claim that by forcing HSTphot to fit a PSF at this position they detect an object of 6.1$\sigma$ significance in the F814W and F450W observations. 

We have repeated the alignment of the pre- and post-explosion HST
observations and find exactly the same transformed SN position as in
\cite{2007ApJ...661.1013L}. We also find the same point source still
visible in the residual image after performing PSF-fitting photometry
using HSTphot. However, we measure the centre of this point source to
be some $\sim$63 mas from the SN position, which given our total
astrometric error is a 2.5$\sigma$ separation. This casts significant
doubt on the identification of this object as the SN
progenitor. Furthermore we are unable to reproduce the photometry
results of \cite{2007ApJ...661.1013L} by forcing HSTphot to fit at the
transformed SN position. Rather we find detections of the highest
significance ($\sim$6.0$\sigma$ in F814W and $\sim$4.4$\sigma$ in
F450W) when we force a fit at our own measured position of this point
source, which as we have already said is not coincident with the SN
position. A more detailed discussion of this analysis are
presented in \cite{Crock04et}.

Since we cannot confirm that this object is the progenitor of
SN~2006ov we derive a 3$\sigma$ detection limit of $m_{F814W}$ =
24.2. 

\cite{2007ApJ...661.1013L} derive a mean distance modulus for NGC4303 (M61) 
of
$\mu = 30.5 \pm0.4$ (or $d = 12.6 \pm2.4$ Mpc) from two Tully-Fisher
distance estimates, and that value is adopted here. We apply only a
correction for the Galactic extinction of $E(B-V)=0.022$. Again
assuming that the progenitor star was a red supergiant we derive a
luminosity limit of $\log L/L_{\odot} = 4.09 \pm0.2$ and an 84 per
cent confidence limit of $\log L/L_{\odot} = 4.29$. From
Fig.~\ref{fig:mass-logL} this implies an upper mass limit of 10\msol.

\subsection{2007aa}

NGC4030 is an Sbc spiral with no study of its \Hii\ regions published. 
The same arguments as in Section\,\ref{sec:1999an} can be used (NGC4030 has 
$M_{B}=-20.7$) to infer an oxygen abundance at the galactocentric radius of 
SN2007aa ($0.91r_{\rm 25}$) of approximately  8.4\,dex. 

The pre-explosion site of SN~2007aa was imaged using WFPC2 (2001 July
30) with exposure times of 460s in each of the F450W and F814W
filters. We determined the position of SN~2007aa on these images
using a ground-based image of 0.8 arcsec resolution taken with the 
AUX Port camera on the William Herschel Telescope on 2007 March 11
in the $I$-band filter. 
Alignment of this image with the pre-explosion frames produced
an error circle of 0.07 arcsec for the SN position on the 
F450W and F814W images. No object was detected within this 
region and  hence a 3$\sigma$ detection limit of
$m_{F814W}$ = 24.44 was determined. A kinematic distance estimate of
$d=20.5 \pm 2.6$\,Mpc for NGC 4030 was calculated from its recessional
velocity (Virgo infall corrected) as recorded in LEDA. Extinction due
to the Milky Way is estimated to be $E(B-V)=0.026$
\citep{1998ApJ...500..525S}. Assuming the progenitor was a red
supergiant we find a luminosity limit of $\log L/L_{\odot} = 4.38
\pm0.15$ and an 84 per cent confidence limit of $\log L/L_{\odot} =
4.53$. From Fig.~\ref{fig:mass-logL} this implies an upper mass limit
of 12\msol.

\subsection{2008bk}

SN2008bk was recently discovered in NGC7793 a nearby galaxy with two
distance modulus estimates of $\mu = 27.96\pm0.24$ \citep{2003A&A...404...93K}
determined from the tip of the red giant branch and $\mu =28.01$ from a Tully
Fisher distance in $LEDA$.  We will adopt the distance modulus  
of $\mu = 27.96\pm0.24$ ($d = 3.9\pm0.4$\,Mpc). 
The galaxy is part of the oxygen abundance
gradient study of \cite{2004A&A...425..849P}.  The galactocentric
radius $0.66r_{\rm 25}$ and the abundance gradient imply an oxygen
abundance of between 8.2-8.4\,dex at the SN position (Mattila et al. 
2008), hence we adopt the LMC-type metallicity. 

The galaxy has a wealth of prediscovery images available from the Very
Large Telescope (VLT), with optical $BVI$ images from FORS1 and NIR
$JHK$ images from HAWK-I and ISAAC. We have shown in a recent letter
that SN2008bk is exactly coincident with a bright, red, source
detected in the $IJHK$ bands, using high resolution $K_{s}$-images
from the NACO system (Mattila et al. 2008). Although the foreground
extinction toward the galaxy is low and the early observations of the
SN appear to show no signs of significant extinction, Mattila et al. (2008)
show that the progenitor SED can be fit with a late-type M4I spectral 
type with a visual extinction of  $A_{V} = 1.0\pm0.5$. Using methods which
are entirely consistent with our approach in this paper, 
 Mattila et al. (2008) have 
estimated the luminosity and mass of this red supergiant. 
The distance of 3.9$\pm$0.4 Mpc and
$A_{V} = 1.0\pm0.5$ results in
$M_{K}$ = -9.73 $\pm$0.26. Levesque et al. (2005) show
that using $M_{K}$ to determine M$_{\rm bol}$ is preferable to
using the optical bands. The best fit SED of around M4I would correspond
to $Teff \simeq 3550\pm50$K and a bolometric correction
$BC_{K} = +2.9 \pm 0.1$ (both from the Levesque et al. scale).
This results in $logL/L_{\odot} = 4.63 \pm 0.1$. 
and from Fig.\,\ref{fig:mass-logL} (at LMC
metallicity) this corresponds to a mass of $9^{+4}_{-1}$\msol.

\begin{figure}
    \centering
    \epsfig{file = 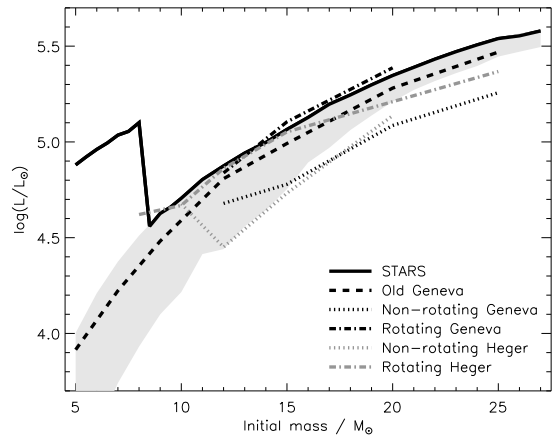, width=80mm}
    \caption[]{ The initial mass compared with the final luminosity of the STARS and
Geneva stellar models. For each mass we plot the luminosity at the end
of the model, just before core-collapse. For the STARS models this is
up to the beginning of neon burning. The old Geneva 
models end after core carbon burning. For the newer Geneva models both
end at silicon burning. The grey
shaded region represents the range of luminosity for the STARS models
from the end of core-helium burning to the luminosity at the on-set of
core neon burning (see Sect.\,\ref{section:models} and Fig.\,\ref{fig:mass-logL})}
\label{fig:massLcomparison}
\end{figure}

\begin{figure}
    \centering
    \epsfig{file = 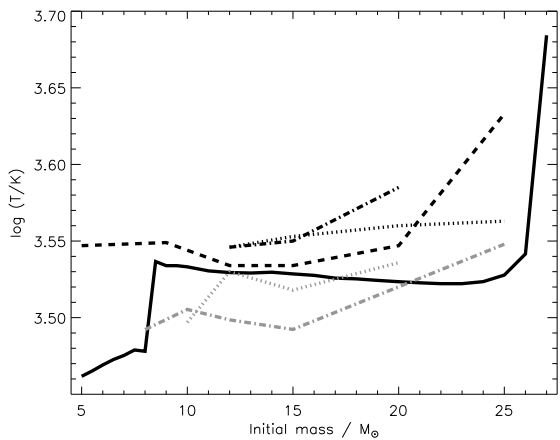, width=80mm}
    \caption[]{Similar to Figure \ref{fig:massLcomparison} but with initial mass versus effective temperature at the end of the stellar model.}
\label{fig:massTcomparison}
\end{figure}

\begin{figure}
    \centering
    \epsfig{file = 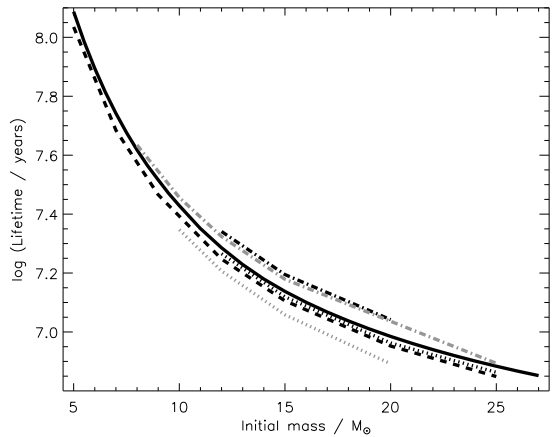, width=80mm}
    \caption[]{Similar to Figure \ref{fig:massLcomparison} but with initial mass versus lifetime to the end of the stellar model.}
\label{fig:masstimecomparison}
\end{figure}

\section{Systematic uncertainties in the determination of stellar mass}
\label{sect:extinct}
\subsection{Stellar evolution models}
\label{subsect:modelerror}

The luminosity estimates and limits used to determine stellar masses
are obviously dependent on the stellar model used. Hence 
it is necessary to 
compare our stellar models to other contemporary models of massive
stars. The constituent physics of modern codes is mostly identical,
using the same nuclear reaction rates and opacity tables. The
differences come from the adopted mass-loss rates, the
numerical schemes employed to solve the stellar structure equations
and the treatment of mixing, convection and rotation in the codes
\citep[see][for example]{red2blue,2006MNRAS.370.1817S} Here we
illustrate the differences between our models and those of
\citet{schaller}, \citet{2004A&A...425..649H} and \citet{2000ApJ...544.1016H}. We
consider three details of the end points of the stellar models, these
are the final luminosity, the final effective temperature and the
stellar lifetimes in Figures \ref{fig:massLcomparison},
\ref{fig:massTcomparison} and \ref{fig:masstimecomparison}.

One detail to note first is that the Geneva rotating models
\citep{2004A&A...425..649H} predict a smaller maximum initial mass for
red supergiant progenitors of 22 \msun\ rather than the 27 \msun\ from
the STARS models, while the non-rotating Geneva models
\citep{schaller} predict a maximum initial mass for red supergiant
progenitors of 34 \msun.  Beyond these masses the codes predict the
stars will end as H-deficient WR stars (depending on the mass-loss
recipe employed).  One other noticeable feature is that only the STARS
models follow the process of second dredge-up and produce massive AGB
stars at low masses. This is because the other models have been stopped
before it could occur. Second dredge-up is found at similar masses 
in other codes specifically designed to follow this stage
\citep[e.g.][]{2008ApJ...675..614P,2006A&A...448..717S,2007A&A...476..893S}.

In Figure \ref{fig:massLcomparison} the difference in final
luminosities between models is illustrated. The model sets with the
greatest difference with our $M - \logl$ relation are the non-rotating
Geneva and Heger \& Langer models, while their rotating models have
reasonable agreement with the STARS models and with the older
\citet{schaller} models. The main reason for the relationships not
being exactly similar is because of different assumptions of mixing in
the stellar models and also the tracks end at different points in the
stars' evolution. For example the old Geneva models have lower
luminosities than the STARS models because they end after core carbon
burning while the STARS models progress slightly further to the
beginning of neon burning and we find the luminosity grows 
after core carbon burning. Also the Geneva $9M_{\odot}$ models end
after core helium burning and therefore it underestimates the final
core mass and luminosity.

The newer Geneva models differ in the treatment of mixing and
convection in the models which affects the vigour of the nuclear
burning in the stars and therefore the luminosity. For example they
use a smaller overshooting parameter than the older Geneva models as
mixing is now also provided by the rotation. Thus the rotating models
agree with our final luminosities but the non-rotating star
luminosities are 0.3 dex lower. We emphasise that the new non-rotating 
models are artificially pushed to lower luminosities as the mixing
efficiencies (from overshooting) have been significantly reduced. Otherwise
employing the same mixing parameters as previously employed {\em and}
adding rotational mixing would push all the luminosities too high
to be consistent with observed HRDs. 

In general the uncertainty in final luminosity due to the assumption
of a certain set of stellar models is typically 0.1 dex between our
STARS models and the most up to date rotating models.  However the
issue of how much mixing is included and by which mechanism can lead
to an uncertainty of up to 0.3 dex.  This does not pose a major
problem to our estimates as we are using the luminosity at the end of
core helium burning to estimate the 84\%-confidence upper mass
limits. In Fig.\,2 one can see that the grey area (which highlights
the region in the STARS code between end of core He burning and the
end of the model as discussed in Sect.\,\ref{section:models}) brackets
nearly all the tracks.

While the initial mass to final luminosity is uncertain the final
helium core mass to final luminosity relationship is much
tighter. This is because the size of the helium core is the major
factor in determining a red supergiants luminosity. To remain
consistent with progenitor masses measured from cluster turn-off ages
and with previous studies we have determined {\em initial masses}
($M_{\rm i}$) for our
progenitors. Helium core masses ($M_{\rm He-core}$) can then be estimated from these by
using the following relation determined from the STARS stellar models:
\begin{eqnarray}
M_{\rm He-core}=(1.14\pm0.05) M_{\rm i} - (0.016\pm0.002) M_{\rm i}^{2} \nonumber\\
 - (0.50\pm0.29).\nonumber\\
\end{eqnarray}

The surface temperatures in Figure \ref{fig:massTcomparison} show that
the final predicted effective temperatures are all within 0.05 dex
with the \citet{2000ApJ...544.1016H} models being coolest. At higher masses the temperature
sharply increases as the hydrogen envelopes in these cases are low mass ($<0.5M_{\odot}$)
as the star is stripped due to mass loss. These temperatures are highly sensitive to the boundary
conditions in the stellar models as well as the opacities used, so it
is not easy to simply identify the reason for the 
differences between the models.  But the uncertainty ($\pm200$\,K)
is well below the uncertainty in the surface temperature implied from
spectral types of observed SN progenitors (typically ($\pm500$\,K, from the
colour-spectral type estimates). 

The stellar lifetimes in Figure \ref{fig:masstimecomparison} also show
close agreement. The most discrepant are the rotating Geneva
models. Rotation increases the hydrogen burning lifetime considerably
by mixing fresh hydrogen into the core and extending the hydrogen
burning lifetime of the star.  The increase, however, is less than 0.1 dex
and therefore masses derived from lifetimes (i.e. turn-off masses
for 2004am and 2004dj) are consistent between stellar models.

Hence we conclude that the use of different stellar models
are unlikely to have a significant effect on the estimated
masses and mass limits we have derived, especially if a 
single method is employed and all masses are derived on a 
homogeneous scale. Furthermore while the initial masses 
may be somewhat dependent on the choice of  single star models, 
 the final helium core masses that our initial mass correspond to 
should be reliable.

\subsection{Extinction determinations}
\label{subsect:extinction}
It is likely that our largest source of error comes from the extinction that 
we assume is applicable to the line of sight toward each SN. This is not 
likely to be a simple systematic effect that would change all 
the mass estimates and limits by a constant. However we need to
consider if we are consistently underestimating the extinction toward the 
progenitor stars and by what magnitude. 

\begin{figure}
    \centering
    \epsfig{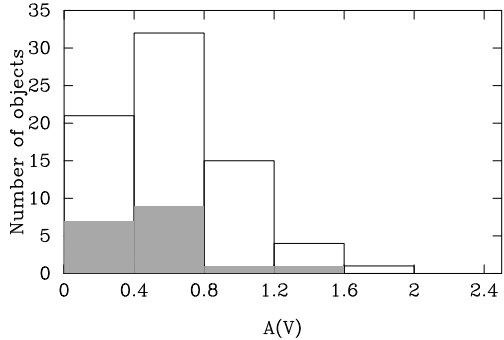}
    \caption[]{Histogram of the $A_{V}$ values adopted for the
progenitor stars (shaded bars) 
compared to $A_{V}$ values of red supergiants in the
LMC and the SMC from \cite{2006ApJ...645.1102L} (open bars). 
There are 18 SN progenitors plotted here
in the shaded bars (2004am and 2002hh were excluded
as explained in the
text), and 73 RSGs from the combined SMC and LMC samples.}
\label{fig:Av-rsg-comp}
\end{figure}

The extinctions which have been estimated for each of the progenitor 
stars come from several methods. All of these suffer from their
own uncertainties
and problems and in general we favour taking the mean of different results. 
The rationale is that no single method is clearly superior to the others
and a mean of several, possibly problematic, estimates is better than 
adopting one. The extinctions have been estimated by some of the following
techniques: measurements of the 
Na\,{\sc i} ISM absorption lines and calibrating this using 
\cite{2003fthp.conf..200T} ; 
comparing the early continuum slopes to the well observed and
reliably modelled 1999em and also to unphysically hot black body continua ; 
fitting stellar spectral energy distributions (SED) to
the surrounding massive star population within about 10-100\,pc; and 
if the SN exploded within an H\,{\sc ii} region or compact cluster then 
using the value determined from the nebular emission lines or the cluster 
SED. An example of the applications
of all of these methods applied to SN2001du can be found in 
\cite{2003MNRAS.343..735S}. In the latest case of SN2008bk there is no 
accurate extinction measurement toward the SN yet and \cite{mattila08bk}
have employed $A_{V} = 1.0$ to fit a late type M4I to the observed 
$IJHK$ progenitor colours. An extinction of less than this results in a 
star which is intrinsically too red to be compatible with known massive
red supergiants. 
We do not revisit the reliability of every method applicable
to each event as this is dealt with in the relevant references 
cited in the subsections for the events above. However we
should consider if these methods as a whole are applicable and
what are the likely sources of error. 

The primary concern is our assumption that the extinction toward the
SN (and surrounding stellar population) is directly applicable toward
the line of sight to the progenitor. Two methods probe the
intervening line of sight directly to the SN. However 
the early soft X-ray, UV-optical flash of the explosion
could conceivably have photo-evaporated 
substantial circumstellar dust close to the progenitor \citep{1983ApJ...274..175D}. 
\cite{2000ApJ...537..796W} have suggested that the X-ray and UV afterglow of a
GRB could photo-evaporate
a large cavity surrounding the progenitor star. 
Scaling the GRB energy to the observed 
flux from recent shock breakouts observed for type II-P and Ibc
SNe, Botticella et al. (in prep)
 have estimated how much dust could conceivably be
destroyed in a dense circumstellar envelope. It would appear
that it is quite possible for such a UV, soft X-ray flash 
to destroy dust masses that could provide several tens of
magnitudes of extinction in the optical $V$-band.

This is of obvious concern when one considers that the observations of
luminous red supergiants in the Magellanic Clouds, the Galaxy and the
Local Group are known to produce large quantities of dust
\citep{2005A&A...438..273V,2005ApJ...634.1286M}.  
A histogram of extinctions toward optically selected red
supergiants in the LMC and SMC clusters by \cite{2006ApJ...645.1102L} 
suggests that RSGs
tend to be redder (by on average $A_{V} \simeq 0.4^{\rm m}$) compared
to the extinctions toward the other OB-stars in their stellar
associations. The mean extinction toward LMC and SMC RSGs is 0.60$^{m}$ and
0.73$^{m}$  respectively.  The mean extinction that has been determined
toward our SN progenitors is $0.7\pm1.1$ (from Table\,\ref{table:main}) 
and the large standard deviation is due to 2002hh with $A_{V} = 5.2$. 
In this calculation we have left out 2004am which clearly suffers 
from high extinction in line of sight in M82; its host cluster
is heavily reddened hence its high $A_{V}$ is unlikely to be due
to CSM dust shells (see Sect.\ref{subsect:2004am}). 
This simple comparison would suggest there is no clear
difference in the extinctions of the two samples. 
If we leave further exclude 2002hh (as an anomalously high extinction 
object), we have a mean extinction toward the 18 
progenitors of $0.44\pm0.34$
In Fig. \ref{fig:Av-rsg-comp} we show the histogram
of our $A_{V}$ estimates towards the likely red supergiant progenitors
and compare them to the LMC and SMC combined population (disregarding
the two highest values for 2004am and 2002hh). 
There is
some evidence to suggest that we have more progenitors in the 
 lowest bin $0-0.2^{m}$ than would be typical for RSG progenitors.
This is not unexpected
as for several of our SNe we have been forced to adopt the extinction
towards the host galaxy alone due to lack of additional information. 
Given the low numbers of objects and the differences in the 
sample size, the distribution between $0.2 - 4.0$ does not appear
to be a major cause
for concern. There are five events for which we adopt a low (foreground
Milky Way component only) extinction of $A_{V} < 0.1$. If these have
been underestimated by  $A_{V} \simeq 0.3^{\rm m}$, that would
bring the mean $A_{V}$ of the progenitor sample into line with the 
SMC/LMC RSG populations. In doing so the luminosity 
and mass limits for each event would increase by :
1999br (\logl =4.88,    $<16$\msol) ; 
2003ie, (\logl =5.49,   $<27$\msol)  ; 
2006my,  (\logl =4.55,   $<13$\msol) ; 
2006ov   (\logl =4.35,   $<11$\msol)
2007aa  (\logl =4.6,  $<13$ \msol). 
This does not affect the lower mass limit that we derive below for the 
sample from the maximum likelihood analysis and has a minimal affect 
on the maximum mass as we shall see (Sect.\,\ref{section:imfs}).

The extinction remains the major source of uncertainty and there exist
populations of dusty red supergiants which are obscured in the visual 
and near-IR (often by $\sim10^{m}$ in the $V$ band) and are mid-IR
bright as their optically thick dust shell is heated by the stellar
luminosity and this light is preprocessed into thermal mid-IR emission
from dust grains \citep{2006A&A...447..971V,2005A&A...438..273V,1997A&AS..125..419L}. 
These would not appear in the \citet{2003AJ....126.2867M} and \citet{2002ApJS..141...81M} sample as they are too faint optically. However
the relative numbers of red supergiants ({\em excluding} AGB stars,
which are below the mass threshold to produce SNe) which are visually
obscured \citep[e.g. objects similar to those
in][]{1998A&A...337..141V, 2005A&A...438..273V} and those which suffer
moderate extinctions 
\citep[the optically detectable stars in][]{2003AJ....126.2867M} is unknown. Such a study to quantify the
latest stages in stellar evolution in a complete manner would be
highly desirable and the Magellanic Clouds would appear to be an
excellent laboratory. Clearly we do see a large population of RSGs with
low-moderate extinctions as shown by \cite{2006ApJ...645.1102L} and \citet{2003AJ....126.2867M}, and we suggest that our
progenitors are part of this population. How many dust obscured RSGs
which are missing from optical and near-IR surveys remains to be seen.

Additionally if a mass-loss mechanism (such as pulsations) occurs 
during the final stages of evolution of most massive stars as
core-collapse approaches one might envisage that the progenitors 
become systematically more obscured. This would invalidate the 
comparison with the LMC RSG population. Such severe mass-loss
is not well constrained observationally or theoretically but 
if it occurred frequently one would expect to see signatures of 
circumstellar gas as well as dust. The type II-P tend to be 
low-luminosity radio and x-ray emitters and tend not to show
narrow hydrogen or helium lines suggestive of CSM shells
\citep{2006ApJ...641.1029C}.

While there is no clear evidence that dense dust shells form around
II-P progenitors, we at  present 
cannot rule out some visual
obscuration due to an optically thick dust shell which was then 
evaporated by a soft X-ray, UV and optical flash at shock breakout. 
This has been suggested as a possible mechanism for SN2008S
\citep{2008arXiv0803.0324P}, see Sect.\,\ref{section:08S}.

\section{Maximum likelihood analysis of the masses of progenitor stars}
\label{section:imfs}

Using the measurements of progenitor masses in Table \ref{table:main} 
 it is possible to estimate parameters that describe the progenitor
population. 
The three parameters of the progenitors that we are interested in are the
minimum initial mass for a type II-P SN, the maximum initial mass and the
initial mass function (IMF) of the population.

Estimating these from a small sample is
not difficult but the relatively small number of data points can 
restrict the accuracy with which one can constrain the most probable 
values. We therefore use the unbinned maximum likelihood method,e.g. 
\citet{1996PhRvD..54.1194J}. For a large number of objects 
this effectively becomes a
$\chi^{2}$ method. The likelihood is defined to be,

\begin{equation}
{\cal L} = \prod  P_{i}(m),
\end{equation}

where $P_{i}$ is the probability of the $i$th event to have mass
$m$. We must define a function for the probability of each event,
$P_{i}$, and then maximize the likelihood to find the
parameters that give the most probable set of events.

To make the maximization more straight-forward we first take the natural logarithm of
the likelihood function and so we are required to calculate a sum
rather than a product,

\begin{equation}
\ln {\cal L} = \sum \ln P_{i}(m).
\end{equation}

The probability function that describes the probability that a progenitor 
will have mass $m$ within a certain mass range is essentially the IMF.
We do need to treat the detections and the non-detections  differently. 
For non-detections we adopt the probability function,

\begin{equation}
P_{i} =\int_{m_{\rm min}}^{m_{i,{\rm limit}}} \frac{m^{\Gamma-1}}{(m_{\rm min}^{\Gamma}-m_{\rm max}^{\Gamma})} dm,
\end{equation}

where $\Gamma$ is the value of the IMF, Salpeter being $-1.35$,
$m_{\rm min}$ is the minimum mass for a type II-P SN, $m_{\rm max}$ is
the maximum mass for a type II-P SN, and $m_{i,{\rm limit}}$ is the
upper mass limit for the $i$th non-detection. If $m_{\rm max}$ is less
than $m_{i,{\rm limit}}$ we set $P_{i} = 1$. If $m_{\rm min}$ is
greater than $m_{i,{\rm limit}}$ we set $P_{i} = 0.16$ because our
mass limits are 84 percent confidence limits so there is a chance that
the progenitor could be more massive. This is only important for the
mass limit from SN 2004et.

\begin{figure}
    \centering
    \epsfig{file = 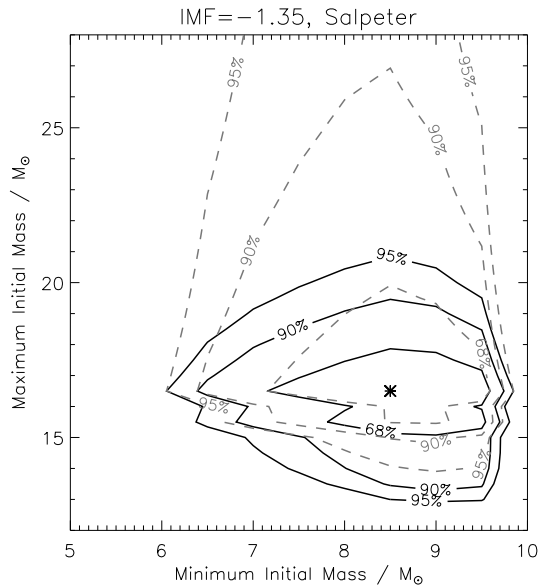, width=80mm}
    \caption[]{Plot of the likelihood function for the mass ranges of type II-P progenitors. The star 
indicates the parameters 
with the highest likelihood and the contours the confidence regions. The dotted grey lines show the results using the six
detections only, which results in a lower mass limit of 8.5\msol. The solid black lines show the contours using the 
fixed lower limit and allowing the maximum mass to vary.}
\label{fig:maxlikelihood}
\end{figure}

For detections the case is more complicated. The errors for the
progenitor luminosity are roughly Gaussian, but converting to an
initial mass affects the distribution. We first take the best
estimated mass, $m_{j}$, as the most probable value for each
detection. Then above this mass we integrate the IMF up to the upper
uncertainty on the mass estimate, 
$m_{j,{\rm high}}$. Below $m_{j}$ we assume the
probability distribution is a straight line going to zero at the lower
uncertainty on the 
mass limit, $m_{j,{\rm low}}$. While these error distributions are somewhat
arbitrary they avoid skewing the overall distribution to higher or
lower masses as happens when using a Gaussian distribution to 
describe the uncertainties. We have
experimented with different probability functions for detection, 
for example using triangular and rectangular error 
functions at the low and high uncertainty limits.  The
$m_{\rm max}$ and $m_{\rm min}$ would typically 
vary by $\approx 1M_{\odot}$ which 
is within the uncertainty we derive for these parameters.  
We feel our chosen method for describing the probability 
function in the case of detections is the best representation of
the asymmetric errors on the mass estimates. Hence for detections we
use the following probability distribution 

\begin{eqnarray}
P_{j}& =   \int_{m_{j,{\rm low}}}^{m_{j}}   \frac{(m -  m_{j,{\rm low}})  m_{j}^{\Gamma-1}}{(m_{j}-m_{j,{\rm low}})(m_{\rm min}^{\Gamma}-m_{\rm max}^{\Gamma})} dm \nonumber \\
     & +    \int_{m_{j}}^{m_{j,{\rm high}}}  \frac{m^{\Gamma-1}}{(m_{\rm min}^{\Gamma}-m_{\rm max}^{\Gamma})} dm.\nonumber  \\
\end{eqnarray}

If $m_{j,{\rm low}}$ is lower or higher than $m_{\rm min}$ and $m_{\rm
max}$ then the integral is truncated within these limits.

 We calculate the likelihood using the masses for the SNe progenitors
listed in Table \ref{table:main} and allow $m_{\rm min}$ and $m_{\rm max}$
to vary, while fixing the IMF slope to Salpeter ($\Gamma=-1.35$). 
Originally  we  attempted to let the IMF vary as well as the
maximum  and minimum mass values but find it constrains the IMF only 
very weakly and we chose to fix it at three different values, as
justified below. 

Furthermore we estimate the confidence regions from,

\begin{equation}
\ln {\cal L}_{max}- \ln {\cal L} = \frac{1}{2} \chi,
\end{equation}

where for two parameters when $\chi=2.3,$ $4.6$ and $6.2$ we have the
68, 90 and 95 percent confidence regions \citep{1992nrfa.book.....P}.

\begin{figure*}
    \centering
\parbox[t]{8cm}{
\psfig{file=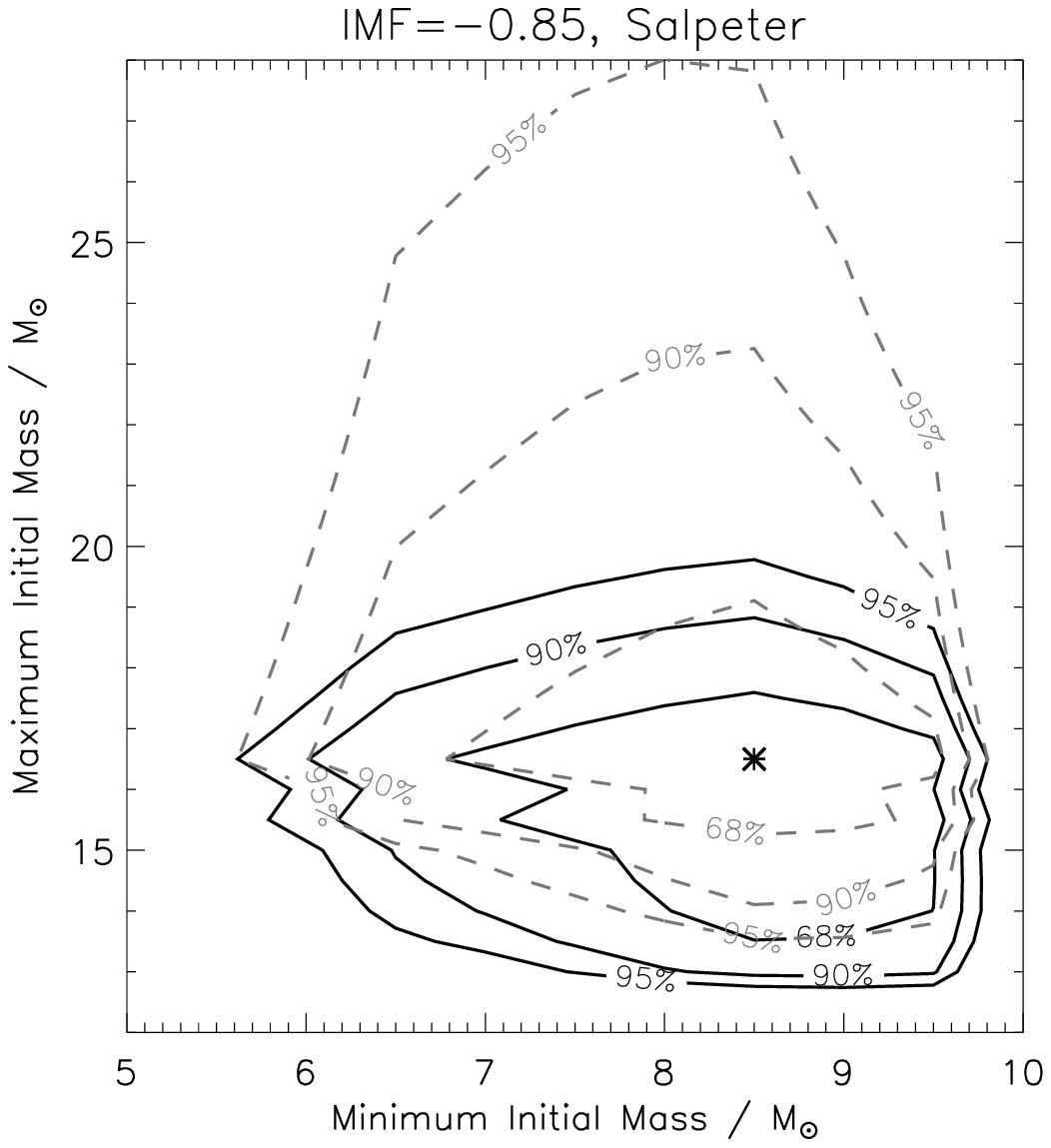,angle=0,width=8.0cm}
}
\parbox[t]{8cm}{
\psfig{file=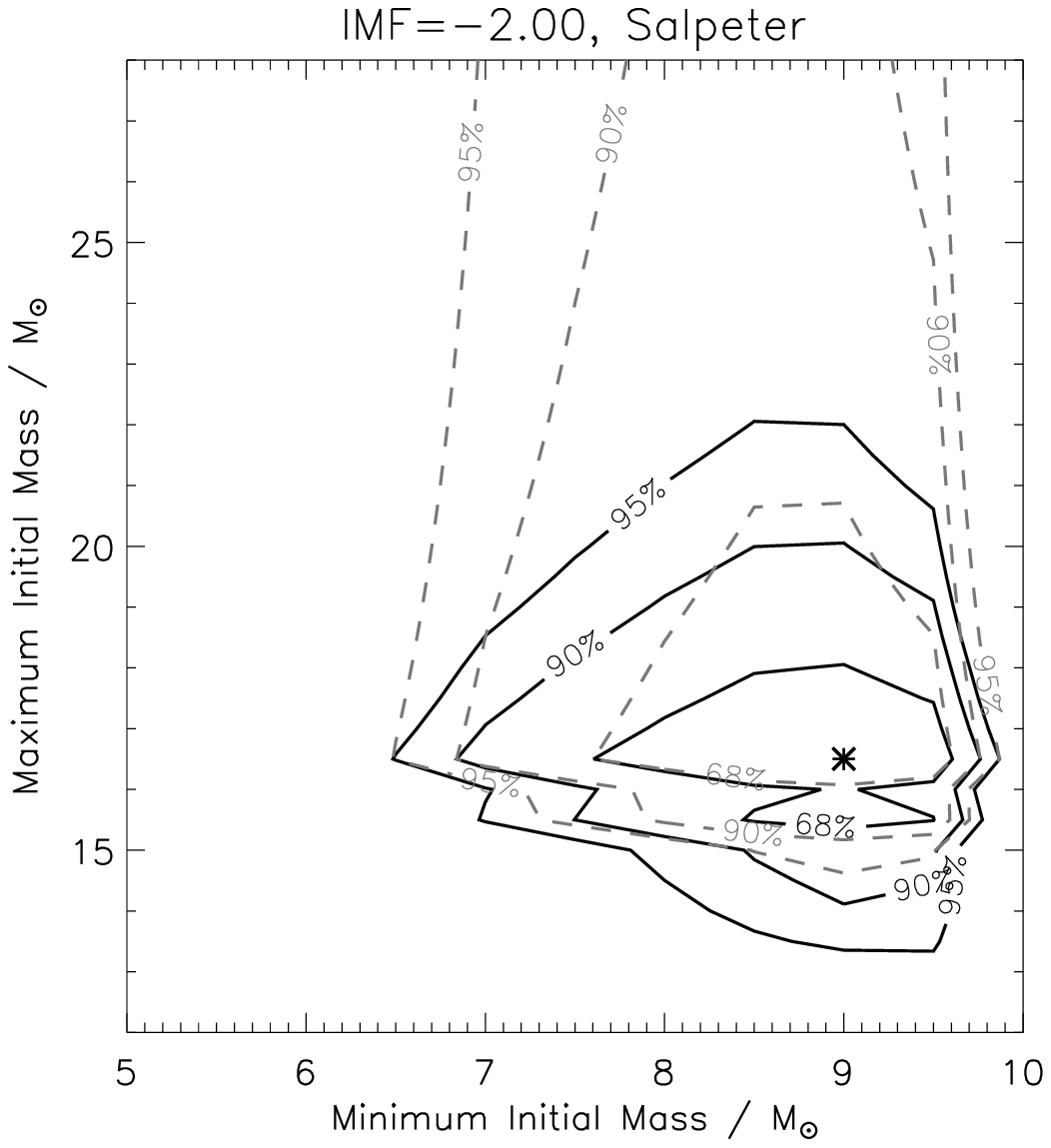,angle=0,width=8.0cm}
}
    \caption[]{As in Figure \ref{fig:maxlikelihood} but with a shallower (left) and steeper (right) IMF.}
\label{fig:maxlikelihood2}
\end{figure*}

We first estimate $m_{\rm min}$ and $m_{\rm max}$ using the six
detections only (without incorporating the upper limits).  The results
can be seen in dashed contours in Figure \ref{fig:maxlikelihood}. The
parameters we estimate are $m_{min}=8.5^{+1}_{-1.5} M_{\odot}$ and
$m_{max}=16.5^{+4}_{-1.0} M_{\odot}$.  We then recalculate the
likelihoods using both the detections and upper limits in the analysis
but fix the minimum initial mass to $8.5M_{\odot}$ as derived from the
detections alone. This is because the non-detections only provide
meaningful information on the maximum initial mass $m_{\rm max}$. In
contrast they provide only a weak constraint on $m_{\rm min}$ as when
combined with the IMF they would simply favour a low mass due to the
rising probability of having more low mass stars. We suggest it is
more reasonable to calculate $m_{\rm min}$ from the detections and
effectively this converges toward the lowest masses detected in the
progenitor sample. 
The upper mass limits have a strong impact on the uncertainty 
 on $m_{\rm max}$ and we
determine that $m_{max}=16.5\pm1.5 M_{\odot}$. With the error on
$m_{\rm max}$ reduced significantly, this suggests that at 95\%
confidence level the maximum initial mass to produce a
type II-P is 21\msol.

In all of the above we have assumed a Salpeter IMF and it is reasonable
to question the validity of this assumption. 
\cite{2008arXiv0803.3154E} has recently reviewed evidence for the 
variation in the IMF slope in local star forming environments
such as Galactic and Local Group Galaxy clusters and field populations
within the Galaxy and the Magellanic Clouds. In clusters and OB associations with total 
masses between $10^{2} - 10^{4}$\msol\ there is little evidence
for strong and real deviations from the Salpeter slope of $\Gamma = -1.35$
above and beyond the RMS measurement errors determined in each region
of high mass stars (with typical uncertainties of order $\pm0.1$ to $\pm0.3$). 
That is not to say that such real variations do not exist, only that 
stochastic affects mean that determining the true IMF in localised
regions can at best reach the accuracy of a few tenths. There is some
evidence that flatter IMF slopes exist in very dense starforming regions 
such as NGC3603 \citep[$\Gamma = -0.9\pm0.15$;][]{2006AJ....132..253S}
and the Galactic centre \citep[$\Gamma = -1.05\pm0.05$;][]{2006ApJ...653L.113K}. 
Also there is evidence that the field population may show much steeper
slopes, with $\Gamma = -1.80\pm0.09$ applicable for a large sample of 
field stars lying outside clusters and associations in the
 LMC \citep{1998AJ....116..180P}. Extreme values of 
around 3-4 have 
even been found \citep{2002ApJS..141...81M,2002A&A...381..862G}
though it is unclear to what extent this is simply 
due to stellar drift out of low mass clusters. 
\cite{2008arXiv0803.3154E} surmises that $\Gamma = -1.35$
appears to be fairly typical in moderate mass clusters and starforming
regions and variation around this is, on the whole, limited to 
approximately $\pm$0.5. As our SNe, and their progenitors, seem 
to reside in typical star forming regions and the field of their 
host spirals there is no compelling evidence to favour an IMF too dissimilar
to Salpeter. In Fig.\,\ref{fig:maxlikelihood2} we have recalculated
the maximum 
likelihood values with the extreme 
IMFs suggested in \cite{2008arXiv0803.3154E} of $\Gamma = -0.85$ and 
$\Gamma = -2.0$. For  the shallow IMF slope of $\Gamma = -0.85$ the 
best estimates of the minimum and maximum initial mass are
unchanged but the 
uncertainties increase slightly  to $m_{min}=8.5^{+1}_{-2}$\msol\
and  $m_{max}=16.5^{+1}_{-3}$\msol. 
This shallow IMF is unlikely to be
representative of our progenitor environments as they are not 
(apart from perhaps 2004am and 2004dj) in dense clusters such as 
seen in NGC3603 and the Arches cluster at the Galactic centre. 
For the steeper IMF, the most likely 
minimum initial mass increases to 
$m_{min}=9^{+0.5}_{-1.5}$\msol\
and the 95\% confidence limit for $m_{max}$ is pushed to the higher value
of 22\msol. 
In summary there is no strong evidence
\citep[from Local Group studies as reviewed by][]{2008arXiv0803.3154E}
 that our progenitor population
should come from a massive stellar population with an IMF slope 
significantly different (i.e. by more than $\pm0.5$) than 
Salpeter, and adoption of the either of those extreme values 
does not significantly affect the values of 
$m_{\rm min}$ and $m_{\rm max}$.

Two remaining uncertainties are extinction and the value of $H_{0}$.
As discussed in Sect.\,\ref{subsect:extinction}, if we have underestimated the
extinctions toward five events with non-detections and replace
them with the slightly higher masses, $m_{min}$ and $m_{max}$ change
by less than $0.1 M_{\odot}$.  
As discussed in Sect.\,\ref{section:masses}, if we employ 
 $\mathrm{H_{0} = 65 km\,s^{-1}\,Mpc^{-1}}$, then the luminosity
differences of the five progenitors for which we employ only
a kinematic host galaxy distance
(see Table\,\ref{table:main})
 would change by  $+0.22$\,dex. 
This corresponds to approximately 2-3\msol\ in the ZAMS estimate. 
The value of $m_{min}$ does not change, but the the maximum mass 
increases to $m_{max} = 18.5\pm2$. This is due to SN1999ev having the most 
massive progenitor estimate and the host of SN1999ev has a kinematic distance 
only.  Similarly, using $\mathrm{H_{0} = 85 km\,s^{-1}\,Mpc^{-1}}$
keeps $m_{min}$ within 8-9\msol (as all the SNe which determine this number have
distances from other methods), but  the maximum mass reduces to 
$m_{max} = 15.5^{+1}_{-0.5}$\msol. 
This illustrates that in the future it is important to try to find
the type II-P SNe from the highest mass progenitors to tie down 
$m_{max}$ as        reliably as possible.

\section{Discussion}
With our analysis of the progenitor observations and mass estimates 
we are able to consider some outstanding questions on the nature of 
supernovae progenitors from a firm observational footing. 
There has recently been much discussion on the initial masses of progenitor
stars of SNe of all types \citep{2007ApJ...656..372G,2007ApJ...661.1013L,2003MNRAS.343..735S}
and the nature of faint type II-P SNe 
\citep{2006MNRAS.370.1752P,2003astro.ph..8136N,2003MNRAS.338..711Z,1998ApJ...502L.149Z}

Our maximum likelihood analysis reveals that the progenitors of type
II-P arise from stars with initial masses between $8.5 ^{+1}_{-1.5}$ and
$16.5\pm1.5$\msun. The derivation of the mass range assumes that
the stars are red supergiants, in that to transform the optical or near infra-red
limiting magnitudes to a luminosity and mass we must assume a stellar 
progenitor spectrum with a suitable photospheric temperature (or range
of temperatures).  This is well justified in that
four of the detections have colours consistent with them being late
K-type to mid M-type supergiants and the requirement that a II-P
lightcurve results from the explosion of a star which has an extended
H-rich envelope. 
\citep[$R_{\ast} \sim 500-1000$R$_{\odot}$][]{1980ApJ...237..541A,1976ApJ...207..872C,1993ApJ...414..712P}.
The maximum initial mass is important to 
constrain what the final evolutionary stage of the most massive
stars and the lowest initial mass that could produce a type Ib/c, or 
perhaps II-L and IIn, explosion. The minimum initial mass that can 
support a SN explosion is of great interest for explosion models, 
stellar evolution, comparing with massive WD progenitor masses
and galactic chemical evolution. 

The mass range that we find for the progenitors is much lower
than ejecta masses of a sample of
II-P SNe suggested by \cite{2003ApJ...582..905H}.
This study estimated ejecta
mass of between 14-56\msol\  from the 
application of the \cite{2003MNRAS.346...97N} formulae to determine
energy of explosion, radius of progenitor and ejected mass. Even
though the error bars on the masses are large there is a clear
discrepancy between our results. The determination of the ejecta
masses is very sensitive to how the mid-point in the lightcurve is
defined to determine $V_{\rm 50}$ (the visual magnitude at 50 days)
and $v_{\rm 50}$ (the ejecta velocity at the same point). The
measurement of the latter is also highly dependent on which ionic
species is used to measure the photospheric velocity and
\cite{2003MNRAS.346...97N} suggests that the bolometric lightcurve
should be used to define the plateau mid-point.  It appears to us that
the choice of the point at which to define the measured parameters has
a critical effect on the physical values determined and caution should
be employed when applying this method.  We note that
\cite{2003MNRAS.346...97N}, with similar data to \cite{2003ApJ...582..905H}
has determined ejecta masses in the range 10-30\msol, closer to our
progenitor mass range but still systematically higher. It is likely
that the the mid-points of the plateau lightcurves defined by 
\cite{2003ApJ...582..905H} (and the parameters thus arising) were not
exactly compatible with those required for the \cite{2003MNRAS.346...97N}
equations to be applied. The lower ejecta masses of \cite{2003MNRAS.346...97N}
are probably more reliable in that they are estimated with the 
appropriate input parameters and are a better match to the progenitor
masses we determine. 

One may ask if a ZAMS mass of 8.5\msol\ is large enough for a 
long plateau phase to be sustained. In our model the star would
loose 0.5\msol\ due to stellar winds and with 
a neutron star remnant of 1.5\msol, this leaves about 6.5\msol\
for the ejected mass. \cite{2006MNRAS.369.1303H,2005MNRAS.359..906H} 
showed that the low progenitor masses of SN2004A and SN2003gd (8-9\msol)
were consistent with the observed recombination powered plateau 
duration, but only just within the error bars of both model
estimates \citep[see also][]{2003MNRAS.343..735S}
In a future paper we will analyse the lightcurves of 
a large subset of the SNe presented here to determine if their 
progenitor mass estimates are consistent with the ejected masses
required to produce their plateau phases.

\subsection{The minimum mass of II-P progenitors}
\label{II-P:mass-min}

Theory predicts that a few of the low-mass progenitors should be
massive AGB stars, sometimes referred to as Super-AGB stars 
\citep{2004MNRAS.353...87E,2007A&A...476..893S,2008ApJ...675..614P}. 
The cores of these objects never
reach high enough temperatures to produce iron, rather the 
oxygen-neon core grows to the Chandrasekhar mass and an 
electron capture SN is triggered. These explosions have been 
predicted to produce less luminous SN than in normal iron-core
collapse \citep{2006A&A...450..345K}, and perhaps this signature
could be used to find real candidates and to identify progenitor
stars at the lowest mass range
(see the Sect.\,\ref{subsect:faintSNe} and references therein for a discussion of
the lowest luminosity SNe). 
From their models of super-AGB stars, 
\cite{2008ApJ...675..614P}
suggest that the number of these stars at solar
metallicity would result in them producing 3 percent of 
the local core-collapse SNe. 
This increases to greater than 10 percent at metallicities below a tenth 
solar. From the observational properties, one of 
the best studied examples of a low-luminosity, 
low ejecta velocity event is SN2005cs, and indeed we do suggest it 
had a low progenitor mass of $8\pm2$\msun.  
However Eldridge, Mattila \&
Smartt (2007) show that there is a clear observational signal 
for AGB stars, in that these progenitors should be much cooler than higher
mass red supergiants and hence be quite bright at near infra-red (NIR) bands. 
Deep NIR pre-discovery images were available for SN2005cs, 
and we showed that it was unlikely to be a massive AGB star. 
Thus we suggest that all of the 20 progenitors were
genuine Fe core-collapse events, and we have no evidence for 
any of them being electron-capture events in ONe cores. 
We also have no evidence to support the idea
that stars in the range $\sim$7-9\msol\
go through 2nd dredge-up and end as quite high luminosity progenitors, 
either as S-AGB stars with ONe cores or genuine Fe core-collapse events.
Our models \cite[and those of][]{2008ApJ...675..614P} would suggest that 
the luminosity of these events can reach $4.6\leq$ \logl $\leq 5.1$ which
is significantly more luminous than any of the progenitors detected 
so far. \cite{2004MNRAS.353...87E} and \cite{2008ApJ...675..614P}
point out that this evolutionary phase is very dependent on the treatment
of semiconvective mixing and convective overshooting. The fact that we
don't see luminous progenitors with \logl$\gtrsim 4.6$\,dex
(the highest of our sample : SN2005cs) 
would apparently disfavour the scenario in which 
$\sim$7-9\msol\ stars increase their luminosity due to 2nd dredge-up before
collapse. The apparent
luminosity of the progenitors (around $4.3-4.6$) favours a lower limit than is
normally assumed for core-collapse with no luminosity spike.

The lower mass limit we derive from Sect.\,\ref{section:imfs} of $8.5^{+1}_{-1.5}$\msol\ 
is interesting to compare with the maximum stellar
mass that produces white dwarfs. A compilation of mass estimates of
white dwarfs by \cite{2006MNRAS.369..383D} suggests that, in Milky Way
intermediate age clusters, stars up to 6.8-8.6\msol\ produce white
dwarfs and they suggest this as the initial mass range for
core-collapse SNe. \cite{2008AJ....135.2163R} suggest that a
homogeneous analysis of WDs in their Lick-Arizona White Dwarf Survey
(LAWDS) confidently determines the maximum mass to be no less
than 6\msol. Further recent evidence 
 suggests that the mass limit is
no less than 7.1\msol\ \citep{2008arXiv0811.1577W}. 
A slightly higher 
mass limit is not ruled out as there is ongoing work on younger
clusters to find WDs and determine their progenitor age (C. Williams, 
private communication). 
Hence the two, very different approaches, of SN progenitor mass and 
WD progenitor masses appear to be  converging toward $8\pm1$\msol.  
Unless both methods are
significantly in error it would seem unlikely that the lower mass for
a core-collapse SN is outside this range.  Theoretically several
mass-limits have been determined, ranging from 6 to 11\msol\
\citep[][ and references
therein]{1999ApJ...515..381R,2003ApJ...591..288H,2004MNRAS.353...87E,2008ApJ...675..614P}
depending critically on the amount of convective overshooting
employed. In our analysis we have used models with convective
overshooting as there is growing evidence that extra mixing is
required above that predicted by mixing-length theory
\citep[e.g.][]{overshooting1,overshooting2}.  We suggest that a
minimum initial mass of 10\msol\ or more can now be ruled out for two
reasons. Firstly we detect four progenitors with best estimated
masses below 10\msol\, although admittedly the individual uncertainties
would not rule out a higher mass progenitor.  Secondly in our maximum
likelihood analysis masses at 10\msol\ and above are ruled out at over 95
percent confidence, even with a steep IMF of $\Gamma = -2$.
This value is supported by the fact that type II-P SNe are not 
always associated with underlying H\,{\sc ii} emission line regions
in their host galaxies \cite{2008MNRAS.390.1527A}
which would suggest their progenitors
are from a population of less than $\sim$10\,\msol. 

We suggest that $8.5^{+1}_{-1.5}$\msol\ is the current best estimate, 
based on observational constraints, for the lower limit to 
produce an Fe core-collapse driven SN of type II-P. This is in 
agreement with the mass range for the most massive progenitors
of WDs.

\begin{figure}
    \centering
    \epsfig{file = 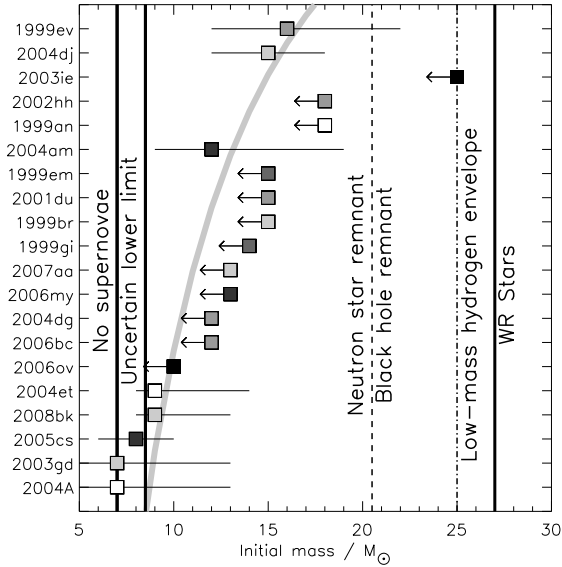, width=80mm}
    \caption[]{The initial masses of all our type II-P progenitor stars, compared with our 
theoretical limits for production of supernovae of different types and type of compact 
remnant. The box symbols are shaded on a metallicity scale, the lighter the shade the lower
the metallicity, with the values taken from Table\,\ref{table:main}.}
\label{fig:heger}
\end{figure}

\subsection{The maximum mass of II-P progenitors}
\label{II-P:mass-max}

The maximum mass of a star that can produce a II-P supernova 
is an important threshold to constrain. 
In Fig. \ref{fig:heger} we summarise the initial masses
of all the progenitors. Similar plots were first shown by 
\cite{2003ApJ...591..288H} and \cite{2004MNRAS.353...87E} 
with a large range of metallicity 
plotted on the vertical axis, from super-solar to metal free. 
As our progenitor stars cover a relatively small range in 
metallicity, we have removed the axis scale and instead 
flagged the points with a metallicity coded grey scale. 
Clearly the highest mass of a detected progenitor is
 16$^{+6}_{-4}$\msol\ (SN1999ev) with one upper limit above 20\msol, 
due to shallow pre-explosion images (SN2003ie). Our
estimated maximum initial mass for a II-P 
(Sect.\,\ref{section:imfs}) is $m_{max}=16.5\pm1.5$\msol, 
with a 95\% confidence limit (assuming Salpeter IMF $\Gamma=-1.35$)
of 21\msol. Fig.\,\ref{fig:heger} is effectively a cumulative
frequency distribution (CDF) which is constrained at the lower
and upper mass limits and has an IMF with $\Gamma=-1.35$ 
consistent with the limits in between (the CDF
of the Salpeter IMF is plotted as the thick grey line).
This Salpeter IMF is a good fit to the distribution of 
masses and mass limits, if the hard minimum and maximum 
masses for II-P progenitors hold. 
Stars more massive than about 20\msol\ would be easily detectable in 
our archive images, and there is unlikely to be any bias against 
detecting the most massive progenitors. Hence there does appear
to be a real upper limit to the mass of stars that produce 
normal type II-P SNe. The one caveat to this is if
the progenitor stars suffer large circumstellar extinctions
which are photo-evaporated in the explosion. We discussed this
in Sect.\,\ref{sect:extinct} and while we cannot see a compelling case
for such an effect in our population we cannot rule it out. 

We can compare this maximum mass limit with 
the ratios of CCSN types in Table\,\ref{table:rates}. With a 
maximum possible stellar mass of 150\msol\ 
\citep{2005Natur.434..192F}, 
the fraction of stars born with masses between
8.5-16.5\msol\ (for a Salpeter IMF, $\Gamma=-1.35$) is $\simeq60$ per cent,
closely mirroring the type II-P rate. One might immediately conclude 
that the agreement suggests that all stars above $\sim$17\msol\ produce
the other varieties of CCSNe. However this is too simplistic
and ignores our wealth of knowledge of massive stellar populations 
from Local Group studies and interacting binaries. 

\subsubsection{The red supergiant problem}

Massive red supergiants have been frequently surveyed
in the Milky Way and the Magellanic Clouds, and up until recently 
their luminosities as determined from model atmospheres 
implied that they are found at evolutionary masses up to 40-60\msol\ 
\citep{2003AJ....126.2867M,1978ApJS...38..309H}. However using new MARCS atmosphere
models \cite{2006ApJ...645.1102L} have shown that 
the effective temperatures of these stars have been revised 
upwards and they have combined this with revised 
bolometric luminosities based on $K-$band magnitudes. 
The result is that the highest luminosity red supergiants of 
\cite{2003AJ....126.2867M} and 
\cite{2005ApJ...628..973L} now have warmer effective 
temperatures and luminosities that imply masses of between 12-30\msol. 
\cite{2001AJ....121.1050M}  and \cite{2007ARA&A..45..177C} suggest 
stars with an initial mass of around 25\msol\ could evolve to 
the WN phase in Galactic clusters, at solar metallicity. 
The mass estimates generally 
come from the estimated age of the stellar clusters as measured
from the turn-off. Only two out of 11 in the Massey et al. 
study are as low as 20-25\msol\ and one can really only take this as a lower limit. 
The minimum initial mass to form a WR star in the LMC (and SMC)
has  been estimated at 30\msol\ (and 45\msol\ respectively)
 using similar methods 
\citep{2000AJ....119.2214M}. Stars above these masses, if they 
explode as bright SNe, should produce H-deficient (and He deficient)  
SNe like the Ib/c we observe. Hence there is good agreement between the 
maximum observed masses of red supergiants in the Galaxy and 
the LMC and the minimum mass required to produce a WR star, 
from the ages and turn-off masses of coeval clusters. 
\cite{2007ARA&A..45..177C} points out that there are few Milky Way clusters 
that harbour both RSGs and WR stars which would suggest that
there is a definite mass segregation between the two populations.
The metallicity ranges of our progenitor
sample (Table\,\ref{table:main}) range between solar and LMC, 
hence these studies of Local Group stellar populations would suggest
the minimum initial mass for 
a single star to become a WR (probably of type WN) is
25-30\msol. 

The question is what is the fate of the massive red supergiants between
17\msol\ and 25-30\msol\ ? They appear to exist in this mass range and
one would expect them to produce SN of type II-P but they are missing
from our progenitor population.  A single star of initial mass
of 17\msol\ does not have a high enough 
mass-loss rate to strip its outer layers of enough
mass to become a WR star and hence a Ib or Ic SN (either observationally
or theoretically). 
If our sample of 20 progenitor stars were really
sampled from an underlying population of red supergiants, with
initial masses in the region $8.5-25$\msol, then a 
Salpeter IMF would suggest we should have 4 between $17-25$\msol. 
The probability
that we detect none by chance is 0.018 (or 2.4$\sigma$ significance). 
For a steeper IMF of $\Gamma = -2.0$ the numbers are 3 
stars, probability of 0.05 and 2$\sigma$ significance. 
We term this discrepancy the ``red supergiant problem'', in that we 
have a population of massive stars with no obvious channel of 
explosion. 

One could attempt to fill this mass-gap with the other SN types IIn
and II-L and IIb.  The fraction of stars born with masses
between $17 - 25$\msol\
(within an underlying population of $8.5 - 150$\msol) is 18 per cent,
and Table\,\ref{table:rates} suggests the combined rate of II-L, IIn
and IIb is 12 per cent. Hence it is perhaps appealing to account for
the red supergiant problem by saying that at least some of these
stars form  II-L, IIn or IIb SNe. 
But there is evidence arguing against this. 
\cite{1982ApJ...257L..63T} presented a deep photographic plate of
NGC6946 49 days before the maximum of the II-L SN1980K and 
found no progenitor or discernible stellar cluster. He 
suggested an upper mass limit of $<18$\msol\ and using our 
stellar tracks and more recent distance 
we recalculated this as $<20$\msol\ in \cite{2003MNRAS.343..735S}. 
SNe IIb have  
been suggested to be from interacting binary systems and for 
SN1993J a viable model is a close pair of 14 and 15\msol\ stars. 
The  binary companion to SN1993J's 
progenitor was theoretically predicted and observationally
detected \citep{1993Natur.364..509P,1994ApJ...429..300W,2004Natur.427..129M}.
\cite{2006MNRAS.369L..32R} suggest a similar scenario explains their
detection of a stellar source at the position of the IIb SN~2001ig.  A single
28\msol\ WNL star was favoured as the progenitor of the recent SN~IIb
2008ax by \cite{2008MNRAS.391L...5C}, but a binary system cannot yet
be ruled out.  There is also evidence that very luminous type 
IIn arise from very massive LBV type stars, generally thought to 
be $>40$\msol\  and hence too high mass to solve the problem
 (see Sect.\ref{subsubsect:IIn}). 

One could appeal to rotation as a way out and the
rotating Geneva models of 
\cite{2004A&A...425..649H}
predict an upper mass limit of 22\msun\ for a hydrogen rich
progenitor (for stars rotating initially at 300\,\kms). 
Above this mass a single star ends its life as a WR
and hence a Ib/c SN.  This is well above our estimated maximum initial mass
17\msol, but consistent with the 95\% confidence limit. However this would mean
every II-P progenitor would have to be rotating initially with speeds
around 300 km s$^{-1}$. This is clearly not what we see in the
rotational velocity distributions in the Galaxy, or Magellanic Clouds, 
\citep{2008A&A...479..541H,2006A&A...457..265D,2006ApJ...648..580H}, 
which suggest less than 5 per cent of massive stars should be
rotating at such intrinsic rotational velocities.

\subsubsection{Black hole formation}
An intriguing possibility is that the red supergiant problem is due to
the vast majority of 
high-mass stars above 17\msol\ collapsing to form black holes and
either very faint supernovae or no explosion at all. Theoretically
this has been suggested for some time, for example most recently by
\cite{1999ApJ...522..413F,2007PASP..119.1211F} and
\cite{2003ApJ...591..288H}. Our model stars in the 
mass range of 20-27\msun\ end as hydrogen rich red supergiants with helium
core masses of $>8$\msol\, and such masses have been suggested to
result in the formation black holes 
\cite[this line is plotted for
reference in Fig.\,\ref{fig:heger}; ][]{1999ApJ...522..413F}. 
The models of 
\cite{2007AIPC..924..226L,2003ApJ...592..404L} suggest the maximum 
mass to produce a type II-P SN is 30-35\msol\  and a
minimum initial mass for black-hole formation is 25-30\msol. 
Although \cite{1999ApJ...522..413F} notes  that the mass range 
to produce black holes is theoretically quite uncertain. 
For example reducing the mean neutrino energy by 20\% could
reduce the explosion energy by a factor 2 and push the minimum
mass for black hole formation to as low as 15\msol. 

As pointed out by \cite{2008arXiv0802.0456K}, the collapsar model in
which a GRB is produced along with a type Ic SN, is likely to be too
rare to produce the bulk of the black holes seen in our Galaxy
\citep{1999ApJ...524..262M}. Although the collapsar scenario would
have problems within massive hydrogen rich progenitors \citep[the jet
would have difficulty in escaping from a red
supergiant][]{2001ApJ...550..410M}. \cite{2005ApJ...625L..87Y}
suggest that bright II-L SNe may be black hole forming events, in which
the collapsar mechanism occurs within a massive H-rich star.

 Whatever the explanation we have evidence
for a lack of progenitors above 17\msol\ and perhaps the minimum mass
to form a black hole could be as low as this. It could be that stars
in the 17-30\msol\ range produce SNe so faint that they have never
been detected by any survey. In this case they would typically
be fainter than $M_{R} \sim -12$.  
If the limit for black hole formation is low then it bodes
well for surveys for disappearing stars. 
\cite{2008arXiv0802.0456K} have
suggested that a survey of nearby galaxies over several years would
have a chance of detecting massive stars that disappear without an
accompanying SN. From a similar comparison of a Salpeter IMF
with the general progenitor compilation of \cite{2007ApJ...661.1013L}
they also suggest there may be a dearth of massive
star progenitors. Their calculation is somewhat inexact in
that it includes 1999gi and 2001du as possible detections and is
neither volume or time limited to minimise biases on SN and progenitor
selection effects, and the masses come from many inhomogeneous methods. 
But it does support our quantitative mass
range estimate for II-P progenitors.

\subsubsection{Binaries and Ibc SNe}
A further flaw in the argument that the type II-P rates match the mass
range of 8.5-16.5\msol\ is that it ignores the consequence of binary 
evolution. \cite{1992ApJ...391..246P} suggested that around 15\% of SNe
could be from interacting binaries in which mass transfer causes the
primary to loose its H (and He) envelope. This assumes that about 30\%
of all massive stars are in close binaries that will interact in case
A, B or C mass transfer. Recent results suggest that $\ge60$\% of
massive stars could be in close binaries \cite{2007ApJ...670..747K}
leading
\cite{2007PASP..119.1211F} to claim that perhaps all local Ibc SNe
could be formed in binary systems and the progenitors could thus have
initial masses down to our $m_{\rm min}$ limit of 8.5\msol. It appears
very likely that at least some Ibc SNe are formed in moderate mass
interacting binaries \citep{2007MNRAS.381..835C,2008MNRAS.384.1109E}

At around solar metallicity 
\cite{2007PASP..119.1211F} and \cite{2003ApJ...591..288H}
argue that single, massive 
WR stars all have core masses large enough to form black holes and that 
they can't be the progenitors of the local, normal Ibc SN population. 
They suggest that these should give weak SNe or no explosion at all. 
Current observations have not yet confirmed that massive
WR stars are definitely the progenitors of Ibc SNe. The bulk of 
the population may form black holes with no explosion and a
fraction (with low metallicity and 
high rotational velocities) may form black holes
in the collapsar model with an accompanying GRB \citep{2006ARA&A..44..507W}. 
The ejecta masses and pre-explosion limits 
of SNe Ibc (with no associated GRB) are consistent with them being
stars of $\sim$10-20\msol\ stripped of their envelopes through 
close binary interaction 
\citep{2002ApJ...572L..61M,2007MNRAS.381..835C,2008MNRAS.383.1485V}. 
We don't yet have a firm confirmation of a Ibc explosion 
(with no associated GRB) directly associated 
with a massive WR star, or with ejecta masses high enough to 
suggest a WR progenitor. Those broad lined SNe {\em with} a 
GRB associated do have high enough ejecta masses to be 
consistent with LMC type WC stars \citep{2007ARA&A..45..177C}. 
However there is a suggestion of a 28\msol\ WNL progenitor for
the type IIb SN2008ax \citep{2008MNRAS.391L...5C} and this SN appears to be on the 
H abundance continuum between Ib and IIb events
\citep{2008arXiv0805.1914P}.  
We will discuss the Ibc progenitor
scenarios further in the second paper in this series.

\subsubsection{The type IIn population and their progenitors}
\label{subsubsect:IIn}

There is clear evidence now that some very massive stars, above
25-30\msol\ do explode as very bright SN, SN2005gl is a type IIn at
65\,Mpc and has a very luminous progenitor detected at $M_{V} \simeq
-10.3$ \citep{2007ApJ...656..372G}.  The latter is evidence that
luminous blue variables are direct progenitors of some SNe and this is
supported by studies of the energies and spectral evolution of other
events
\citep{2006A&A...460L...5K,2007ApJ...666.1116S,2008A&A...483L..47T,2008arXiv0804.0042S}.
The case of SN2006jc showed that an LBV-like outburst occurred
directly coincident with a peculiar type of hydrogen deficient SN
\citep{2007Natur.447..829P}. SN2006jc resembles a type Ic with narrow
lines of He arising from a circumstellar He-rich shell
\citep{2007ApJ...657L.105F,2008arXiv0801.2277P}.
\cite{2007Natur.450..390W} have suggested that both the 
super-bright IIn events 
\cite[2006gy like;][]{2007ApJ...666.1116S} and the double outburst events 
\cite[2006jc-like;][]{2008arXiv0801.2277P} may not be
the canonical core-collapse mechanism, but be due to pulsational
pair-instability in the cores.

Thus one might venture that above 17\msol\ the vast majority of 
stars form black 
holes at core-collapse and can't produce
bright explosions through the canonical neutrino driven convection
mechanism. A fraction of them form collapsars due to 
a combination of rotation or binarity and
low metallicity \cite[see][]{2006ARA&A..44..507W}. 
And a fraction may form H-rich luminous type IIn SNe through 
the pulsational pair-instability mechanism. 

A caveat to this is the discovery of 
neutron stars in two young clusters 
\citep{2006ApJ...636L..41M,2008ApJ...683L.155M}, which suggests
the progenitors had initial masses greater than 40\msol\
and 20-30\msol\ respectively. These stars should perhaps have
formed black holes but \cite{2008ApJ...685..400B}
suggests that under certain conditions, binary evolution could
result in stars as massive as 50-80\msol\ ending up as neutron stars. 
 A further argument
against is that  the locations of Ibc SNe tend to be more closely
associated with H\,{\sc ii} 
regions than II-P SNe, suggesting 
a higher initial mass range for the progenitors of Ibc \citep{2008MNRAS.390.1527A}

The nature of the deaths of the most massive stars, whether in black-hole
forming events, or other explosion mechanisms, still remains to be
determined. The combination of studies of direct progenitor detections, 
environment evaluation, SN ejecta and remnant properties will be a
fertile field for discovery for many years to come. 

\subsubsection{The progenitor of SN~1987a : Sanduleak $-69^{\circ}202$} 

Although the progenitor of SN~1987A is often quoted to be a
20\msol\ star, one needs to be careful with a simple interpretation 
of placing the progenitor on an HRD and taking the closest mass track. 
The spectral type and $UBV$ magnitudes from \cite{1989A&A...219..229W} 
suggest a B3-type supergiant \citep[$T_{eff} \simeq 15750$, from the calibration of LMC B-supergiants in][]{2007A&A...471..625T} and 
hence $\log(L/L_{\odot}) = 5.1 \pm 0.1$. When placed on single-star 
evolutionary tracks this lies close to a 20\msol\ model just after
the end of core-H burning. However it is not valid to assume 20\msol\
as the progenitor initial mass, as the model track is not at its
endpoint and is no where near to having an Fe-core (or at least at the 
point of neon burning within a helium core). The luminosity of the 
He-core of an evolved massive star determines the stars luminosity and
we estimate the corresponding He-core mass to be
 $5.2^{+2}_{-1.0}M_{\odot}$. The initial mass of a single star required
to produce this core mass  
is
$15^{+4}_{-1}M_{\odot}$. The interacting binary model in 
\cite{2004Natur.427..129M} and \cite{1993Natur.364..509P} can 
produce a SN1987A like progenitor with a pair of 14 and 15\msol\ stars. 
And a merger involving a lower mass star of 3-6\msol\ with an evolved
15-16\msol\ primary can also account 
for the luminosity \citep{1992PASP..104..717P,2007Sci...315.1103M}. 
In both scenarios an initially less massive star gains mass to
explode with a final mass of 20\msol. However the helium core mass
was that expected for a 15\msol\ star leading to its position in the
blue part of the HRD. 
The ejecta mass has also been estimated at around 15\msol\ 
\citep{1996snai.book.....A}. 
Hence our suggestion of black holes coming from $\sim$17\msol\ 
stars and above is not directly disproven by the example of 
Sanduleak $-69^{\circ}202$



\subsection{Explosion mechanisms and production of $^{56}$Ni}

The tail phase of type II-P SNe are thought to be powered by the
radioactive decay of \nick\ and recent studies have shown that 
there can be a large range in tail phase luminosities. This 
would imply that a different mass of \nick\ has been ejected in the
explosions. As \nick\ is created by explosive burning of Si and O
as the shock wave destroys the star, it can be used as a probe of 
the explosion mechanism. For example 
\cite{1998ApJ...498L.129T}, 
\cite{2004MNRAS.347...74P}, and 
\cite{2003astro.ph..8136N}
 predict that high
mass stars may undergo fall back in which some of the \nick\ 
falls back onto a proto-neutron star or black hole and, hence, one
might get a fainter supernova. This has lead to suggestions that
plotting initial mass versus ejected mass of \nick\ could lead
to a bimodal population. There are estimates of the mass of \nick\ 
for nine of the SNe in our sample (by ourselves and also from other 
groups, already 
published in the literature) and we can investigate this relation
in a direct way. 

The \nick\ mass can be estimated from the tail phase magnitudes
using several different methods. For example, the bolometric
luminosity of the tail phase \citep{2003ApJ...582..905H},
a direct comparison with SN 1987A \citep{1998ApJ...498L.129T}
and the ``steepness of decline'' correlation \citep{2003A&A...404.1077E}. 
For the two recent SNe 2004A and 2003gd, \cite{2005MNRAS.359..906H,2006MNRAS.369.1303H} 
have compared the three methods and find the first two in 
good agreement, while the mass from the ``steepness of decline'' relation
gives somewhat lower results (at least for these two events). 
It is important that a consistent
method is used to determine all the \nick\ masses if any comparison is to
meaningful, particularly as the uncertainties on the estimates
are often fairly large. Hence we determine \nick\ masses from one 
consistent approach. The values for SNe 2004A, 2003gd, 1999gi and 1999em 
were taken from \cite{2006PhDTHendry} who used the bolometric tail-phase 
luminosity method of \citep{2003ApJ...582..905H}, and the distances and 
reddening already adopted in Table\,\ref{table:main}. 
The mass for 
SN2004dj was taken from \cite{2006AJ....131.2245Z} who found that 
the bolometric luminosity of the tail-phase gave a value very 
similar to that from the ``steepness of decline'' relation. We
take their value from the  bolometric tail-phase 
luminosity to ensure consistency ($0.025\pm0.010$\msol). 
This is very similar to the value determined by 
\cite{2006MNRAS.369.1780V} with a simple radioactive
decay model applied to the bolometric tail luminosity 
($0.02\pm0.010$\msol). By a similar analysis, the value of 
SN2004et was determined with the \cite{2003ApJ...582..905H} formula, 
to give $0.06\pm0.02$\msol, the highest of our estimates. 

The value for SN2004dg has been determined from the late-time, 
tail phase magnitudes from our HST imaging (see Sect.\ref{subsec:04dg}). 
We determine $V=20.8$ and estimate an explosion epoch from spectra
and photometry during the plateau phase.
Applying the same bolometric tail phase method as above results 
in a value of $0.010\pm0.005$\msol. A very 
low \nick\ mass for SN1999br has been reported in  \citep{2003ApJ...582..905H},
although he used a distance of $10.8\pm2.4$\,Mpc, and to keep the 
analysis consistent we scaled his value to our larger adopted distance 
listed in Table\,\ref{table:main}, resulting in a value of 
$0.003\pm0.001$\msol. This is is in good agreement with the
value determined by \cite{2004MNRAS.347...74P} of $\sim$0.002 who used a similar
distance to ours to calculate the mass from a direct comparison with the 
tail phase luminosity of SN1987A. The estimates for  SN2006ov and SN2006my
have been calculated using a similar bolometric tail phase method as
described above to give $0.003\pm0.002$\msol\ and $0.03\pm0.015$\msol\ 
respectively (Maguire et al., Spiro et al. in prep). 

\begin{figure}
    \centering
    \epsfig{file = 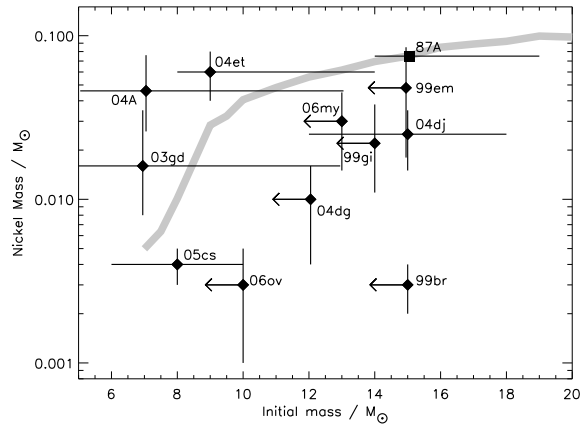, width=80mm}
    \caption[]{Plots of initial mass vs. mass of $^{56}$Ni. The grey line is the $M_{\rm O}/M({\rm CO core})$ normalised to pass from the 1987A nickel mass.}
\label{fig:Ni56}
\end{figure}

Finally we consider the case of SN2005cs. The early time spectra 
have been studied extensively by 
\cite{2006MNRAS.370.1752P}, \citet{2006MNRAS.372.1735T} and \citet{2007ApJ...659.1488B},
who all find it to be a moderately faint II-P with low ejecta
velocities. It appears a similar type of event to the faint
SNe 1999br and 2002gd \citep{2006MNRAS.370.1752P}, and a measure of 
its luminosity in the tail phase is especially interesting 
particularly as it is one of the events with a detected 
progenitor and well determined mass (Table\,\ref{table:main}). 
\cite{2006A&A...460..769T} present several photometric measurements
of SN2005cs in the tail phase and suggest a \nick\ mass of 0.018\msol, 
which would be similar to 2003gd. As the progenitors are likely to 
have been red supergiants of quite similar masses one might be
encouraged by this agreement. However  \citep{ap2005cs}
presents new measurements of the
the tail phase magnitudes of SN2005cs finds it to be 
significantly fainter than reported in \cite{2006A&A...460..769T}. 
The difference is likely due to the differing methods employed
to determine the luminosity in this faint phase. 
Pastorello et al.  have used image subtraction to 
remove contamination of the host galaxy, which can be  
significant as the SNe fades. 
These fainter measured magnitudes 
suggest an ejected \nick\ mass of $0.003\pm0.001$\msol, 
and we believe this to be a more realistic
estimate. For reference, the determined value of 0.075\msol\
for SN1987A from \cite{Arnett} is also plotted in Fig. \ref{fig:Ni56}

\subsection{The nature of faint II-P SNe}
\label{subsect:faintSNe}
In Figure \ref{fig:Ni56} we see how the nickel mass created in II-P
SNe compares to the initial mass of the progenitor star. Similar diagrams
have been produced before but these derive the initial mass
by a model dependent two step process. First of all the ejected 
mass is estimated from modelling the SN lightcurve and velocity evolution
of the ejecta, and then an initial main-sequence mass is inferred
by assuming an estimated remnant mass and accounting for the effects
of mass-loss during stellar evolution
\citep{2003astro.ph..8136N,2003MNRAS.338..711Z}. Such 
models have given satisfactory fits to the luminosity and 
velocity measurements, but our direct pre-explosion measurements
can provide valuable independent information.
It is certainly clear
that there is a population of low luminosity II-P SNe, which have
lower ejecta velocities throughout the photospheric stage and
very low tail phase bolometric luminosities. The first one 
recognised was SN1997D 
\citep{1998ApJ...498L.129T,1998ApJ...502L.149Z,2001MNRAS.322..361B}
and in our sample 1999br and 2005cs
are similar 
\citep{2004MNRAS.347...74P,2006MNRAS.370.1752P}.
The clear implication of the low tail phase luminosity is that 
a low mass of \nick\ is produced in the explosion. Hence one might
hope to relate this to the explosion mechanism. 

Two alternative suggestions for the faint II-P SNe have been
proposed. One is that the SNe formed black holes at core-collapse and
the \nick\ produced fell back into the black hole rather than being
ejected \citep{2003MNRAS.338..711Z,2003astro.ph..8136N}. An
alternative is that they are intrinsically less energetic explosions
of lower mass stars. Stars of 9-11\msol\ can have 
large density gradients in the
O-Si layers around the proto-neutron star and the shock may
produce lower temperatures than in higher
mass counterparts. Statistical equilibrium is only reached in a
thinner shell of O and Si rich gas, hence low amounts of 
\nick\ is produced \citep{1988ApJ...334..909M}. 
\cite{2000A&A...354..557C}  have also favoured this low-mass
star scenario in their fitting of the 
nebular spectra of SN1997D with a hydrodynamic model. 
The high mass stellar origin of
\cite{2003MNRAS.338..711Z} and \citet{2003astro.ph..8136N}
predicts hydrogen rich progenitors of greater than 25\msol, 
but as shown above we do not detect any red supergiant stars of 
such high masses. These types of stars, would be the easiest to 
detect and in particular for 1999br and 2005cs we favour 
the low mass scenario. 
Figure\,\ref{fig:Ni56} argues against the high mass scenario
for the low \nick\ mass SNe, and we find no evidence of the 
branching of the figure at high masses into low and high energy SNe
as suggested by both \cite{2003MNRAS.338..711Z} and \cite{2003astro.ph..8136N}
The low \nick\ SNe have initial masses well below the limit 
required to produce a massive enough core for black
hole formation. In fact rather than being two separate populations
there appears to be a continuous trend with lower initial mass
producing lower masses of nickel in core-collapse \citep{jrmthesis}.

One could advance an argument that the progenitors were actually 
of higher mass and not detected in our images for some 
reason. The most obvious reason is high dust extinction as
we discussed in Sect.\,\ref{subsect:extinction}. We argued
that it is unlikely that we have systematically underestimated
the masses due to large extinctions. In addition, 
the fact that there was no detection of the 2005cs progenitor 
in deep NIR $JHK$ bands argues against a massive progenitor surrounded
in dust, of anything up to $A_{V}\sim5$ \citep{2005MNRAS.364L..33M}. 
Also SN2008bk was detected in the the NIR ($JHK$ bands) and 
hence even a visual extinction of $A_{V}\sim10$ would have only a 
$\sim1^{m}$ affect in $K$, increasing the \logl\ by 0.25\,dex. 
Another possibility is that the
stars were not red supergiants, but hotter, blue stars. One could
invoke this explain the non-detections and also argue that the
detected red stars are heavily reddened bluer objects. But in 
this case they would be more compact and the lightcurves would 
then tend to resemble SN1987A, and there is no evidence such peculiarities
in any of the SNe presented here. If all, or most, II-P
progenitors are not red supergiants this would pose serious problems
for our stellar evolution theory and models of II-P lightcurves. We
consider this possibility unlikely. As this paper was 
in the review stage, \cite{2008A&A...491..507U} proposed that the 
ejecta mass of SN2005cs, from hydrodynamic modelling of the lightcurve
and velocity evolution, was around $17.3\pm1$\,\msol. They also 
suggest a  progenitor mass of 20-25\msol\ for SN1999em which
is again significantly higher than our stellar evolutionary mass. 
This severe discrepancy between two methods is interesting and 
should be explored further in the future.

As the \nick\ is produced by explosive silicon and oxygen burning of
the mantle material \citep{2002RvMP...74.1015W}, one might
imagine that the amount of nickel formed in a SN is related to the
amount of oxygen and silicon in the progenitors core. 
In Figure \ref{fig:Ni56} we over-plot the mass of oxygen in the
carbon-oxygen (CO) core divided by the mass of the CO core from our
stellar models against the initial stellar mass. The line is scaled to
fit SN 1987A. This simple model follows the general trend of the
observations, 
but with some large scatter. Other factors will effect the amount of
nickel produced in a SN but the relation could suggest 
 that the amount of
source material is  a primary factor in determining how much nickel is
produced.
Alternatively it could be a reflection of the density and temperature
in the explosive burning region in the stars. Whatever the physical 
reason, we have not detected high mass stars as progenitors of the 
faint II-P SNe. Hence we favour the lower mass star progenitor as the origin
of these events.

\section{Lessons learned and future possibilities}
This decade long effort to systematically search and detect SN 
progenitors provides an opportunity to reflect on lessons learned and 
ways to increase discovery potential. It is perhaps somewhat surprising
that unambiguous detections in multi-band images 
have been difficult. As discussed above this could imply that the
high mass, H-rich, bright supergiants up to around 30\msol\ are not the
progenitors of type II-P or II-L SNe. Whatever the reason, it
seems clear that pursuing this project needs a different approach to 
make firm detections of progenitors
rather than providing many more upper limits. 
The latter can be interesting if enough are available, but ultimately it
is discovery and characterisation of objects that will advance the field. 

As discussed in Sect.\,5, the three progenitors which are reliably and
unambiguously detected (2003gd, 2005cs and 2008bk) and have
high-quality multi-colour images (giving colour information) are all
closer than 10\,Mpc.  Although the global image archives of nearby
galaxies have been {\em steadily} increasing over the last 10 years,
the total number of galaxies and the quality of the images will not
rapidly improve in the near future.  e.g. it has taken 15 years of
post-SM1 HST operations to get to the current archive content.  It
seems clear that the current HST archive does not contain deep enough
images of galaxies to enable routine detection of progenitors beyond
about 10\,Mpc.  This does not mean we should give up on searching for
progenitors of SNe between $\sim$10-25\,Mpc as one may be fortunate to
detect higher mass, and brighter, progenitors than we so far have done
\cite[e.g.][]{2007ApJ...656..372G}. Each event should certainly be scrutinized
for the potential of fortuitous progenitor discovery but when considering
how to actively improve the possibilities the following should be 
noted. In
Cycle\,10 of HST operations (2001 June 01 - 2002 May 31) we were allocated
a SNAP proposal (SNAP9042) to enhance the image archive of starforming
galaxies within about 20\,Mpc and then to wait for future SNe to
occur.  We observed approximately 160 galaxies with typical exposure
times of 460s in F450W, F814W and F606W. These set of images produced
pre-explosion environments for SNe 2003jg, 2004A, 2004dg, 2005V,
2006ov, 2007aa, 2007gr and SN2008ax.  The progenitors of SN2004A and
SN2008ax have been detected \cite[see
Sect.\ref{subsect:04A},][]{2006MNRAS.369.1303H,2008MNRAS.391L...5C}
and the limits on the non-detections have proved very useful. This
provides a benchmark for any future studies dedicated to
detecting progenitors retrospectively. The depth reached in these
images (typically $m_{\rm F814W} \simeq 25$, 3$\sigma$) at a distance
of $\sim$20\,Mpc results in sensitivities down to $M_{\rm F814W} \sim
-6.8$ (for a typical $E(B-V) \simeq 0.2$). 

The most fruitful method of detecting progenitors in the future would
be to carry out a deep, wide-field survey of starforming galaxies
within about 10\,Mpc with the revived ACS and the new WF3 on HST after
servicing mission SM4.  Reaching AB magnitudes of around 26 would be
required to ensure the images go deep enough to detect progenitors
down to around 8\msol. For example SN2003gd was discovered at $V=25.8$
($M_{V} = -4.4$) and SN2003gd, 2005cs and 2008bk were all
discovered at $M_I \sim -6.5 \pm 0.2$ (all Vega based magnitudes).
By restricting the distance limit to 10\,Mpc would of course result
in a lower rate of CCSNe (1-2
per year) in the sample of around 100 major starforming galaxies
\citep{2008arXiv0807.2035K}, but the discovery potential would be
excellent. One further issue is the size of the starforming disks of these
galaxies which are significantly larger than the 3.4 arcmin diameter of ACS
Wide-Field-Channel. The limited size of the WFPC2 and ACS cameras has 
resulted in many galaxies having been observed by HST before a CCSN occurs
but the position falling outside the camera footprints (43 per cent : see
Sect.\ref{section:rates}). Hence a carefully planned future survey of nearby
galaxies must use multiple HST pointings to cover the full optical extents of the 
galaxy disks. A campaign of this extent is already in the arena of 
``multi-cycle treasury proposals''. 

A {\em systematic} survey beyond 10\,Mpc that would
significantly increase detection probability is probably not the best
strategy. A distance limit increase to $r_{\rm 10Mpc}$ (where $r_{\rm
10Mpc}$ is the distance limit in units of 10\,Mpc) would increase the
number of galaxies by a factor of approximately $100(r_{\rm 10Mpc}^{3}- 1)$ 
and the time required would increase more than linearly
(depending on the $r_{\rm 10Mpc}$ distribution and galaxy sizes). The
latter is not feasible in any reasonable allocation of time and given
the difficulty in detecting progenitors beyond 10\,Mpc is not an 
optimal strategy.

We have found it essential that astrometry of the SNe employs 
images of the resolution of the HST pre-explosion images to avoid 
large astrometric uncertainties and ambiguous (or spurious) detection of progenitors. 
Ground-based (natural seeing) images are normally not adequate for 
providing geometric transformations between pre and post-explosion
images to the 10-12 milli-arcsec level required. Between 2001-2006
our group used HST to take post-explosion images at the same 
resolution (or higher, depending the camera pixel scales) as the pre-explosion
HST images. This proved to be essential as there were several announcements
(in IAU Circulars and Central Bureau Electronic Telegrams) of erroneous
detections of progenitors using ground-based 
astrometry \citep{2003PASP..115....1V,2005IAUC.8555....2R,2007ApJ...661.1013L}. 
Using unchecked and absolute astrometry to identify progenitors on ground-based
images has also lead to erroneous claims of progenitor detections
\citep[SN~2002ap and SN~2004dj:][]{2002IAUC.7816....3S,2004IAUC.8388....2L,2004IAUC.8384....4W}
Images with adaptive optics systems on large ground based telescopes (e.g. 
the VLT, Gemini and Keck) routinely deliver diffraction limited images
in the $K-$band. These images are of sufficient resolution to use instead of 
HST follow-up images \citep[e.g.][]{2007ApJ...656..372G,2008MNRAS.391L...5C}
Rapid analysis of images after SNe discoveries and reporting of possible progenitors
before extensive analysis has been completed is understandable in a competitive field
and helps the community to prioritise potentially interesting events. However
claims of spatial coincidence require 
detailed {\em differential} astrometry between post and pre-explosion
images to precisions of around 10 milli-arcsec and  comprehensive
error analyses in order to be acceptable. Our experience has shown if this
rule of thumb is not followed then erroneous results often follow.  
The flexibility and rapid reaction time of ground-based AO systems
make them an excellent facility to provide precision differential astrometry 
(at the 10 milli-arcsec level), and this has been our preferred strategy
since 2007. 
 
Finally, we are now at the point at which, for some progenitors, 
we can return to see if the stars have disappeared. This has been done for
only SN~1987A \citep{1987Natur.328..318G} and deep late-time images of 
SN~2004et alerted us to likely spurious detection of the progenitor. This area
of late-time observations may lead to further surprises, but it is essential it
is pursued to verify the results claimed here, and elsewhere, for 2003gd, 2005cs
and 2008bk.

\section{Conclusions}
This paper presents a consistent and homogeneous analysis of the constraints
on progenitor stars of II-P SNe, within a volume and time limited search. 
The work builds and enhances the discoveries and limits presented
in the literature. There are now enough data that a statistical study of
progenitor properties is feasible. This represents the culmination of a
a 10.5 year search for progenitor stars and the conclusions are as follows:

\begin{itemize}
\item We have compiled all SNe discovered within a strict volume and time limit
and reviewed the SN types reported in the literature. This gives an very
good estimate of the relative rate of CCNSe in the Local Universe, at metallicities
between LMC and solar. 
\item Of the 55 type II-P SNe found, 20 have HST pre-explosion (or excellent ground-based)
images available to search for progenitor stars. We summarise the data presented in the
literature to date on these SNe and carry out a consistent and homogeneous analysis
of all events. Three groups of events are discussed - those with probable single star
progenitors detected, those with upper limits to their luminosities and masses, and
those falling on compact, unresolved coeval clusters
\item The masses and mass limits of each are determined using our STARS evolutionary models and a statistical analyses
presented of the final masses. A maximum likelihood analysis suggests that 
the minimum mass for a type II-P SN is $m_{min}=8.5^{+1}_{-1.5}$\msol. This is 
consistent with current estimates of the maximum mass that will produce 
a white dwarf in young clusters. 
\item We have not detected any progenitors above 16\msol. 
Assuming a Salpeter IMF, the most likely 
maximum mass for a II-P SN is $m_{max}=16.5\pm1.5$\msol.This is not 
particularly sensitive to the IMF slope within the typical variations 
known in the Local Universes of $\pm0.7$.
\item We suggest that there is a discrepancy between this maximum mass and
our knowledge of massive star evolution. Red supergiants between 
17 - 30\msol\ are not detected as progenitors but are predicted by theory
to exist and this is supported by stellar population studies. We term this 
discrepancy the ``red supergiant problem''. It is unlikely to be due 
to IMF variations and possible explanations 
include the possibility that we have systematically under estimated the
stellar luminosities and masses due to foreground extinction or that 
the gap is 
filled with the other flavours of type II SNe (e.g. II-L, IIn and IIb). 
However neither of these solutions are supported by current data.  
We suggest that these
objects may be forming black-holes with faint, or non-existent SN explosions. 
\item Although low-luminosity SNe are already known (e.g. 1999br, 2005cs)
we suggest that these events are
not the missing SNe - rather our analysis supports the interpretation that
they are stars at the minimum mass limit for SNe.  
\item We review the information on extinctions toward these SNe and their
progenitors and compare it with that toward red supergiant populations in the
LMC. We suggest that high extinction toward the SNe progenitors is unlikely to 
be the cause of the lack of detections of massive supergiants. 
\item The search for progenitor stars should continue for every nearby
SN that 
has deep, high resolution pre-explosion images. In particular the missing 
high mass red supergiants should be sought in both optical, near infra-red
and mid infra-red images. If the limit for black hole formation is as low 
as 17\msol\ then it bodes
well for surveys for disappearing stars as suggested by 
\cite{2008arXiv0802.0456K}. 

\end{itemize}

\section*{Acknowledgments}

This work, conducted as part of the award "Understanding the lives of
massive stars from birth to supernovae" (S.J. Smartt) made under the
European Heads of Research Councils and European Science Foundation
EURYI (European Young Investigator) Awards scheme, was supported by
funds from the Participating Organisations of EURYI and the EC Sixth
Framework Programme. RMC and SJS than the ESF and Leverhulme Trust for
financial support.  The research of JRM is supported in part by NSF
grant AST-0406740 and NASA grant NNG04GL00G.  This paper is Based on
observations made with the following telescopes : NASA/ESA Hubble
Space Telescope, obtained from the data archive at the Space Telescope
Institute. STScI is operated by the association of Universities for
Research in Astronomy, Inc. under the NASA contract NAS 5-26555
(Programs GO 10187, 10498, 10803 and SNAP9042).  the William Herschel
Telescope and data obtained from the Isaac Newton Group Archive which
is maintained as part of the CASU Astronomical Data Centre at the
Institute of Astronomy, Cambridge.  We acknowledge the usage of the
HyperLeda database, and This research has made use of the NASA/IPAC
Extragalactic Database (NED) which is operated by the Jet Propulsion
Laboratory, California Institute of Technology, under contract with
the National Aeronautics and Space Administration. We thank Seppo
Mattila, Andrea Pastorello, John Danziger, Avishay Gal-Yam,
Chris Kochanek for
discussions and debate; Mike Irwin and Annette Ferguson for early
access to the INT images of NGC6946; Martin Mobberley for access to an
image of SN2003ie ; John Beacom for discussions on statistical methods
to use, recommending the maximum likelihood approach and discussions
on local SNe rates;  Mario Hamuy
for providing several SNe classifications from private spectra.
We thank the referee, David Branch for comments and suggestions
that improved the paper.

\bibliographystyle{mn2e}


\appendix
\section{Complete list of SNe employed in the survey for progenitors}
\small
\onecolumn
\setlongtables
\begin{longtable}{cccccc}
\caption{SNe discovered between 1998-2008.5 in galaxies with recessional velocities less than 2000 kms$^{-1}$ \label{table:fullsample}. The HST FOV column notes if the host galaxy
has been observed by HST prior to explosion and if the position of the SN is ``in'' 
or ``out'' of the camera field-of-view.}\\
\hline\hline
Supernova & Galaxy & $V_{vir}$  & Type & HST FOV & Comments \cr\hline
\endfirsthead
\multicolumn{6}{l}{Table~\ref{table:fullsample} continued}\\
\hline\hline
Supernova & Galaxy & $V_{vir}$  & Type & HST flag & Comments \cr
\hline
\endhead
\hline
\multicolumn{6}{r}{\small\sl continued on next page}
\endfoot
\hline
\endlastfoot
\noalign{\smallskip}
{\bf Core-collapse} \\
1998A  &  IC2627      &   1975.7   &     IIpec  & ...  &  \cite{2005MNRAS.360..950P} SN1987A-like  \\ %
1998S  &  NGC3877     &   1114.8   &     IIn    & ...  &  \cite{2001MNRAS.325..907F} \\
1998bm &  IC2458      &   1809.7   &     II     & ...  &   \cite{1998IAUC.6882....1L}, unknown subtype  \\ %
1998bv &  PGC2302994  &   1793.7   &     II-P   & ...  &   \cite{1998IAUC.6900....1K}\\
1998dl &  NGC1084     &   1298.2   &     II-P   & ...  &   \cite {1998IAUC.6994....3F}\\
1998dn &  NGC 337A    &   997.1    &     II     & ...  &  \cite{2000AJ....120..367G}, unknown subtype \\
1999B   &  UGC7189    & 1962.1    &    II    & ... &    \cite{1999IAUC.7089....3F}, unknown subtype\\ 
1999an  & IC 755      & 1607.2    &    II    & in  &   \cite{1999IAUC.7124....1W}, unknown subtype\\ 
1999bg  &  IC758      & 1530      &    II-P  & ... &   \cite{1999IAUC.7137....1A}, amateur LC \\   
1999br  & NGC4900     & 1012.6    &    II-P  & in  &   \cite{2004MNRAS.347...74P} Low luminosity II-P \\ 
1999bw  & NGC3198     & 850.9     &    LBV   & out &   \cite{1999IAUC.7152....2F}\footnote{http://etacar.umn.edu/etainfo/related/}\\ 
1999el  & NGC6951     & 1704.2    &    IIn   & out &   \cite{2002ApJ...573..144D} \\ 
1999em  & NGC1637     & 615.2     &    II-P  & out &    \cite{2002PASP..114...35L}\\
1999eu  & NGC1097     & 1065.8    &    II-P  & out &    \cite{2004MNRAS.347...74P} Low luminosity II-P  \\ 
1999ev  & NGC4274     & 1089.5    &    II-P  & in  &    \cite{1999IAUC.7306....2G}, amateur LC.\\ 
1999ga  & NGC2442     & 1149.6    &    II-L  & ... &    Peculiar II-L (Pastorello et al. in prep)  \\ 
1999gi  & NGC3184     & 765.8     &    II-P  & in  &    \cite{2002AJ....124.2490L}\\ 
1999gn  & NGC4303     & 1616      &    II-P  & ...   &  \cite{2004MNRAS.347...74P} possible low luminosity II-P   \\ 
1999gq  & NGC4523     & 360.3     &    II-P  & ...   &  \cite{1999IAUC.7339....2J}, amateur LC  \\ 
2000ch  & NGC3432     & 778.6     &    LBV   & out &    \cite{2004PASP..116..326W} \\ 
2000db  & NGC3949     & 1022.8    &    II-P  & out &    \cite{2000IAUC.7481....2P}, amateur LC \\ 
2000ds  & NGC2768     & 1620.5    &    Ib    & in  &    \cite{2000IAUC.7511....2F}\\
2000ew  & NGC3810     & 1061.3    &    Ic    & in  &    \cite{2002PASJ...54..905G} \\
2001B   & IC 391      & 1815.9    &    Ib    & in  &   \cite{2001IAUC.7577....2C,2006PZ.....26....3T} \\
2001X   & NGC5921     & 1581.6    &    II-P  & ...   &  \cite{2006PZ.....26....3T}  \\
2001ac  & NGC3504     & 1668.9    &    LBV   & out &    \cite{2001IAUC.7597....3M}\\
2001ci  & NGC3079     & 1340.5    &    Ic    & in  &    \cite{2001IAUC.7638....1F}\\
2001du  & NGC1365     & 1407.8    &    II-P  & in  &    \cite{2003MNRAS.343..735S,2003PASP..115..448V} \\
2001fv  & NGC3512     & 1505.5    &    II-P  & ...   &  \cite{2001IAUC.7756....4M}, amateur LC \\ 
2001fz  & NGC2280     & 1720.9    &    II-P  & ...   &  \cite{2001IAUC.7759....2M}, amateur LC  \\ 
2001gd  & NGC5033     & 1080.6    &    IIb   & out &   \cite{2007ApJ...671..689S} \\
2001ig  & NGC7424     & 754.4     &    IIb   & out &    \cite{2006MNRAS.369L..32R,2007ApJ...671.1944M} \\
2002E   & NGC4129     & 1152      &    II    & ...   &  \cite{2002IAUC.7800....4M}, unknown subtype \\ 
2002ao  & UGC 9299    & 1614.9    &    Ic   & ...   &  \cite{2008arXiv0801.2277P}: narrow He lines\footnote{SNe 2002ao and 2006jc have been termed Ibn as they show narrow He lines due to circumstellar He rich shells. We discuss in Sect.\,\ref{section:defintion} why we use Ic rather than Ibn} \\
2002ap  & NGC 628     & 685.7     &    Ic    & out &   \cite{2002ApJ...572L..61M}  \\
2002bu  & NGC4242     & 736.6     &    IIn   & out &   \cite{2002IAUC.7864....4A}, LC appears like II-L \\ 
2002hc  & NGC2559     & 1371      &    II-L  & ...   & \cite{2002IAUC.7999....1W}, similar to 1979C \\
2002hh  & NGC6946     & 318.5     &    II-P  & out &   \cite{2006MNRAS.368.1169P} \\
2002ji  & NGC3655     & 1525.4    &    Ib/c  & ...   &  \cite{2002IAUC.8028....3R}   \\
2002jz  & UGC 2984    & 1529      &    Ic    & ...   &  \cite{2002IAUC.8037....2H}, somewhat uncertain \\
2002kg  & NGC2403     & 370       &    LBV   & out    & \cite{2006MNRAS.369..390M}\\
2003B   & NGC1097     & 1065.8    &    II-P  & ...   &    M. Hamuy (priv. communication) \\
2003J   & NGC4157     & 1002.7    &    II-P  & ...   &  \cite{2003IAUC.8048....2A}, amateur LC \\ 
 2003Z & NGC2742     & 1517.7     &   II-P    & ...   &   \cite{utrobin07}, Spiro et al. in prep.; low luminosity II-P\\
2003bg  & M-05-10-15  & 1151      &    Ic    & ...   &   \cite{2006ApJ...651.1005S}, evidence of H during evolution\\
2003bk  & NGC4316     & 1325.5    &    II    & ...   &   \cite{2003IAUC.8086....2P}, subtype uknown \\
2003ed  & NGC5303A    & 1841      &    IIb   & ...   &   \cite{2003IAUC.8144....2L} \\ 
2003gd  & NGC 628     & 685.7     &    II-P  & in  &     \cite{2005MNRAS.359..906H} \\        
2003gm  & NGC5334     & 1433.8    &    LBV   & in  &    \cite{2006MNRAS.369..390M} \\
2003hn  & NGC1448     & 911.3     &    II-P  & out &     M. Hamuy (priv. communication) \\      
2003ie  & NGC4051     & 917       &    II-P  & ...   &   See Sect. 5.9, \cite{2008arXiv0804.1939H}  \\ 
2003jg  & NGC2997     & 914.1     &    Ib/c  & in  &     \cite{2003IAUC.8241....2H} \\
2004A   & NGC6207     & 1090.3    &    II-P  & in  &     \cite{2006MNRAS.369.1303H} \\
2004C   & NGC3683     & 1944.3    &    Ic    & ...   &   \cite{2004IAUC.8269....2M}    \\
2004G   & NGC5668     & 1672.7    &    II    & out &     \cite{2004IAUC.8273....2E}, subtype unknown \\  
2004am  & NGC3034     & 487.1     &    II-P  & in  &      See Sect. 5.11 \\
2004ao  & UGC10862    & 1811.6    &    Ib    & ...   &    \cite{2004IAUC.8304....4M}  \\
2004bm  & NGC3437     & 1379.6    &    Ic    & ...   &    \cite{2004IAUC.8339....2F}   \\
2004cm  & NGC5486     & 1648.2    &    II-P  & ...   &     \cite{2004IAUC.8359....1C}, probable low-luminosity II-P\\ 
2004cz  & E407-G09    & 1399.8    &    II-P  & ...   &    \cite{2004IAUC.8374....2F} \\ 
2004dg  & NGC5806     & 1439.1    &    II-P  & in    &  \cite{2004IAUC.8376....2E}, and Sect. 5.12\\
2004dj  & NGC2403     & 370       &    II-P  & out &     \cite{2006MNRAS.369.1780V,2006AJ....131.2245Z} \\
2004dk  & NGC6118     & 1645.4    &    Ib    & ...   &   \cite{2004IAUC.8404....1F} \\
2004ep  & IC2152      & 1687.1    &    II    & ...   &   \cite{2004IAUC.8420....2F}, subtype unknown \\
2004et  & NGC6946     & 318.5     &    II-P  & ...   &    \cite{2007MNRAS.381..280M,2006MNRAS.372.1315S} \\
2004ez  & NGC3430     & 1731.3    &    II-P  & ...   &    \cite{2004IAUC.8420....2F}, amateur LC  \\ 
2004fc  & NGC 701     & 1727.2    &    II-P  & ...   &    \cite{2004IAUC.8432....2S}  \\
2004gk  &  IC3311     & -79       &    Ic    & ...   &     \cite{2004IAUC.8446....1Q} \\ 
2004gn  & NGC4527     & 1780.3    &    Ic    & out &     1990B-like, SNIFS spectrum \footnote{http://www.supernovae.net/sn2004/sn2004gn.jpg} \\ 
2004gq  & NGC1832     & 1779.7    &    Ib/c  &  -  &      \cite{2005IAUC.8461....3M,2007AIPC..924..297G}\\
2004gt  & NGC4038     & 1570.4    &    Ib/c  & in  &       \cite{2005ApJ...630L..29G,2005ApJ...630L..33M}\\
2005V   & NGC2146     & 1158.4    &    Ib/c  & in  &       \cite{2005IAUC.8474....3T}\\
2005ad  & NGC941      & 1519.6    &    II-P  & ... &    M. Hamuy (priv. communication), probable II-P\\ 
2005ae  & E209-G09    & 860.9     &    IIb   & in  &      \cite{2005IAUC.8486....3F}\\
2005af  & NGC4945     & 373.5     &    II-P  & ... &    Pereyra et al. (2006) \\
2005at  & NGC6744     & 617.6     &    Ic    & out &     \cite{2005IAUC.8496....2S} \\
2005ay  & NGC3938     & 1016.8    &    II-P  & ...   &   \cite{2008ApJ...685L.117G}\\
2005cs  & NGC5194     & 702.1     &    II-P  & in  &     \cite{2006MNRAS.370.1752P}, low-luminosity II-P \\ 
2005cz  & NGC4589     & 2283      &    Ib    & in  &     \cite{2005IAUC.8579....2L} \footnote{Although $V_{vir}$ outside limit, the TF/SBF distance from 
\cite{2000ApJ...530..625T} puts it within our distance limit and it will be included in the stripped progenitor study of Crockett et al. in prep}\\ 
2005kl  & NGC4369     & 1248.2    &    Ic    & ...   &    \cite{2005CBET..305....1T}   \\
2006bc  & NGC 2397    & 1059.7    &    II-P  & in  &     \cite{2006CBET..450....1P}, amateur LC\\ 
2006bp  & NGC 3953    & 1285.3    &    II-P  & ...   &   \cite{2008ApJ...675..644D}   \\ 
2006fp  & UGC12182    & 1807.5     &    LBV   &  ..   &   \cite{2006CBET..636....1B} \\
2006jc  & UGC 4904    & 1818       &    Ic    &  ..    &   \cite{2007Natur.447..829P,2007ApJ...657L.105F}: narrow He\footnote{SNe 2002ao and 2006jc have been termed Ibn as they show narrow He lines due to circumstellar He rich shells. We discuss in Sect.\,\ref{section:defintion} why we use Ic rather than Ibn} \\
2006my  & NGC 4651    & 912.1      &    II-P  &  in   &    \cite{2007ApJ...661.1013L} \\
2006ov  & NGC 4303    & 1616       &    II-P  &  in   &    \cite{2007ApJ...661.1013L}\\
2007C    &  NGC4981   &  1686      &    Ib    &  ...  &   \cite{2007CBET..800....2B} \\
2007Y    &  NGC1187   &  1213      &    Ib/c  &  ...  &    \cite{2007CBET..862....1F}\\
2007aa   &  NGC4030   &  1475.7    &    II-P  &  in   &    \cite{2007CBET..850....1F}, our own LC\\
2007av   &  NGC3279   &  1435.5    &    II-P    &  ...  &  \cite{2007CBET..903....1H}, amateur LC\\
2007gr   &  NGC1058   &  633.8     &    Ic    &  in   &  \cite{2008ApJ...673L.155V}\\
2007it   &  NGC5530   &  1045.3    &    II    &  ...  &   \cite{2007CBET.1068....1C}, unknown subtype \\
2007oc   &  NGC7418A  &  1938.3    &    II-P  &  ...  &   \cite{2007CBET.1120....1O}, 1999em like \\
2007od   & UGC12846   &  1810.5    &   II-P   &  ...  &   \cite{2007CBET.1119....1B}, 1999em like\\
2007sv   & UGC 5979   &  1377.0    &   LBV    &  ...  &   \cite{2007CBET.1184....1H}, our own data \\
2008M   &  E121-G26  &  1979       &   II-P      & ...   &   \cite{2008CBET.1227....1F}   \\
2008ax  &  NGC4490   &  797        &   IIb       & in    &  \cite{2008arXiv0805.1914P}    \\
2008bk  &  NGC7793   &  60         &   II-P      & out   & Sect 5.2, Mattila et al. (2008)      \\
2008bo  &  NGC6643   &  1778       &   Ib      &  ...    & \cite{2008CBET.1325....1N}     \\
\\
\\
{\bf Type Ia} \\
1998aq  &  NGC3982     &  1388.4  &    Ia     & in   & \cite{2003AJ....126.1489B}\\
1998bn  &  NGC4462     &  1717.8  &    Ia     &	...  & \cite{1998IAUC.6888....1P} \\
1998bu  &  NGC3368     &  940.2   &    Ia     &	out  & \cite{2000MNRAS.319..223H} \\
1998dm  &  PGC 005341  &  1881.9  &    Ia     & ...  & \cite{1998IAUC.6997....2F}\\
1999by & NGC2841       &   830.9   &  Ia     & ...     & \cite{2004ApJ...613.1120G} \\
2000E  & NGC6951       &   1704.2  &  Ia     & ...     &  \cite{2003ApJ...595..779V} \\
2001dp & NGC3953       &   1285.3  &  Ia     & out   &    \cite{2001IAUC.7683....2A}      \\
2001el & NGC1448       &   911.3   &  Ia     & out   &     \cite{2003AJ....125..166K}    \\ 
2001fu & M-03-23-11    &   1555    &  Ia     & ...     &  \cite{2001IAUC.7748....3M}       \\
2002bo & NGC3190       &   1396.1  &  Ia     & out   &   \cite{2004MNRAS.348..261B}        \\  
2002cv & NGC3190       &   1396.1  &  Ia     & out  &    \cite{2008MNRAS.384..107E}     \\
2002fk & NGC1309       &   1986.7  &  Ia     & ...    &    \cite{2002IAUC.7976....3A}      \\
2003cg & NGC3169       &   1236.7  &  Ia     & in	  &  \cite{ 2006MNRAS.369.1880E}     \\
2003gs & NGC 936       &   1274.3  &  Ia     & in	  &   \cite{2003IAUC.8173....3G}\\
2003hv & NGC1201       &   1490.3  &  Ia     & in    &        \cite{2007ApJ...661..995G}  \\
2003hx & NGC2076       &   1967.1  &  Ia     & ...    &  \cite{2008MNRAS.tmp..896M} \\
2003if & NGC1302       &   1504.8  &  Ia     & ... 	  &    \cite{2003IAUC.8206....2M}   \\
2004W  & NGC4649       &   1197.8  &  Ia     & in    &   \cite{2004IAUC.8286....2M} \\
2004ab & NGC5054       &   1706.6  &  Ia     & in 	  &    \cite{2004IAUC.8293....2M}     \\
2004ea & M-03-11-19    &   1791.3  &  Ia     & ... 	  &    \cite{2004IAUC.8409....2G}     \\
2005cf & M-01-39-03    &   1976.8  &  Ia     & ...     &    \cite{2007MNRAS.376.1301P} \\
2005cn & NGC5061       &   1946.6  &  Ia     & out  & \cite{2005IAUC.8553....2M} \\
2005df & NGC1559       &   1004.4  &  Ia     & in	  &    \cite{2007ApJ...661..995G}      \\
2005ke & NGC1371       &   1272.2  &  Ia     & ...	  &   \cite{2006ApJ...648L.119I}      \\
2006E  & NGC 5338      &   895.4   &  Ia     & ...   &  \cite{2006ATel..690....1A}   \\ 
2006X  & NGC4321       &   1679.1  &  Ia     & in    &  \cite{2008ApJ...675..626W}  \\
2006ce & NGC 908       &   1345.3  &  Ia     & ...   &  \cite{2006CBET..541....1B}   \\
2006dd & NGC 1316      &   1542.2  &  Ia     & in    &  \cite{2006CBET..557....1S} \\
2006mq & ESO 494-G26   &   779.4   &  Ia     &  .. & \cite{2006CBET..731....1P} \\
2006mr & NGC 1316      &   1542.2  &  Ia     & in  &  \cite{2006CBET..729....1P}  \\
2007af &  NGC5584    &   1704.7   &     Ia   & in &  \cite{2007ApJ...671L..25S}   \\   
2007bm &  NGC3672    &   1821.5   &     Ia   & ... & \cite{2007CBET..939....1N}   \\
2007gi &  NGC4036    &   1629     &     Ia   & in  &  \cite{2007CBET.1021....1H} \\
2007le &  NGC7721    &   1972.1   &     Ia   & ... &  \cite{2007CBET.1101....1F}   \\
2007on &  NGC1404    &   1716.7   &     Ia   & in  &  \cite{2008Natur.451..802V}  \\
2007sa &  NGC3499    &   1762.2   &     Ia   & ... &  \cite{2007CBET.1163....1A}   \\
2007sr &  NGC4038    &   1570.4   &     Ia   & in  & \cite{2007CBET.1174....1U}  \\
\\
{\bf Unclassified} \\
1998cf &  NGC3504     &   1668.9  &  ...    &    in  \\ %
1999gs & NGC4725      &  1206 &   ..   & out &   \\
\\
{\bf Uncertain} \\
M85 OT2006-1     &  M85       &  957        &   SN or OT?  & in  & Nature uncertain, See Sect. 2.4  \\
2008S             &  NGC6946   &  318        &   IIn or OT? & out  & Nature uncertain, See Sect. 2.4  \\
NGC300 OT2008-1   &  NGC300    &  100        &   SN or OT?  & in  & Nature uncertain, See Sect. 2.4  \\
\end{longtable}
\label{lastpage}

\end{document}